\documentstyle[12pt,fullpage]{article}

\newcommand{\bra}{\langle}
\newcommand{\ket}{\rangle}
\def\L{{\cal L}}
\def\K{{\cal K}}

\def\P{{\cal P}}
\def\D{{\cal D}}
\def\Z{{\cal Z}}
\def\N{{\cal N}}

\def\T{{\cal T}}

\begin{document}
\parindent=0cm 
\parskip=10pt
\baselineskip 7mm
\input feynman
\parskip=10pt
\begin{flushright}
TAUP-2227-95
\end{flushright}
\vskip 3 true cm

\begin{center}
{\bf Off-Shell Quantum Electrodynamics}

M. C. Land  and L. P. Horwitz

School of Physics and Astronomy  \\
Raymond and Beverly Sackler Faculty of Exact Sciences  \\
Tel Aviv University, Ramat Aviv, Israel 

\end{center}
\vskip .5 true cm
\parindent=0cm 

\begin{abstract}
\parindent=0cm 
\parskip=10pt
More than twenty years have passed since the threads of the `proper 
time formalism' in covariant classical and quantum mechanics were
brought together to construct a canonical formalism for the relativistic 
mechanics of many particles.  Drawing on the work of Fock, Stueckelberg, 
Nambu, Schwinger, and Feynman, the formalism was raised from the status
of a purely formal mathematical technique to a covariant evolution theory
for interacting particles. In the context of this theory, solutions have
been found for the relativistic bound state problem, classical and quantum
scattering in relativistic potentials, as well as applications in
statistical mechanics.

It has been shown that a generalization of the Maxwell theory is required
in order that the electromagnetic interaction be well-posed in the theory.
The resulting theory of electromagnetism involves a fifth gauge field
introduced to compensate for the dependence of the gauge transformation
on the invariant time parameter; permitting such dependence relaxes the
requirement that individual particles be on fixed mass shells and allows
exchange of mass during scattering. In this paper, we develop the quantum
field theory of off-shell electromagnetism, and use it to calculate certain
elementary processes, including Compton scattering and M{\o}ller scattering.
These calculations lead to {\em qualitative} deviations from the usual
scattering cross-sections, which are, however, small effects, but may be
visible at small angles near the forward direction. The familiar IR
divergence of the M{\o}ller scattering is, moreover, completely regularized.
\end{abstract}
\newpage
\section{Introduction}
\setcounter{equation}{0}
{\bf Covariant Quantum Mechanics}  

In 1941, Stueckelberg \cite{Stueckelberg} (see also Fock
\cite{Fock}) proposed a Poincar\'e invariant Hamiltonian
mechanics in which particle worldlines are traced out by the
evolution of events according to an invariant parameter $\tau$.
Stueckelberg's purpose was to describe pair annihilation as the
evolution of a single worldline, first forward and then backward
in time. Since the Einstein time coordinate $x^0 = t$ does not
increase monotonically as the system evolves in such a model (the
basis of the Feynman-Stueckelberg interpretation of
anti-particles \cite{S-F}), it was necessary to replace $x^0$
with a new order parameter. The Poincar\'e invariant parameter
$\tau$ is formally similar to the Galilean invariant time in
Newtonian theory, and its introduction permits the adaptation of
many techniques from non-relativistic classical and quantum
mechanics. In the symplectic mechanics which follows from this
formulation, one writes classical Hamilton equations in the form
\begin{equation}
\frac{dx^\mu}{d\tau} = \frac{\partial K}{\partial p_\mu}
\qquad \frac{dp^\mu}{d\tau} = -\frac{\partial K}{\partial x_\mu}
\label{eqn:1.1}
\end{equation}
where
\begin{equation}
g_{\mu\nu} = {\rm diag} (-1,1,1,1) \qquad {\rm and} \qquad
\mu,\nu = 0, \cdots ,3 \ ,
\label{eqn:1.2}
\end{equation}
and $K$ is the analog of the Hamiltonian which generates system
evolution according to $\tau$.  For the free particle, one may
choose the Hamiltonian
\begin{equation}
K=\frac{p^\mu p_\mu}{2M} \ ,
\label{eqn:1.3}
\end{equation}
find the equations of motion
\begin{equation}
\frac{dt}{d\tau} = \frac{E}{M} \qquad \qquad \frac{d\vec{x}}{d\tau} =
\frac{\vec{p}}{M} \qquad \qquad p^\mu = {\rm constant} \ ,
\label{eqn:1.4}
\end{equation}
and recover $d\vec{x}/dt = \vec{p}/E$ in the usual form.  Since 
$m^2$ is constant for the free particle, the
proper time of the motion
\begin{equation}
ds^2 = d\vec{x} \ ^2 - dt^2 = \frac{p^2}{M^2} d\tau^2
= -\frac{m^2}{M^2} d\tau^2
\label{eqn:1.5}
\end{equation}
is proportional to the invariant time $\tau$ in this case.

As in the non-relativistic case, one makes the transition from
classical to quantum mechanics by regarding the Hamiltonian as
the Hermitian generator of unitary $\tau$ evolution and writing
\begin{equation}
i \frac{\partial}{\partial \tau} \psi_\tau (x) = K \psi_\tau (x)
\label{eqn:1.6}
\end{equation}
as a covariant Schr\"odinger equation with first order $\tau$
evolution.  The squared magnitude of the wavefunction, $\Bigl|
\psi_\tau (x) \Bigr| ^2$, may be interpreted as a probability
density, at $\tau$, of finding the event at $x$.  For the
Hamiltonian (\ref{eqn:1.3}), this density satisfies
the conservation law
\begin{equation}
\partial_\tau \Bigl| \psi_\tau (x) \Bigr| ^2 = \partial_\mu
\left\{ \frac{i}{2M}[\psi^{*} \partial^{\mu} \psi -
\psi\partial^{\mu}\psi^{*}]\, \right\} \ ,
\label{eqn:1.7}
\end{equation}
and so the integral of $\Bigl| \psi_\tau (x) \Bigr| ^2$ over
spacetime is conserved in $\tau$.  Eq. (\ref{eqn:1.7})
corresponds to the current conservation law
\begin{equation}
\partial_\mu j^\mu + \frac{\partial \rho}{\partial \tau} =0
\ ,
\label{eqn:1.7b}
\end{equation}
where
\begin{equation}
j^\mu = - \frac{i}{2M}[\psi^{*} \partial^{\mu} \psi -
\psi\partial^{\mu}\psi^{*}] \qquad {\rm and} \qquad \rho
= \Bigl| \psi_\tau (x) \Bigr| ^2 \ .
\label{eqn:1.7c}
\end{equation}
The Green's function for the
Schr\"odinger equation propagates the wavefunction monotonically
from $\tau_1$ to $\tau_2$, for $\tau_2 > \tau_1$ --- although
$x^0(\tau_2)$ need not be greater than $x^0(\tau_1)$.

In order to exploit the advantages of a covariant canonical
formalism with invariant evolution, and the familiar methods
of non-relativistic mechanics,
both Schwinger and Feynman introduced an invariant
time parameter --- as a formal technique --- in their work
on quantum electrodynamics. In 1951, Schwinger \cite{Schwinger}
represented the Green's functions of the Dirac field as a
parametric integral and formally transformed the Dirac problem
into a dynamical theory in which the integration parameter acts
as a proper time according to which a Hamiltonian operator
generates the evolution of the system through spacetime.  This
method was the basis for his calculation of the vacuum
polarization in an external electromagnetic field.  Applying
Schwinger's method to the Klein-Gordon equation, one obtains an
equation for the Green's function (we take $\hbar =1 $ in the
following)
\begin{equation}
G=\frac{1}{(p-eA)^{2}+ m^{2} - i\epsilon}
\label{eqn:1.8}
\end{equation}
given by
\begin{equation}
G(x,x')=\bra x|G|x'\ket =i \int_{0}^{\infty} ds
e^{-i (m^{2} - i\epsilon) s}
\bra x|e^{-i (p-eA)^{2} s}|x'\ket  \ .
\label{eqn:1.9}
\end{equation}
The function
\begin{equation}
G(x,x';s)=\bra x(s)|x'(0)\ket = \bra x|e^{-i (p-eA)^{2} s}|x'\ket
\label{eqn:1.10}
\end{equation}
satisfies
\begin{equation}
i \frac{\partial}{\partial s}\bra x(s)|x'(0)\ket =
(p-eA)^{2}\bra x(s)|x'(0)\ket
\label{eqn:1.11}
\end{equation}
with the boundary condition
\begin{equation}
\lim_{s \rightarrow 0} \bra x(s)|x'(0)\ket  = \delta ^{4} (x-x')
\ .
\label{eqn:1.12}
\end{equation}
Schwinger regarded $x^{\mu}(s)$ and
$\pi^{\mu}(s)=p^{\mu}(s)-eA^{\mu}(s)$
as operators, in a Heisenberg picture, which satisfy
\begin{equation}
[x^{\mu},\pi^{\nu}]=ig^{\mu \nu} \qquad \qquad
[\pi^{\mu},\pi^{\nu}]=ieF^{\mu \nu}
\label{eqn:1.13}
\end{equation}
$$i[x^{\mu},K]=-\frac{\partial x^{\mu}}{\partial s} \qquad \qquad
i[\pi^{\mu},K]=-\frac{\partial \pi^{\mu}}{\partial s} \ . $$
Equivalently, DeWitt \cite{DeWitt} regarded (\ref{eqn:1.11}) as
defining the Green's function for the Schr\"odinger equation
\begin{equation}
i \frac{\partial}{\partial s}\psi_{s}(x)=K\psi_{s}(x)=
(p-eA)^{2}\psi_{s}(x) \ ,
\label{eqn:1.14}
\end{equation}
which (with the inclusion of a local metric tensor) he used for
quantum mechanical calculations in curved space. The mini
superspace formulation under investigation recently are of this
type. Notice that (\ref{eqn:1.14}) is in the form written by
Stueckelberg in (\ref{eqn:1.6}), using the Hamiltonian of
(\ref{eqn:1.3}) with $2M=1$ and the minimal substitution $p^\mu
\rightarrow p^\mu -eA^\mu$.

Feynman \cite{Feynman} used an expression identical to
(\ref{eqn:1.14}) in his derivation of the path integral for the
Klein-Gordon equation. He regarded the integration of the Green's
function with the weight $e^{-im^{2} s} $, as the requirement
(see also Nambu \cite{Nambu}) that asymptotic solutions of the
Schr\"odinger equation be stationary eigenstates of the mass
operator $i\partial_\tau$. From this point of view, one picks the
mass eigenvalue by extending the lower limit of integration in
(\ref{eqn:1.9}) from $0$ to $-\infty$, and adding the requirement
that $G(x,x';s)=0$ for $s<0$. It is worth noting \cite{Feynman}
that the usual Feynman propagator $\Delta_{F} (x-x')$ emerges
naturally from the classical causality condition of retarded
propagation (in which $s$ is the order parameter).

To Schwinger, the principal advantage of the proper time method
is that the physical interactions are independent of the
evolution parameter and so the symmetries of the system are
preserved.  Therefore \cite{Schwinger}, the method provides a
natural approach to perturbation theory and regularization.  In
deriving the vacuum polarization, Schwinger found that all
quantities remained finite as long as the lower limit of
integration in (\ref{eqn:1.9}) is taken to be $s_0 > 0$.  Ball
\cite{ball} demonstrates that the proper time method includes the
usual regularization techniques as subcases and provides the
specific correspondence with Pauli-Villars, point-splitting,
zeta function, and dimensional regularization.
L\"uscher \cite{luscher} used the proper time method to obtain
the dimensional regularization in the presence of large
background fields.  The Casimir effect \cite{casimir} and the
harmonic oscillator \cite{osc} partition function have recently
been studied by inserting the factor $s^\nu$ into (\ref{eqn:1.9})
for $\nu$ large enough to regularize the integral.

{\bf Covariant Two-Body Mechanics With Interactions}  

In 1973, Horwitz and Piron \cite{H-P} constructed a canonical
formalism for the relativistic classical and quantum mechanics of
many particles. In order to formulate a generalized Hamilton's
principle, they introduce a Poincar\'e invariant evolution
parameter $\tau$, which we shall call the {\it world time} and
regard as corresponding to the ordering relation of successive
events in spacetime. This $\tau$ is therefore similar to the
parameter in the formalisms of Stueckelberg, Schwinger, and
Feynman, except that it is regarded as a true physical time, with
the status of the Newtonian time in non-relativistic mechanics.
Moreover, by introducing two-body potentials between particles,
Horwitz and Piron implicitly relaxed the requirement of a
constant proportionality between the {\it world time} and the
proper time of the classical particle motions. Thus, particles
will not remain on mass-shell.

The two-body Hamiltonian in the Horwitz-Piron theory is
\begin{equation}
K = {p_{1\mu}p_1^{\mu} \over {2M_1}} + {p_{2 \mu} p_2^{\mu}
\over {2M_2}} + V(x_1 , x_2)
\label{eqn:1.15}
\end{equation}
and in the case of non-trivial interaction, the $p^\mu$ will not
be constant and the particle masses become dynamical quantities.
Non-relativistic central force problems may be generalized to
covariant form by taking
\begin{equation}
V(x_1 , x_2) =  V(\rho) \qquad {\rm where} \qquad
\rho = \sqrt{({\bf x}_1 - {\bf x}_2)^2 - (t_1 -t_2)^2} \ .
\label{eqn:1.16}
\end{equation}
Since $t_1 \rightarrow t_2 $ in the Galilean limit
($\frac{dt_i}{d\tau} =\frac{E_i}{m_ic^2} \sim 1 $), $V(\rho)
\rightarrow V(r)$ for the corresponding non-relativistic
problem.  As in the non-relativistic problem, the two-body
Hamiltonian is quadratic in the momenta, and one may
separate variables of the center of mass motion and relative
motion,
\begin{equation}
K = {P^{\mu} P_{\mu} \over 2M} + {p^{\mu} p_{\mu}
\over 2m} + V(\rho)
\equiv {P^{\mu} P_{\mu} \over 2M} + K_{rel},
\label{eqn:1.17}
\end{equation}
where
\begin{equation}
P^\mu = p_1^\mu + p_2^\mu \qquad\qquad M = M_1 + M_2
\label{eqn:1.18}
\end{equation}
$$p^\mu = (M_2p_1^\mu - M_1 p_2^\mu)/M \qquad\qquad m=M_1M_2/M.$$
Solutions of the Schr\"odinger equation have been found for the
relativistic bound state \cite{I,II} and for scattering
potentials \cite{scattering}.
For bound states, the mass spectrum coincides with the
non-relativistic Schr\"odinger energy spectrum, and so it follows
that for small excitations, the corresponding energy spectrum is
that of the non-relativistic Schr\"odinger theory with
relativistic corrections.  To obtain these spectra, one must
choose a spacelike unit vector $n_\mu$ and restrict the support
of the eigenfunctions in spacetime to the subspace of the
Minkowski measure space for which the component of the relative
coordinate $x^\mu_1 - x^\mu_2$ normal to $n_\mu$ is spacelike.
The restricted space is transitive and invariant under the O(2,1)
subgroup of O(3,1) leaving $n_\mu$ invariant and translations
along $n_\mu$.
Mathematically, this restriction is related to the existence of
discrete unitary representations of the Lorentz group in this
subspace \cite{Zmuidzinas}.  Physically, this restriction leads to a
lowering of the mass spectrum (compare \cite{Cook}).
It was shown in \cite{I} that the eigenfunctions
of $K_{rel}$ form irreducible representations of SU(1,1) --- in
the double covering of O(2,1) --- parameterized by the spacelike
vector $n_\mu$ stabilized by this particular O(2,1).  In
\cite{II}, an induced representation of SL(2,C) was constructed,
by applying the Lorentz group to the coordinates $x^\mu$ of the
restricted space and the frame orientation $n_\mu$, and studying
the action on these wavefunctions.

In \cite{selrul}, the selection rules for dipole
radiation from these states are calculated and shown to be
identical with those of the usual non-relativistic theory, but
with manifestly covariant interpretation.  Moreover, it is shown
that the change in the magnetic quantum number corresponds to a
change in the orientation of $n_\mu$ with respect to the
polarization of the emitted or absorbed photon.  The group
theoretical aspects of this bound state recoil are discussed
in the context of the induced representation of SL(2,C).

In \cite{zeeman}, we provide a derivation of the normal Zeeman
effect for the bound state, which requires $n_\mu$ to become a
dynamical quantity. We begin with a discussion of the classical
O(3,1) in the induced representation and obtain the group
generators, which coincide with those of \cite{II}, when the
momenta are understood as derivatives in the Poisson bracket
sense. We construct a classical Lagrangian, in which $n_\mu$
plays an explicit dynamical role, and show that the generators
are conserved. We then construct the Hamiltonian, which may be
unambiguously quantized and made locally gauge invariant.
Finally, it is shown that an external gauge field representing a
constant magnetic field induces, as a first order perturbation, a
mass level splitting corresponding to the usual non-relativistic
expression. 

{\bf Local Gauge Theory}

In the context of the covariant mechanics of Horwitz and Piron,
Saad, Horwitz, and Arshansky have argued \cite{saad} that the
local gauge function of the field should include dependence on
$\tau$, as well as on the spacetime coordinates. This requirement
of full gauge covariance leads to a theory of five gauge
compensation fields, which differs in significant aspects from
conventional electrodynamics, but whose zero modes coincide with
the Maxwell theory (see also \cite{others}). 

Under local gauge transformations of the form
\begin{equation}
\psi(x,\tau) \rightarrow e^{i e_{0} \Lambda (x,\tau)} \psi (x,\tau)
\label{eqn:1.19}
\end{equation}
the equation
\begin{equation}
(i \partial_{\tau} +e_{0}a_{5})\psi(x,\tau)
=\frac{1}{2M}(p^{\mu}-e_{0}a^{\mu})(p_{\mu}-
e_{0}a_{\mu})\psi(x,\tau)
\label{eqn:1.20}
\end{equation}
is invariant, when the compensation fields transform as
\begin{equation}
a_{\mu}(x,\tau) \rightarrow
a_{\mu}(x,\tau)+\partial_{\mu}\Lambda(x,\tau)
\qquad
a_{5}(x,\tau) \rightarrow a_{5}(x,\tau) + \partial_{\tau}
\Lambda(x,\tau)  \ ,
\label{eqn:1.21}
\end{equation}
where the potentials $a_\mu (x,\tau)$ and $a_{5}(x,\tau)$ must
now also depend explicitly on the {\it world time} $\tau$. This
Schr\"odinger equation (\ref{eqn:1.20}) leads (as for
(\ref{eqn:1.7b})) to the five dimensional conserved current
\begin{equation}
\partial_{\mu}j^{\mu}+\partial_{\tau}j^{5}=0
\label{eqn:1.22}
\end{equation}
where
\begin{equation}
j^{5} \equiv \rho = \Bigl| \psi(x,\tau) \Bigr| ^{2} \qquad
j^{\mu} = \frac{-i }{2M} \Bigl\{ \psi^{*}(\partial^{\mu}-i
e_{0}a^{\mu})\psi - \psi(\partial^{\mu}+i e_{0}a^{\mu})\psi^{*}
\Bigr\} \ ,
\label{eqn:1.23}
\end{equation}
so that, as in Stueckelberg's formulation, $\Bigl| \psi(x,\tau)
\Bigr| ^{2}$ may be interpreted as a probability density.  The
current conservation law may be written as
$\partial_{\alpha}j^{\alpha}=0$, where the index convention
is now
\begin{equation}
\lambda,\mu,\nu = 0,1,2,3 \qquad\qquad {\rm and} \qquad\qquad
\alpha,\beta,\gamma=0,1,2,3,5
\label{eqn:1.24}
\end{equation}
and the parameter $\tau$ is formally designated $x^5$, so that
$\partial_\tau = \partial_5$.

In \cite{emlf}, we obtain the classical Lorentz force,
by using (\ref{eqn:1.20}) to write the classical Hamiltonian and 
Lagrangian.  The result,
\begin{equation}
M \: \ddot x^\mu = e_0 \ f^\mu_{\:\;\;\alpha}(x,\tau) \, \dot
x^\alpha \qquad \qquad \frac{d}{d\tau} (-\frac{1}{2} M \dot x^2) =
e_0 \ f_{5\alpha} \dot x^\alpha \ ,
\label{eqn:1.25}
\end{equation}
where $f_{\alpha\beta}$ is given by the gauge invariant quantity
\begin{equation}
f_{\alpha\beta} =\partial_\alpha a_\beta - \partial_\beta
a_\alpha \ ,
\label{eqn:1.26}
\end{equation}
and where we use the fact that $\dot x^5 \equiv 1$, shows that
the new field strength tensor components $f_{5\alpha}$ act on the
classical particle motions in this theory. It follows from the
second of (\ref{eqn:1.25}) that mass need not be conserved in
this theory, even at the classical level, and that pair
annihilation is classically permitted \cite{Stueckelberg}. It was
shown in reference \cite{emlf}, by a study of the
energy-momentum-mass tensor for the classical motions, that the
total energy, momentum, and mass of the particles plus fields are
conserved.

The Schr\"odinger equation (\ref{eqn:1.20}) may be derived by
variation of the action
\begin{equation}
{\rm S} = \int d^4 x d\tau \left\{ 
\psi^* (i\partial_\tau + e_0 a_5 ) \psi - 
\frac{1}{2M} \psi^* (p_\mu - e_0 a_\mu )
    (p^\mu - e_0 a^\mu )\psi 
- \frac{\lambda}{4} f_{\alpha\beta}f^{\alpha\beta} \right\}
\label{eqn:1.27}
\end{equation}
to which Sa'ad, et.\  al.\ have added a kinetic term for the fields,
formed from the gauge invariant quantity $f_{\alpha\beta}$.
In writing the kinetic term, one must formally raise the index
$\beta = 5$ in the term
\begin{equation}
f_{\mu 5} =\partial_\mu a_5 - \partial_\tau a_\mu \ ,
\label{eqn:1.28}
\end{equation}
and Sa'ad, et.\ al.\ argue that this term suggests a higher symmetry.
Since this symmetry must contain O(3,1), it can be O(4,1) or
O(3,2), which correspond respectively to a signature
$g^{55} = \sigma = \pm 1$, and they write the metric for the free
field as
\begin{equation}
g^{\alpha\beta} = {\rm diag}(-1,1,1,1,\sigma) \ .
\label{eqn:1.29}
\end{equation}
Varying the action (\ref{eqn:1.27}) with respect to the gauge
fields, the equations of motion are found to be
\begin{eqnarray}
\partial_{\beta} f^{\alpha \beta}
&=&\frac{e_{0}}{\lambda}j^{\alpha}=ej^{\alpha}
\label{eqn:1.30}
\\
\epsilon^{\alpha \beta \gamma \delta
\epsilon}\partial_{\alpha}f_{\beta \gamma}&=&0
\label{eqn:1.31}
\end{eqnarray}
where $j^{\alpha}$ is given by (\ref{eqn:1.23}). As we show
below, $\lambda$ and $e_0$ are dimensional constants; one
identifies $e_0 / \lambda $ as the dimensionless Maxwell charge
$e$. The sourceless gauge field equations inherit the formal five
dimensional symmetry of the free gauge field Lagrangian, while
the {\it physical} Lorentz covariance of the matter currents
breaks the O(4,1) or O(3,2) symmetry of the free fields to
O(3,1). Nevertheless, the wave equation associated with the
fields is \cite{saad}
\begin{equation}
\partial_{\alpha}\partial^{\alpha}f^{\beta
\gamma}=(\partial_{\mu}\partial^{\mu}+\partial_{\tau}
\partial^{\tau})f^{\beta \gamma}=
(\partial_{\mu}\partial^{\mu}+ \sigma \; \partial_{\tau}^2)
f^{\beta \gamma}
= -e(\partial^{\alpha}j^{\beta} - 
\partial^{\beta}j^{\alpha}) \ ,
\label{eqn:1.32}
\end{equation}
and the causal properties of the (free) Green's functions for the
operator on the left hand side of (\ref{eqn:1.32}) reflect
the higher symmetry.  The dependence of the wave equation on
the signature $\sigma$ of $\partial_\tau$ implies that these
causal properties will be different for the symmetry groups
O(3,2) and O(4,1).

In \cite{green}, we derive the Green's
functions; the O(3,2) case supports spacelike and lightlike
correlations through spacetime, without permitting superluminal
transmission of information.  The O(4,1) case contains timelike
and lightlike correlations; the timelike correlations similarly
do not carry information in the usual sense. 

Under the boundary conditions $j^5 \rightarrow 0$, pointwise, as
$\tau \rightarrow \pm \infty$, integration of (\ref{eqn:1.23})
over $\tau$, leads to $\partial_{\mu}J^{\mu}=0$, where
\begin{equation}
J^{\mu}(x)=\int_{-\infty}^{\infty} d\tau j^{\mu}(x,\tau)
\label{eqn:1.33}
\end{equation}
so that we may identify $J^{\mu}$ as the source of the Maxwell
field.  Similarly, under the boundary conditions $f^{5\mu}
\rightarrow 0$ pointwise in $x$ as $\tau \rightarrow \pm \infty$, 
integration of (\ref{eqn:1.30}) over $\tau$ leads to the Maxwell
equations, in the form,
\begin{equation}
\partial_{\nu}F^{\mu \nu}=eJ^{\mu} \qquad\qquad
\epsilon^{\mu \nu \rho \lambda }\partial_{\mu}F_{\nu \rho}=0
\label{eqn:1.34}
\end{equation}
where
\begin{equation}
F^{\mu \nu}(x)=\int_{-\infty}^{\infty} d\tau f^{\mu \nu}(x,\tau)
\qquad\qquad {\rm and} \qquad\qquad
A^{\mu}(x)=\int_{-\infty}^{\infty} d\tau a^{\mu}(x,\tau)
\label{eqn:1.35}
\end{equation}
so that $a^{\alpha}(x,\tau)$ has been called the pre-Maxwell
field.  It follows from (\ref{eqn:1.35}) that $e_0$ and
$\lambda$ have dimensions of length. 

In the pre-Maxwell theory, interactions take place between events
in spacetime rather than between worldlines. The resulting system
of equations is integrable. Each event, occurring at $\tau$,
induces a current density in spacetime which, for free particles
(and hence asymptotically), disperses for large $\tau$, and the
continuity equation (\ref{eqn:1.22}) states that these current
densities evolve as the event density $j^{5}$ progresses through
spacetime as a function of $\tau$. As noted above, if
$j^{5}\rightarrow 0$ as $|\tau| \rightarrow \infty$, then
$j^{\mu}$ may be identified with the Maxwell current when
integrated over $\tau$. This integration has been called
concatenation \cite{concat} and provides the link between the
event along a worldline and the notion of a particle, whose
support is the entire worldline. Concatenation is evidently
related to the integration performed in the Schwinger proper time
method, and following Feynman's interpretation, imposes the
requirement that the Maxwell electromagnetic field be the zero
mode with respect to the conjugate mass variable. The Maxwell
theory thus has the character of an {\it equilibrium limit} of
the microscopic pre-Maxwell theory. Shnerb and Horwitz
\cite{nadav} have given an interpretation of the dimensional
constant $\lambda$ as a coherence length. In this sense, the
Maxwell theory is a correlation limit of the pre-Maxwell theory,
in which it is properly contained. Frastai and Horwitz
\cite{jaime} have shown that close to this limit, the strongest
singularities of one-loop diagrams are regularized in a manner
similar to that of Pauli and Villars \cite{P-V}.

{\bf Feynman's Approach to the Foundations of Gauge Theory}  

In 1948, Feynman showed Dyson how the Lorentz force law and
homogeneous Maxwell equations could be derived from commutation
relations among Euclidean coordinates and velocities, without
reference to an action or variational principle. When Dyson
published the work in 1990 \cite{Dyson}, several authors
\cite{comments,vaidya,H-S,Hughes} noted that the derived
equations have only Galilean symmetry and so are not actually the
Maxwell theory. In a more recent paper, Tanimura \cite{Tanimura}
generalized Feynman's derivation to a Lorentz covariant form with
scalar evolution parameter, and obtained an expression for the
Lorentz force which appears to be consistent with relativistic
kinematics and relates the force to the Maxwell field in the
usual manner. However, Tanimura's derivation does not lead to the
usual Maxwell theory either, but rather to the off-shell
pre-Maxwell theory described here \cite{beyond}
Hojman and Shepley \cite{H-S}
proved that the existence of commutation relations is a strong
assumption, sufficient to determine the corresponding action,
which for Feynman's derivation is of Newtonian form. In
\cite{beyond}, we examine Tanimura's derivation in the framework
of the proper time method, and use the technique of Hojman and
Shepley to study the unconstrained commutation relations. In this
context, we explain Tanimura's observations that the invariant
evolution parameter cannot be consistently identified with the
proper time of the particle motion, and that the derivation
cannot be made reparameterization invariant. Using the techniques
of Tanimura and of Hojman and Shepley, we obtain the form of the
pre-Maxwell theory in a background curved space and in the
presence of a classical non-Abelian gauge field. 

{\bf Off-Shell Quantum Electrodynamics}  

Shnerb and Horwitz \cite{nadav} have presented a consistent
canonical quantization of off-shell electromagnetism which
preserves the theory's formal five dimensional symmetry, using a
bosonic gauge fixing method. Frastai and Horwitz \cite{jaime}
used a path integral quantization to study the theory in the near
on-shell limit. In this paper, we develop the methods required to
apply off-shell quantum electrodynamics as a scattering theory,
and calculate certain elementary processes. In Section 2, we
perform a canonical quantization of the interacting
off-shell theory, which
takes advantage the fact that it is a parameterized evolution
theory. We use a procedure advocated by Jackiw \cite{jackiw} for
theories in which the action can be made linear in time
derivatives, and obtain results in complete agreement with those
of Shnerb and Horwitz.
For conventional quantum theories, in which the
momentum is related to a $t$-derivative, putting the action into
first order form may break manifest Lorentz covariance. For the
off-shell theory, in which the classical velocity is a
$\tau$-derivative, Jackiw's approach is natural and manifestly
covariant throughout. In Section 3, we show how Jackiw's approach
leads to a path integral quantization in a equally natural
way.  Having identified the essential degrees of freedom, we
perform the detailed Fourier expansion of the operators of the
free fields, required in perturbation theory.
In Section 4, we carry out a canonical quantization of the free
off-shell matter field, and provide an expansion of the field
operators in terms of annihilation operators in momentum space.
We use these expansions to show that the vacuum expectation value of
$\tau$-ordered fields is precisely the Stueckelberg-Schr\"odinger
equation Green's function, with Schwinger-Feynman boundary
conditions. In Section 5, we carry out the canonical quantization
of the five dimensional electromagnetic field and the expansion
of field operators in momentum space. This field has three
independent polarizations, one of which decouples in the Maxwell
(zero mode) limit. We show that vacuum expectation value of
products of two $\tau$-ordered fields is the Feynman Green's
function for the wave equation, and that it preserves the gauge
condition. In Section 6, we develop the perturbation theory for
the S-matrix in the interaction picture. In order to provide the
link with the usual formulas for the perturbation expansion of
Green's functions, we adapt the LSZ reduction formulas to the
off-shell theory. We then derive the Feynman rules. In Section 7,
we derive the scattering cross-section in terms of the transition
amplitude, and demonstrate the relationship of this expression to
the usual cross-section. In Section 8, we use the machinery of
the previous sections to calculate the amplitudes for Compton and
M{\o}ller scattering. We analyze the M{\o}ller scattering
amplitude in detail and examine the qualitative deviation from
the corresponding amplitude calculated in conventional QED.
In Section 9, we consider the problem of renormalization of
the off-theory.  We derive the Ward identity associated with
current conservation and show that it connects the matter
field propagators with both the 3-particle and the
4-particle vertex functions.  We conclude that the off-shell
theory is renormalizable, when a cut-off is placed on the
mass of the off-shell photons.  This cut-off nevertheless
preserves the invariances of the original theory.

%
\section{Canonical Quantization of the \newline Interacting Theory}
\setcounter{equation}{0}

Following \cite{saad} we take the action for off-shell
electromagnetism to be
\begin{equation}
{\rm S} = \int d^4 x d\tau \left\{ 
\psi^* (i\partial_\tau + e_0 a_5 ) \psi - 
\frac{1}{2M} \psi^* (-i\partial_\mu - e_0 a_\mu )
    (-i\partial^\mu - e_0 a^\mu )\psi 
- \frac{\lambda}{4} f_{\alpha\beta}f^{\alpha\beta} \right\}
\label{eqn:2.1}
\end{equation}
where
\begin{equation}
\mu,\nu = 0,\cdots,3 \qquad {\rm and}
\qquad \alpha,\beta = 0,\cdots,3,5
\label{eqn:2.2}
\end{equation}
and
\begin{equation}
a_5 = \sigma a^5 \qquad \qquad g^{\alpha\beta} = 
{\rm diag}(-1,1,1,1,\sigma).
\label{eqn:2.3}
\end{equation}
We now show that the action in (\ref{eqn:2.1}) is hermitian, up
to surface terms which may be discarded from the action.  We
regard 4-divergences as vanishing under spacetime integration
because of their support properties, while $\tau$-derivatives
vanish {\em independently} under $\tau$-integration because of
Riemann-Lebesgue oscillations (the asymptotic mass dependence of
the fields $exp\{-i(m^2/2M) \, \tau\}$ varies arbitrarily rapidly
for large $\tau$).

Thus, for the $\tau$-derivative terms in
(\ref{eqn:2.1}),
\begin{eqnarray}
\frac{1}{2} \left[ \psi^* (i\partial_\tau + e_0 a_5 ) \psi +
{\rm h.c.} \right] &=& \frac{i}{2} \left[ \psi^* \dot \psi - 
\dot \psi^* \psi \right] + e_0 a_5 \; \psi^* \psi
\nonumber \\
&=& i \psi^* \dot \psi + e_0 a_5 \; \psi^* \psi + \partial_\tau
(\psi^* \psi )
\label{eqn:2.4}
\end{eqnarray}
where $\dot \psi = \partial\psi/\partial\tau $.
%
%
%
%
We will henceforth regard (\ref{eqn:2.1}) as hermitian.  

In Dirac's method \cite{Dirac} of quantization for gauge theories,
we would form the Hamiltonian from (\ref{eqn:2.1}), including a
momentum $\pi_5$ conjugate to $a^5$ and a Lagrange multiplier to
enforce the primary constraint $\pi_5 =0$ (in \cite{nadav}, this is
accomplished through the gauge fixing term).  The secondary
constraint (that the primary constraint commute with the Hamiltonian)
would lead to the Gauss Law for the off-shell theory, which is the
$\alpha=5$ term of (\ref{eqn:1.30}).

We will follow a quantization scheme advocated by Jackiw
\cite{jackiw}, in which one first eliminates the constraint from
the Lagrangian, and then constructs the Hamiltonian from the
unconstrained degrees of freedom.  In order to perform the
transformation which eliminates the constrained degrees of
freedom, it is necessary that the action be made linear in the time
derivatives.  For conventional quantum theories, in which
the field's $t$-derivative carries the $0^{\rm th}$ Lorentz index,
the first order form of the
action will not be manifestly covariant, and transformation
properties must be checked after quantization (see also \cite{B-D}).
In
off-shell electromagnetism, this method singles out the
$\tau$-derivatives of the fields, so that the method is
O(3,1)-covariant.  The loss of the formal five-dimensional
symmetry of the free off-shell electromagnetic field does not
constitute a loss of generality for the interacting theory, since
this higher symmetry is not a property of the matter field.
Lorentz covariance is maintained throughout, in a manner
consistent with the original spirit of the proper time method.

In Jackiw's method, a choice of gauge is made implicitly by
solving the constraints and introducing a decomposition of the
fields consistent with that solution. The elimination of the
longitudinal polarizations is carried out as an application of
the Darboux theorem, diagonalizing the Hamiltonian. This
procedure is evidently related to the quantization method
proposed by Fermi in 1932 \cite{sakurai}. Applying Fermi's method
to (\ref{eqn:2.1}), we would perform a gauge transformation
which guarantees the condition $\partial_\mu a^\mu =0$, and
leads
to the Gauss law for the component $a_5$. The presence of sources
in the action would require the decomposition of the fields into
longitudinal terms induced by the sources and transverse
propagating terms with vanishing our-divergence.  We could then
argue that the longitudinal
fields do not satisfy wave equations and need not be quantized;
this would leave the unconstrained transverse fields in the
action, and we would obtain a consistent canonical quantization.
Elements of such a procedure may be recognized in the method
presented below.

We now put the action into the required first order form.  The
kinetic term for the matter field is linear in $\partial_\tau$
by construction.  In order to put the kinetic term for the gauge
field into explicitly canonical form, we rewrite $f^{5\mu}f_{5\mu}$
as $f^{5\mu}(\partial_\tau a_\mu - \partial_\mu a_5)$ and take
the quantity $f^{5\mu}$, to be independent of the fields $a_\alpha$
(this is a variant of the first order Lagrangian form for the
usual electromagnetic field).  Expanding the electromagnetic
term,
\begin{eqnarray}
f_{\alpha\beta}f^{\alpha\beta} &=&
f_{\mu\nu}f^{\mu\nu} + 2 f^{5\mu}f_{5\mu}
\nonumber \\
&=& f_{\mu\nu}f^{\mu\nu} + 2 \sigma f^{5\mu}f^5 \ _{\mu}
\nonumber \\
&=& f_{\mu\nu}f^{\mu\nu} + 2 \sigma \left[ 2
(\sigma \partial_\tau a^\mu - \partial^\mu a^5) f^5 \ _{\mu}
- f^{5\mu}f^5 \ _{\mu} \right] \ ,
\label{eqn:2.10}
\end{eqnarray}
when we vary the action with respect to $f^{5\mu}$, we recover
its relationship to the $a^\alpha$.  Now, performing the
integration by parts, we obtain
\begin{equation}
(\sigma \partial_\tau a^\mu - \partial^\mu a^5) f^5 \ _{\mu}
= \sigma (\partial_\tau a^\mu) f^5 \ _{\mu} +
a^5 \partial^\mu f^5 \ _{\mu} - {\rm divergence} \ ;
\label{eqn:2.12}
\end{equation}
we drop the divergence and introduce the notation
\begin{equation}
\epsilon^\mu = f^{5\mu} \ .
\label{eqn:2.11}
\end{equation}
Using (\ref{eqn:2.12}) and (\ref{eqn:2.11}), the action becomes
\begin{eqnarray}
{\rm S} &=& \int d^4 x d\tau \Bigl[ 
\psi^* (i\partial_\tau + e_0 a_5 ) \psi - 
\frac{1}{2M} \psi^* (-i\partial_\mu - e_0 a_\mu )
    (-i\partial^\mu - e_0 a^\mu )\psi \nonumber \\
& & \mbox{\qquad\qquad} - \frac{\lambda}{4} f_{\mu\nu}f^{\mu\nu} 
+ \frac{\lambda\sigma}{2} \epsilon^\mu \epsilon_\mu 
- \lambda \epsilon_\mu \partial_\tau a^\mu 
- \lambda a_5 \partial^\mu \epsilon_\mu \Bigr]
\nonumber \\
&=& \int d^4 x d\tau \Bigl[ i \psi^* \dot \psi  
- \lambda \epsilon_\mu \dot a^\mu 
- \frac{1}{2M} \psi^* (-i\partial_\mu - e_0 a_\mu )
(-i\partial^\mu - e_0 a^\mu )\psi
\nonumber \\
& & \mbox{\qquad\qquad} 
- \frac{\lambda}{4} f_{\mu\nu}f^{\mu\nu} + \frac{\lambda\sigma}{2} 
\epsilon^\mu \epsilon_\mu +
 a_5 ( e_0 \psi^* \psi - \lambda \partial^\mu \epsilon_\mu
 )\Bigr]
\label{eqn:2.13}
\end{eqnarray}
where in the second line we have collected terms in $a_5$.
Following Jackiw, we regard (\ref{eqn:2.13}) as an action 
for the conjugate pairs $\{i\psi^* , \psi \}$ and $\{\epsilon_\mu ,
a^\mu \}$, with $a_5$ playing the role of a Lagrange multiplier for
the constraint\footnote{It will been shown below that the
resulting quantized fields satisfy equations corresponding
to the classical theory in the $a_5=0$ gauge.}
\begin{equation}
e_0 \psi^* \psi - \lambda \partial^\mu \epsilon_\mu = 0 \quad \Rightarrow 
\quad \partial^\mu \epsilon_\mu = \frac{e_0 }{\lambda}  \psi^* \psi
= e \rho,
\label{eqn:2.14}
\end{equation}
which is just the Gauss law for the off-shell theory.
The constraint equation (\ref{eqn:2.14}) can be solved through the 
decomposition
\begin{equation}
\epsilon^\mu = (\epsilon_\perp)^\mu +
e \partial^\mu [G\rho]
\label{eqn:2.15}
\end{equation}
where
\begin{equation}
\partial_\mu (\epsilon_\perp)^\mu = 0 
\label{eqn:2.16}
\end{equation}
and where $G\rho$ is a shorthand for the functional 
\begin{equation}
[G\rho](x,\tau) = \int d^4 y G(x-y) \ \rho(y,\tau)
\label{eqn:2.17}
\end{equation}
in which we specify the Green's function 
\begin{equation}
G(x-y) = \delta \left( (x-y)^2 \right) \quad \Rightarrow \quad
\Box G = 1.
\label{eqn:2.18}
\end{equation}
Performing a similar decomposition of $a^\mu$,
\begin{equation}
a^\mu = (a_\perp)^\mu + \partial^\mu [G\Lambda] \qquad 
\partial_\mu (a_\perp)^\mu =0
\label{eqn:2.19}
\end{equation}
(by which we implicitly choose the gauge condition
$\partial_\mu a^\mu =\Lambda$), the remaining terms of the
theory are expressed as
\parskip=0 pt
\begin{equation}
\dot a^\mu = (\dot a_\perp)^\mu + \partial^\mu [G\dot \Lambda],
\label{eqn:2.21}
\end{equation}
\begin{equation}
f^{\mu\nu} =
\partial^\mu (a_\perp)^\nu - \partial^\nu (a_\perp)^\mu =
(f_\perp)^{\mu\nu},
\label{eqn:2.22}
\end{equation}
\begin{equation}
 -i\partial^\mu - e_0 a^\mu = -i\partial^\mu - e_0 (a_\perp)^\mu 
- e_0 \partial^\mu [G\dot \Lambda],
\label{eqn:2.23}
\end{equation}
\begin{eqnarray}
\epsilon^\mu \epsilon_\mu &=& 
[\ (\epsilon_\perp)^\mu + e \partial^\mu [G\rho] \ ]
[\ (\epsilon_\perp)_\mu + e \partial_\mu [G\rho] \ ] \nonumber \\
&=& (\epsilon_\perp)^\mu (\epsilon_\perp)_\mu 
- e^2 \rho [G\rho] + {\rm divergence},
\label{eqn:2.24}
\end{eqnarray}
%

%
%
%
%
%
\begin{equation}
\epsilon_\mu \dot a^\mu = (\epsilon_\perp)_\mu (\dot a_\perp)^\mu 
- e \rho [G\dot\Lambda] \ .
\label{eqn:eqn:2.25}
\end{equation}

\parskip=10 pt
In terms of this decomposition, the action becomes
\begin{eqnarray}
{\rm S} &=& \int d^4 x d\tau \Bigl\{ i \psi^* \dot \psi  
- \lambda (\epsilon_\perp)_\mu (\dot a_\perp)^\mu
+ \lambda e \rho [G\dot\Lambda]
\nonumber \\
& & \mbox{\qquad\qquad} 
- \frac{1}{2M} \psi^* (-i\partial_\mu - e_0 (a_\perp)_\mu
- e_0 \partial_\mu [G\dot\Lambda] )
(-i\partial^\mu - e_0 (a_\perp)^\mu - e_0 \partial^\mu
[G\dot\Lambda])\psi
\nonumber \\
& & \mbox{\qquad\qquad} 
- \frac{\lambda}{4} (f_\perp)_{\mu\nu}(f_\perp)^{\mu\nu}
 + \frac{\lambda\sigma}{2} 
(\epsilon_\perp)^\mu (\epsilon_\perp)_\mu -
\frac{\lambda\sigma}{2} e^2 \rho [G\rho]
\Bigr\} \ .
\label{eqn:2.26}
\end{eqnarray}
We now perform the gauge transformation
\begin{equation}
\psi \longrightarrow e^{i e_0 [G\Lambda]} \psi 
\label{eqn:2.27}
\end{equation}
which entails
\begin{equation}
\dot \psi \longrightarrow e^{i e_0 [G\Lambda]} [\dot \psi
+ i e_0 [G\dot \Lambda] \psi]
\qquad \qquad i\psi^*\dot\psi \longrightarrow i\psi^*\dot\psi
- e_0 \rho [G\dot\Lambda]
\label{eqn:2.28}
\end{equation}
and
\begin{eqnarray}
( -i\partial^\mu - e_0 (a_\perp)^\mu 
- e_0 \partial^\mu [G\dot \Lambda])
\psi &\longrightarrow& ( -i\partial^\mu - e_0 (a_\perp)^\mu 
- e_0 \partial^\mu [G\dot \Lambda]) e^{i e_0 [G\Lambda]} \psi
\nonumber \\
&=& e^{i e_0 [G\Lambda]}
(-i\partial^\mu - e_0 (a_\perp)^\mu 
- e_0 \partial^\mu [G\dot \Lambda] + e_0 \partial^\mu [G\dot
\Lambda]) \psi \nonumber \\
&=&  e^{i e_0 [G\Lambda]}(-i\partial^\mu - e_0 (a_\perp)^\mu )
\psi \ .
\label{eqn:2.29}
\end{eqnarray}
This transforms the action to the form
\begin{eqnarray}
{\rm S} &=& \int d^4 x d\tau \Bigl\{ i \psi^* \dot \psi  
- \lambda (\epsilon_\perp)_\mu (\dot a_\perp)^\mu
- \frac{1}{2M} \psi^* (-i\partial_\mu - e_0 (a_\perp)_\mu)
(-i\partial^\mu - e_0 (a_\perp)^\mu)\psi
\nonumber \\
& & \mbox{\qquad\qquad} 
- \frac{\lambda}{4} (f_\perp)_{\mu\nu}(f_\perp)^{\mu\nu}
 + \frac{\lambda\sigma}{2} 
(\epsilon_\perp)^\mu (\epsilon_\perp)_\mu -
\frac{\lambda\sigma}{2} e^2 \rho [G\rho]
 \Bigr\} \ .
\label{eqn:2.30}
\end{eqnarray}
Notice that (\ref{eqn:2.30}) is an unconstrained functional (only
the unconstrained field degrees of freedom are present) and so may
be canonically quantized.  Henceforth, we may drop the subscript
$_{\perp}$ from the quantized gauge field variables and assume the
transversality conditions (\ref{eqn:2.16}) and (\ref{eqn:2.19})
for the field operators.

The conjugate momenta are found from (\ref{eqn:2.30}) to be
\begin{eqnarray}
\pi_{\psi} &=& \frac{\partial \L}{\partial \dot \psi} = i \psi^*
\label{eqn:2.31}\\
\pi_{a_\mu} &=& \frac{\partial \L}{\partial \dot a_\mu} = -\lambda
\epsilon^\mu
\label{eqn:2.32}
\end{eqnarray}
from which we compute the Hamiltonian
\begin{eqnarray}
\K &=& \pi_{\psi} \dot \psi + \pi_{a_\mu} \dot a_\mu - \L
\nonumber \\
&=& i\psi^*\dot \psi -\lambda \epsilon^\mu \dot a_\mu -
\Bigl\{i \psi^* \dot \psi  
- \lambda \epsilon_\mu \dot a^\mu
- \frac{1}{2M} \left[(i\partial_\mu - e_0 a_\mu)\psi^* \right]
\left[(-i\partial^\mu - e_0 a^\mu)\psi\right]
\nonumber \\
& & \mbox{\qquad\qquad} 
- \frac{\lambda}{4} f_{\mu\nu}f^{\mu\nu}
 + \frac{\lambda\sigma}{2} 
\epsilon^\mu \epsilon_\mu -
\frac{\lambda\sigma}{2} e^2 \rho [G\rho]
 \Bigr\} \ .
\nonumber \\
&=&
\frac{1}{2M} \left[(i\partial_\mu - e_0 a_\mu)\psi^* \right]
\left[(-i\partial^\mu - e_0 a^\mu)\psi\right]
+ \frac{\lambda}{4} f_{\mu\nu}f^{\mu\nu}
- \frac{\lambda\sigma}{2} \epsilon^\mu \epsilon_\mu
+ \frac{\lambda\sigma}{2} e^2 \rho [G\rho].
\nonumber \\
\label{eqn:2.33}
\end{eqnarray}

We may now decompose (\ref{eqn:2.33}) into the free Hamiltonian for the
matter and the gauge fields and the interaction terms.
\begin{equation}
\K = \K_{\rm photon} + \K_{\rm matter} + \K_{\rm interaction}
\label{eqn:2.34}
\end{equation}
where
\begin{eqnarray}
\K_{\rm photon} &=& \frac{\lambda}{4} f_{\mu\nu}f^{\mu\nu}
- \frac{\lambda\sigma}{2} \epsilon^\mu \epsilon_\mu
\label{eqn:2.35}\\
\K_{\rm matter} &=& \frac{1}{2M} \left[\partial_\mu \psi^* \right]
\left[\partial^\mu \psi\right]
\label{eqn:2.36}\\
\K_{\rm interaction} &=& \frac{ie_0}{2M} a_\mu (\psi^*
\partial^\mu \psi - \psi \partial^\mu \psi^* )
+ \frac{e_0^2}{2M} a_\mu a^\mu |\psi|^2 + \frac{\lambda\sigma}{2} e^2
\rho [G\rho]
\label{eqn:2.37}
\end{eqnarray}

The last term in (\ref{eqn:2.37}) has the form of a c-number
energy density which represents the mass-energy equivalent
required to assemble the matter field (see also \cite{nadav}).
  From equations (\ref{eqn:1.25}) and (\ref{eqn:1.30}), one
sees that the coupling $e_0 \ e = \lambda \ e^2$, which
appears in this mass-energy density, is characteristic of
the classical Lorentz force due to a charge distribution.
On the other hand, we demonstrate below that the remaining
terms in (\ref{eqn:2.37}), which lead to 3-particle and
4-particle interactions in the Feynman diagrams, are
connected by the Ward identity for the conserved current
(\ref{eqn:1.22}).

%
\section{Path Integral Quantization}
\setcounter{equation}{0}

The first order action used in the previous section is naturally
suited to path integral quantization.  We write the path
integral as
\begin{equation}
\Z  = \frac{1}{\cal N} \int \D \psi^*
\ \D \psi \ \D a_\mu \ \D a_5 \ \D \epsilon_\mu e^{iS}
\label{eqn:3.1}
\end{equation}
where we use (\ref{eqn:2.13}) for the action S. The integration
over $a_5$ places the constraint (\ref{eqn:2.14}) into the
measure in the form of $\delta(e_0 \psi^* \psi - \lambda
\partial^\mu \epsilon_\mu)$; we may, furthermore, insert the
gauge fixing constraint $\delta(\partial_\mu a^\mu - \Lambda)$.
By integrating over $\D \epsilon_{\parallel}$ and $\D
a_{\parallel}$, the path integral becomes
\begin{equation}
\Z  = \frac{1}{\cal N} \int \D \psi^* \ \D \psi \ \D
(a_\perp)_\mu \ \D (\epsilon_\perp)_\mu \ e^{iS}
\label{eqn:3.2}
\end{equation}
where ${\rm S}$ is now given by (\ref{eqn:2.26}).
Then, by carrying out the
gauge transformation (\ref{eqn:2.27}), which leaves the measure
invariant because we have chosen some specific function $\Lambda$, we
obtain the path integral
\begin{equation}
\Z  = \frac{1}{\cal N} \int \D \psi^* \ \D \psi \ \D (a_\perp)_\mu \ \D 
(\epsilon_\perp)_\mu \ e^{iS}
\label{eqn:3.3}
\end{equation}
in which ${\rm S}$ is the unconstrained action in (\ref{eqn:2.30}).  Since
$\epsilon_\perp$ plays the role of a conjugate momentum, we may perform the
Gaussian integration over $\D(\epsilon_\perp)_\mu$, which puts the
path integral into the form
\begin{equation}
\Z  = \frac{1}{\cal N} \int \D \psi^* \ \D \psi \ \D (a_\perp)_\mu \ e^{iS}
\label{eqn:3.4}
\end{equation}
where the action ${\rm S}$ is given by
\begin{eqnarray}
{\rm S} &=& \int d^4 x d\tau \left\{ i \psi^* \dot \psi
- \frac{1}{2M} \psi^* (-i\partial_\mu - e_0 (a_\perp)_\mu)
(-i\partial^\mu - e_0 (a_\perp)^\mu)\psi \right. \nonumber \\
& & \left. \mbox{\qquad\qquad}
- \frac{\lambda}{4} (f_\perp)_{\mu\nu}(f_\perp)^{\mu\nu}
- \frac{\lambda\sigma}{2} (\dot a_\perp)_\mu (\dot a_\perp)^\mu
- \frac{\lambda\sigma}{2} e^2 \rho [G\rho]
  \right\}.
\label{eqn:3.5}
\end{eqnarray}
We expand
\begin{eqnarray}
- \frac{\lambda}{4} (f_\perp)_{\mu\nu}(f_\perp)^{\mu\nu}
&=& - \frac{\lambda}{4} 
(\partial_\mu a_\nu - \partial_\nu a_\mu)(\partial^\mu a^\nu
- \partial^\nu a^\mu)
\nonumber \\
&=& - \frac{\lambda}{2}[(\partial_\mu a_\nu)(\partial^\mu a^\nu)-
(\partial_\mu a_\nu)(\partial^\nu a^\mu)] 
\nonumber \\
&=& \frac{\lambda}{2} a_\mu [g^{\mu\nu}
\Box - \partial^\mu \partial^\nu ] a_\nu + {\rm divergence}
\label{eqn:3.6}
\end{eqnarray}
and similarly,
\begin{equation}
(\dot a_\perp)_\mu (\dot a_\perp)^\mu = - (a_\perp)_\mu [g^{\mu\nu}
\partial_\tau^2 ] (a_\perp)_\nu + {\rm divergence}
\label{eqn:3.7}
\end{equation}
so that the action takes the form
\begin{eqnarray}
{\rm S} &=& \int d^4 x d\tau \left\{ i \psi^* \dot \psi
- \frac{1}{2M} \psi^* (-i\partial_\mu - e_0 (a_\perp)_\mu)
(-i\partial^\mu - e_0 (a_\perp)^\mu)\psi \right. \nonumber \\
& & \left. \mbox{\qquad\qquad}
+ \frac{\lambda}{2} (a_\perp)_\mu [g^{\mu\nu} \Box + \sigma g^{\mu\nu}
\partial_\tau^2 - \partial^\mu \partial^\nu ] (a_\perp)_\nu 
- \frac{\lambda\sigma}{2} e^2 \rho [G\rho] \right\}.
\label{eqn:3.8}
\end{eqnarray}
In Section 5, we will show that the operator
$[g^{\mu\nu} \Box - \partial^\mu \partial^\nu ]$ 
projects onto the transverse states.
Since only transverse states are present in (\ref{eqn:3.8}),
we may replace the action with 
\begin{eqnarray}
{\rm S} &=& \int d^4 x d\tau \Bigl\{ i \psi^* \dot \psi
- \frac{1}{2M} \psi^* (-i\partial_\mu - e_0 (a_\perp)_\mu)
(-i\partial^\mu - e_0 (a_\perp)^\mu)\psi
\nonumber \\
& & \mbox{\qquad\qquad}
+ \frac{\lambda}{2} (a_\perp)_\mu [\Box + \sigma \partial_\tau^2 ]
(a_\perp)^\mu - \frac{\lambda\sigma}{2} e^2 \rho [G\rho] \Bigr\} ,
\label{eqn:3.9}
\end{eqnarray}
and we recognize the ``inverse propagator''
$[\Box + \sigma \partial_\tau^2 ]$ for
the gauge fields.  From this action, the Feynman rules for Green's
functions, which will be derived in Section 6 by canonical procedures,
may be read-off directly.

\section{Canonical Quantization of the Free Spinless
\newline Matter Field}
\setcounter{equation}{0}

We begin with the Hamiltonian density (\ref{eqn:2.36}) and take
the Hamiltonian operator to be
\begin{equation}
{\rm K} = \int d^4 x \K_{\rm matter}  .
\label{eqn:4.1}
\end{equation}

The fields must satisfy the canonical equal-$\tau$
commutation relations,
\begin{equation}
[\psi (x,\tau), \pi_{\psi} (x',\tau) ] = i \delta^4 (x-x'),
\label{eqn:4.2}
\end{equation}
which together with (\ref{eqn:2.31}) leads to 
\begin{equation}
[\psi (x,\tau), \psi^* (x',\tau) ] = \delta^4 (x-x').
\label{eqn:4.3}
\end{equation}
The field evolves dynamically according to the Heisenberg equation
\begin{equation}
i\partial_\tau \psi = [\psi , {\rm K} ],
\label{eqn:4.4}
\end{equation}
and using equations (\ref{eqn:2.36}), (\ref{eqn:4.1}), (\ref{eqn:4.2}),
and (\ref{eqn:4.4}), we find
\begin{eqnarray}
i\partial_\tau \psi (x,\tau ) &=& \left[ \psi (x,\tau), \frac{1}{2M}
\int d^4 x' \left(\partial'_\mu \psi^* (x',\tau)
 \right) \left(\partial'^\mu \psi (x',\tau) \right) \right]
\nonumber \\
&=& - \frac{1}{2M} \int d^4 x' \left[ \psi (x,\tau), \psi^* (x',\tau)
\partial'_\mu \partial'^\mu \psi (x',\tau) \right]
\nonumber \\
&=& - \frac{1}{2M} \int d^4 x' \left[ \psi (x,\tau), \psi^* (x',\tau)
\right] \partial'_\mu \partial'^\mu \psi (x',\tau)
\nonumber \\
&=& - \frac{1}{2M} \int d^4 x' \delta^4 (x-x')
\partial'_\mu \partial'^\mu \psi (x',\tau)
\nonumber \\
&=& - \frac{1}{2M} \partial_\mu \partial^\mu \psi (x,\tau).
\label{eqn:4.5}
\end{eqnarray}
Notice that (\ref{eqn:4.5}) is the Schr\"odinger equation
(\ref{eqn:1.20}) for the 
field operator $\psi (x,\tau)$ in the absence of interaction,
and because it is satisfied, we may perform the Fourier
expansion
\begin{equation}
\psi (x,\tau) = \int \frac{d^4 k}{(2\pi)^4} b(k) e^{i(k\cdot x -
\kappa \tau)} \qquad
\psi^* (x,\tau) = \int \frac{d^4 k}{(2\pi)^4}
b^*(k) e^{-i(k\cdot x - \kappa \tau)} 
\label{eqn:4.6}
\end{equation}
where from (\ref{eqn:4.5}) we see that $\kappa=k^2 /2M$.  From
(\ref{eqn:4.2}), we find that
\begin{equation}
\left[ b(k), b^*(k') \right] = (2\pi)^4 \delta^4 (k-k')
\qquad \left[ b(k),  b(k') \right] =
\left[ b^*(k),  b^*(k') \right] = 0.
\label{eqn:4.7}
\end{equation}
In terms of the Fourier expansion, the Hamiltonian is given by
\begin{eqnarray}
{\rm K} &=& \int d^4 x \frac{1}{2M} \left[\partial_\mu \psi^* \right]
\left[\partial^\mu \psi\right]
\nonumber \\
&=& \frac{1}{2M} \int d^4 x \int \frac{d^4 k}{(2\pi)^4} (-ik_\mu)
b^*(k) e^{-i(k\cdot x - \kappa \tau)}
\int \frac{d^4 k'}{(2\pi)^4} (ik'_\mu) b(k') e^{i(k'\cdot x - \kappa'\tau)}
\nonumber \\
&=& \frac{1}{2M} \int \frac{d^4 k}{(2\pi)^4} \frac{d^4 k'}{(2\pi)^4}
b^*(k) b(k') (2\pi)^4 \delta^4 (k-k') e^{i\kappa \tau}
e^{-i \kappa'\tau} (k\cdot k') \nonumber \\
&=& \int \frac{d^4 k}{(2\pi)^4} \ \frac{ k^2 }{2M} \ b^*(k) b(k) 
\nonumber \\
&=& \int \frac{d^4 k}{(2\pi)^4} \ \kappa \ b^*(k) b(k) 
\label{eqn:4.8}
\end{eqnarray}
Notice that the ground state mass for this Hamiltonian vanishes
even without the need for normal ordering.

We write the Green's function $G_i(x,\tau)$ for the
Schr\"odinger equation (\ref{eqn:4.5}) in the integral form,
\begin{equation}
G(x,\tau) = \frac{1}{(2\pi)^5} \int_{C_i} d^4 k d\kappa
\frac{e^{i(k\cdot x- \kappa\tau)}}{\frac{1}{2M} k^2-\kappa} 
\label{eqn:4.9}
\end{equation}
where the contour of $\kappa$ integration $C_i$ determines the
boundary conditions.  If $C_i$ includes the interval
$(-\infty,\infty)$, then $G_i(x,\tau)$ will satisfy the
inhomogeneous equation
\begin{equation}
(i\partial_\tau + \frac{1}{2M} \partial^\mu \partial_\mu)
G_i(x,\tau) = -\delta^4 (x) \delta (\tau).
\label{eqn:4.10}
\end{equation}
Recalling Schwinger's parametric representation of the Klein-Gordon
Green's function 
and Feynman's observation that the choice of retarded contour
in (\ref{eqn:4.9}) (by displacing the pole into the lower 
half plane) is equivalent to choosing the Feynman contour for the
$\tau$-integrated propagator,
we will take the Green's function for the matter field to be
\begin{equation}
G(x,\tau) = \frac{1}{(2\pi)^5} \int d^4 k d\kappa
\frac{e^{i(k\cdot x- \kappa\tau)}}
{\frac{1}{2M} k^2-\kappa -i\epsilon} .
\label{eqn:4.12}
\end{equation}
First performing the $\kappa$-integration (the pole is at
$\kappa = k^2/2M -i\epsilon$ in the lower half plane), we obtain
zero for $\tau < 0$ (for which we must close in the upper half
plane) and for $\tau > 0$ (we must close in the lower half
plane), we obtain
\begin{equation}
\int d\kappa \frac{e^{-i \kappa\tau}} 
{\frac{1}{2M} k^2-\kappa -i\epsilon} = 2\pi i \ {\rm Res}
\frac{e^{-i \kappa\tau}} {\frac{1}{2M} k^2-\kappa -i\epsilon}
\times {\rm index \ of \ contour} = 2\pi i e^{-i( \frac{k^2}{2M}
-i\epsilon) \tau} \theta (\tau).
\label{eqn:4.13}
\end{equation}
Thus, we find that
\begin{equation}
G(x,\tau) = \frac{i}{(2\pi)^4} \int d^4 k e^{i(k\cdot x-
\frac{k^2}{2M}\tau + i\epsilon\tau)}\theta (\tau)
\label{eqn:4.14}
\end{equation}
in which we see the retarded nature of $G(x,\tau)$ explicitly through
$\theta (\tau)$.  We easily verify that the $\tau$-integral
of the Green's function,
\begin{eqnarray}
\int_{-\infty}^\infty d\tau e^{-i(m^2/2M)\tau} G(x,\tau) &=&
\frac{i}{(2\pi)^4} \int_{0}^\infty d\tau e^{-i(m^2/2M)\tau}
\int d^4 k e^{i(k\cdot x-\frac{k^2}{2M}\tau + i\epsilon\tau)}
\nonumber \\
&=& i\int_{0}^\infty d\tau \int \frac{d^4 k}{(2\pi)^4}
e^{-i(m^2/2M+\frac{k^2}{2M} -i\epsilon)\tau}
e^{ik\cdot x} \nonumber \\
&=& \int \frac{d^4 k}{(2\pi)^4} \frac{e^{ik\cdot x}}
{\frac{1}{2M} (k^2+ m^2) -i\epsilon} \nonumber \\
&=& 2M \ \Delta_{\rm F} (x)
\label{eqn:4.15}
\end{eqnarray}
goes over to the Feynman propagator for a particle of mass
$m$ with an overall factor of $2M$.

On the other hand, we may consider the Green's function for the
matter field as expressed through the vacuum expectation value of
the ($\tau$-ordered) operator products.  Using the momentum
expansions (\ref{eqn:4.6}), we find that
\begin{eqnarray}
\lefteqn{\langle 0| {\rm T} \psi(x_1,\tau_1)
\psi^*(x_2,\tau_2) |0\rangle =} \nonumber \\
& & = \theta(\tau_1 - \tau_2)  
\langle 0|\psi(x_1,\tau_1) \psi^*(x_2,\tau_2) |0\rangle
+ \theta(\tau_2 - \tau_1)
\langle 0| \psi^*(x_2,\tau_2) \psi(x_1,\tau_1)|0\rangle
\nonumber \\
& & = \theta(\tau_1 - \tau_2) \int \frac{d^4 k}{(2\pi)^4}
\frac{d^4 k'}{(2\pi)^4} 
e^{-i(k\cdot x_2 - \frac{k^2}{2M} \tau_2)}
e^{i(k'\cdot x_1 -\frac{k'^2}{2M}\tau_1)} 
\langle 0|b(k')  b^*(k) |0\rangle 
\nonumber \\
& & = \theta(\tau_1 - \tau_2) \int \frac{d^4 k}{(2\pi)^4}
e^{i[k\cdot (x_1-x_2) - \frac{k^2}{2M} (\tau_1-\tau_2)]}
\nonumber \\
& & = -iG(x_1-x_2,\tau_1-\tau_2)
\label{eqn:4.16}
\end{eqnarray}
where we have used $b(k)|0\rangle = 0$ and (\ref{eqn:4.7}).   
We thus verify that the Green's function we have chosen in
(\ref{eqn:4.12}) is equal to the vacuum expectation value of the
$\tau$-ordered product of the operators, as required for the
application of Wick's theorem in perturbation theory.

%
\section{Quantization of the Free Gauge Field}
\setcounter{equation}{0}

We begin with the Hamiltonian obtained from (\ref{eqn:2.35}) as
\begin{equation}
{\rm K} = \int d^4 x \K_{\rm photon}.
\label{eqn:5.1}
\end{equation}
In order to evaluate the canonical commutation relations for the
transverse fields, we first consider the projection operator
\begin{equation}
\delta^{\mu\nu}_{\perp} (x-y) = \int \frac{d^4 k}{(2\pi)^4}
\left[ g^{\mu\nu} - \frac{k^\mu k^\nu}{k^2} \right]e^{ik\cdot (x-y)}
= \int \frac{d^4 k}{(2\pi)^4} \P^{\mu\nu}(k) e^{ik\cdot (x-y)},
\label{eqn:5.2}
\end{equation}
which projects onto the transverse part of a vector function.   For a
function $f_\nu (x,\tau)$ with Fourier transform
$\tilde f_\nu (k,\tau)$, 
\begin{eqnarray}
\int d^4 y \delta^{\mu\nu}_{\perp}(x-y) f_\nu (y,\tau) &=&
\int d^4 y \frac{d^4 k}{(2\pi)^4} d^4k' 
\left[ g^{\mu\nu} - \frac{k^\mu k^\nu}{k^2} \right]e^{ik\cdot (x-y)}
\tilde f_\nu(k',\tau) e^{ik'\cdot y}
\nonumber \\
&=& \int d^4 k \left[ g^{\mu\nu} - \frac{k^\mu k^\nu}{k^2} \right]
f_\nu(k,\tau) e^{ik\cdot x}
\label{eqn:5.3}
\end{eqnarray}
which clearly satisfies $\partial^\mu f_\mu (x,\tau)=0$.  So we
may represent the transverse fields as
\begin{equation}
a^\mu_{\perp}(x,\tau) = \left[\delta^{\mu\rho}_{\perp}
a_\rho \right] (x,\tau) \qquad\qquad
\epsilon^\mu_{\perp} (x,\tau) = \left[\delta^{\mu\rho}_{\perp}
\epsilon_\rho \right] (x,\tau) .
\label{eqn:5.4}
\end{equation}
Canonical quantization requires the equal-$\tau$ commutation
relations for the field components
\begin{equation}
[a^\mu (x,\tau) , \pi_{a_\nu} (y,\tau) ] = i g^{\mu\nu}\delta^4(x-y) 
\label{eqn:5.5}
\end{equation}
which, using (\ref{eqn:2.32}) for $\pi_{a_\nu}$, becomes
\begin{equation}
[a^\mu (x,\tau) , \epsilon^\nu (y,\tau) ] = -\frac{i}{\lambda}
g^{\mu\nu}\delta^4(x-y).
\label{eqn:5.6}
\end{equation}
So for the transverse field components, we find
\begin{eqnarray}
[a^\mu_{\perp} (x,\tau) , \epsilon^\nu_{\perp} (y,\tau) ] &=&
[ \ [\delta^{\mu\rho}_{\perp} a_\rho] (x,\tau) ,
\ [\delta^{\nu\sigma}_{\perp} \epsilon_\sigma (y,\tau)] \  ] 
\nonumber \\
&=& \frac{-i}{\lambda} g_{\rho\sigma} \left[ \delta^{\mu\rho}_{\perp}
\delta^{\nu\sigma}_{\perp} \delta^4 \right] (x-y)
\nonumber \\
&=& \frac{-i}{\lambda} \delta^{\mu\nu}_{\perp} (x-y) 
\label{eqn:5.7}
\end{eqnarray}
We may drop the subscript $_{\perp}$ from the field variables
provided that we use (\ref{eqn:5.7}) as the canonical
commutation relation.  This
relation insures that the Gauss Law constraint (\ref{eqn:2.14})
commutes with the other variables of the theory, and so, with the
Hamiltonian (\ref{eqn:5.1}).

The Heisenberg equations for the fields $a^\mu$ are
\begin{eqnarray}
i\partial_\tau a^\mu (x,\tau) &=& [a^\mu (x,\tau) , {\rm K}]
\nonumber \\
&=& [a^\mu (x,\tau), \int d^4 y (\frac{\lambda}{4}
f_{\rho\nu}f^{\rho\nu}
- \frac{\lambda\sigma}{2} \epsilon^\rho \epsilon_\rho ) ]
\nonumber \\
&=& - \frac{\lambda\sigma}{2} \int d^4 y \{
[a^\mu (x,\tau), \epsilon^\rho (y,\tau) ] \epsilon_\rho (y,\tau) +
\epsilon^\rho (y,\tau) [a^\mu (x,\tau),
\epsilon_\rho (y,\tau) ] \}
\nonumber \\
&=& - \frac{\lambda\sigma}{2} [-2\frac{i}{\lambda}
\epsilon^\mu (x,\tau) ],
\nonumber 
\end{eqnarray}
and we find that
\begin{equation}
\partial_\tau a^\mu (x,\tau) = \sigma \epsilon^\mu (x,\tau) =
\sigma f^{5\mu} = f_5\; ^\mu 
\label{eqn:5.8}
\end{equation}
which agrees with the classical definition of $f^{5\mu}$ in the gauge
$a_5=0$.  The Heisenberg equation for the fields $\epsilon^\mu$
are
\begin{eqnarray}
i\partial_\tau  \epsilon^\mu (x,\tau) &=& [\epsilon^\mu (x,\tau),
{\rm K} ]
\nonumber \\
&=& [\epsilon^\mu (x,\tau), \int d^4 y
\frac{\lambda}{4} f_{\mu\nu}f^{\mu\nu} ]
\nonumber \\
&=& \frac{\lambda}{4} \int d^4 y \{
[\epsilon^\mu (x,\tau), (\partial_\lambda a_\nu - \partial_\nu
a_\lambda) ] f^{\lambda\nu} +
f^{\lambda\nu} [\epsilon^\mu (x,\tau),
(\partial_\lambda a_\nu - \partial_\nu a_\lambda) ]\}
\nonumber \\
&=& \frac{\lambda}{4} \int d^4 y \{
\partial_\lambda [\epsilon^\mu , a_\nu ] f^{\lambda\nu} -
\partial_\nu [\epsilon^\mu , a_\lambda ] f^{\lambda\nu} +
f^{\lambda\nu} \partial_\lambda [\epsilon^\mu , a_\nu ] -
f^{\lambda\nu} \partial_\nu [\epsilon^\mu , a_\lambda ] \}
\nonumber \\
&=& \frac{i}{4} \int d^4 y \{
2 f^{\lambda\mu} \partial_\lambda \delta^4 (x-y) -
2 f^{\mu\lambda} \partial_\lambda \delta^4 (x-y) \}
\nonumber \\
&=& -i \partial_\lambda f^{\lambda\mu} (x,\tau).
\label{eqn:5.9}
\end{eqnarray}
Since $\epsilon^\mu =f^{5\mu}$, (\ref{eqn:5.9}) can be written
\begin{equation}
\partial_\lambda f^{\lambda\mu} + \partial_\tau f^{5\mu} = 0
\label{eqn:5.10}
\end{equation}
which may be combined with (\ref{eqn:2.16}) to write
\begin{equation}
\partial_\alpha f^{\alpha\beta} = 0
\label{eqn:5.11}
\end{equation}
for $\alpha, \beta = 0,\cdots,3,5$.  These are half of the 
pre-Maxwell field equations; the other pre-Maxwell equations
follow from the definition of $f^{\mu\nu}$ and (\ref{eqn:5.8}).
By virtue of (\ref{eqn:2.19}), (\ref{eqn:5.10}) may be written in
the form
\begin{equation}
0 = \partial_\tau f^{5\mu} +  \partial_\lambda
(\partial^\lambda a^\mu - \partial^\mu a^\lambda ) =
\partial_\tau f^{5\mu} +
\partial_\lambda \partial^\lambda a^\mu - \partial^\mu
\partial_\lambda a^\lambda = \partial_\tau f^{5\mu} +
\partial_\lambda \partial^\lambda a^\mu .
\label{eqn:5.12}
\end{equation}
Combining (\ref{eqn:5.12}) with (\ref{eqn:5.8}), we obtain the
wave equation,
\begin{equation}
0 = \partial_\lambda \partial^\lambda a^\mu
+ \partial_\tau \epsilon^\mu =
\partial_\lambda \partial^\lambda a^\mu
+ \partial_\tau (\sigma \partial_\tau a^\mu) =
(\partial_\lambda \partial^\lambda + \sigma \partial_\tau^2)
a^\mu.
\label{eqn:5.13}
\end{equation}
Therefore, we may perform a Fourier expansion of the field
operator,
\begin{equation}
a^\mu (x,\tau) = \sum_{s={\rm polarizations}} \int \frac{d^4 k}
{2\kappa} \left[ \varepsilon^\mu_s a(k,s)
e^{i(k\cdot x + \sigma \kappa \tau)} +
\varepsilon^{\mu*}_s a^*(k,s)
e^{-i(k\cdot x + \sigma \kappa \tau)} \right]
\label{eqn:5.14}
\end{equation}
where the five-dimensional mass shell condition is
\begin{equation}
\kappa = \sqrt{-\sigma k^2 }.
\label{eqn:5.15}
\end{equation}

Evidently, for free fields,
the condition $-\sigma k^2 > 0$ is required to prevent
divergence at \hbox{$\tau\rightarrow\pm\infty$},
however this condition need not apply for virtual
photons.  Using (\ref{eqn:5.8}), $\epsilon^\mu$ is given by
\begin{equation}
\epsilon^\mu (x,\tau) = i\sum_{s={\rm polarizations}}
\int \frac{d^4 k}{2\kappa}  \kappa
\left[ \varepsilon^\mu_s a(k,s)
e^{i(k\cdot x + \sigma \kappa \tau)} -
\varepsilon^{\mu*}_s a^*(k,s)
e^{-i(k\cdot x + \sigma \kappa \tau)} \right].
\label{eqn:5.16}
\end{equation}

We must choose polarization basis states for the field
vectors.  In momentum space (\ref{eqn:2.19}) becomes $k_\mu a^\mu
= 0$, so that there will be three independent polarizations (see
also \cite{polar}).
To choose the polarizations in a covariant way, we begin by choosing
an arbitrary timelike vector $n^\mu$ ($n^2=-1$).  There are then two
orthonormal spacelike vectors, $\varepsilon_1$ and $\varepsilon_2$,
which satisfy the conditions
\begin{equation}
n\cdot\varepsilon_1 = n\cdot\varepsilon_2 = k\cdot\varepsilon_1 =
k\cdot\varepsilon_2 = 0,
\label{eqn:5.17}
\end{equation}
which may be constructed in the following way.  Since $n^2 =-1$,
\begin{equation}
(\vec{n})^2-(n^0)^2 =-1 \Longrightarrow n^0 = \pm \sqrt{(\vec{n})^2+1}
\label{eqn:5.18}
\end{equation}
so that taking
\begin{equation}
\varepsilon_{1,2}^0 =
\frac{\vec{n}\cdot\vec{\varepsilon}_{1,2}}{\sqrt{(\vec{n})^2+1}}
\label{eqn:5.19}
\end{equation}
guarantees that $n\cdot\varepsilon_{1,2}=0$.  We now require that
\begin{equation}
0=k\cdot \varepsilon_{1,2} = \vec{k}\cdot\vec{\varepsilon}_{1,2} -
(k^0) \varepsilon_{1,2}^0 = \left[ \vec{k}- \frac{k^0}{\sqrt{(\vec{n})^2+1}}
\vec{n} \right]\cdot \vec{\varepsilon}_{1,2}.
\label{eqn:5.20}
\end{equation}
Since two orthogonal 3-vectors can be found which satisfy
(\ref{eqn:5.20}), this establishes that (\ref{eqn:5.17})
can also be satisfied.  We normalize
$\varepsilon_{1,2}$ so that
\begin{equation}
\varepsilon_{1}\cdot \varepsilon_{2} = 0 \qquad\qquad
(\varepsilon_{1})^2 = (\varepsilon_{2})^2 =1
\label{eqn:5.21}
\end{equation}
To form the third polarization, which we will denote
$\varepsilon_5$, we take a linear combination of $n$ and $k$,
which guarantees linear independence from $\varepsilon_{1,2}$.  Writing
\begin{equation}
\varepsilon_5 = A \ k + B \ n
\label{eqn:5.22}
\end{equation}
we apply (\ref{eqn:2.19}) and find that
\begin{equation}
0=k\cdot \varepsilon_5 = k\cdot (A \ k + B \ n) = A k^2 + B(k\cdot n )
\label{eqn:5.23}
\end{equation}
We may take 
\begin{equation}
A = -B \ \frac{k\cdot n}{k^2} = B \sigma \frac{k\cdot n}{\kappa^2} 
\label{eqn:5.24}
\end{equation}
so that
\begin{equation}
\varepsilon_5 = B \left[ n + \sigma \frac{k\cdot n}{\kappa^2} k\right]
\label{eqn:5.25}
\end{equation}
and
\begin{equation}
(\varepsilon_5 )^2 =  B^2 \left[ n^2 +2 \sigma \frac{(k\cdot n)^2}{\kappa^2}
+ \frac{k^2}{\kappa^2} \right]  =  B^2 \frac{\sigma}{\kappa^2}
\left[ k^2 +(k\cdot n)^2 \right] = B^2 \frac{\sigma}{\kappa^2} \left[  k +
n(k\cdot n) \right]^2
\label{eqn:5.26}
\end{equation}
The expression $k +n(k\cdot n)$ is just $k_{\perp}$, and its square is
positive definite (expanding as a quadratic in $k^0$, one sees that
there are no roots).  We thus find that choosing
\begin{equation}
B= \frac{\kappa}{ |k +n(k\cdot n)|} 
\label{eqn:5.27}
\end{equation}
(\ref{eqn:5.26}) becomes
\begin{equation}
(\varepsilon_5 )^2 = \sigma.
\label{eqn:5.28}
\end{equation}
In the simple case that $n=(1,0,0,0)$,
we may take as $\varepsilon_1$ and $\varepsilon_2$ the vectors
\begin{equation}
\varepsilon_{1,2} = \left[ \begin{array}{c}
                             0 \\ \vec{\varepsilon}_{1,2}
                           \end{array} \right]  
\label{eqn:5.29}
\end{equation}
where
\begin{equation}
\vec{k}\cdot\vec{\varepsilon}_{1,2} =0 \qquad
\vec{\varepsilon}_{1} \cdot\vec{\varepsilon}_{2} =0 \qquad
\vec{\varepsilon}_{1} \cdot \vec{\varepsilon}_{1}
 = \vec{\varepsilon}_{2} \cdot \vec{\varepsilon}_{2} = 1
\label{eqn:5.30}
\end{equation}
Thus, the set $\{ \vec{\varepsilon}_{1}, \vec{\varepsilon}_{2},
\hat{k}=\vec{k}/|\vec{k}| \}$ forms an orthonormal basis for the
3-space in this frame.  We find for the third polarization,
$\varepsilon_5$,
\begin{equation}
\varepsilon_5 = \frac{1}{\kappa} \left[ \begin{array}{c}
                                  |\vec{k}| \\ k^0 \hat{k}
                                 \end{array} \right]
\label{eqn:5.31}
\end{equation}
Notice that for $k^2 \rightarrow 0$, $\kappa \rightarrow 0$ and
$\varepsilon_5 \cdot \varepsilon_5 \rightarrow 0/0$, so we will
treat the lightlike case as the limit of the massive case, in
which we take $\kappa \rightarrow 0$ at the end of all other
computations.

In order to evaluate the commutation relations among the
operators $a(k,s)$ and $a^*(k,s)$, we must first invert
the Fourier expansion of $a^\mu (x,\tau)$.  To do this, we write
the orthogonality relations among the polarization vectors in the
form
\begin{equation}
\varepsilon_s \cdot \varepsilon_{s'} =  g_{ss'} = g(s) \ \delta_{ss'} 
\qquad \mbox{for $s,s' = 1,2,5$}
\label{eqn:5.32}
\end{equation}
where
\begin{equation}
g(s) = \left\{ \begin{array}{ll}
                1, & \mbox{if $s=s'=1,2$} \\
                \sigma, & \mbox{if $s=s'=5$}
               \end{array} \right.        
\label{eqn:5.33}
\end{equation}
so that we have from (\ref{eqn:5.14})
\begin{eqnarray}
\lefteqn{\int d^4 x e^{-i(k\cdot x + \sigma \kappa \tau)} \varepsilon_s
\cdot a(x,\tau) =} \nonumber \\
& & =\sum_{s'} \int d^4 x \frac{d^4 k'}{2\kappa'}
\varepsilon_s \cdot \varepsilon_{s'}
\left[a(k',s') e^{i(x\cdot (k'-k) + \sigma (\kappa'-\kappa) \tau)}
+a^*(k',s')e^{-i(x\cdot (k+k')+\sigma(\kappa+\kappa')\tau)}
\right]
\nonumber \\
& & = \frac{(2\pi)^4}{2\kappa} g(s) \left[  a(k,s) +
a^*(k,s) e^{-2i \sigma\kappa\tau)} \right]
\label{eqn:5.34}
\end{eqnarray}
Similarly,
\begin{eqnarray}
\lefteqn{\int d^4 x e^{-i(k\cdot x + \sigma \kappa \tau)} \varepsilon_s
\cdot \dot a(x,\tau) =} \nonumber \\
& & = \sum_{s'} \int d^4 x \frac{d^4 k'}{2\kappa'}
\varepsilon_s \cdot \varepsilon_{s'} (i \sigma\kappa')
\nonumber \\&&\mbox{\qquad}
\left[a(k',s') e^{i(x\cdot (k'-k) + \sigma (\kappa'-\kappa)\tau)}-
a^*(k',s')e^{-i(x\cdot (k+k')+\sigma(\kappa+\kappa')\tau)}
\right]
\nonumber \\
& & = \frac{i \sigma}{2} (2\pi)^4 g(s) \left[  a(k,s) -
a^*(k,s) e^{-2i \sigma\kappa\tau)} \right]
\label{eqn:5.35}
\end{eqnarray}
Combining (\ref{eqn:5.34}) and (\ref{eqn:5.35}), we obtain
\begin{eqnarray}
i\sigma g(s)(2\pi)^4 a(k,s) &=& \int d^4 x e^{-i(k\cdot x +
\sigma \kappa \tau)} \varepsilon_s \cdot [\dot a(x,\tau) +
i\sigma\kappa a(x,\tau) ] \nonumber \\
&=& \int d^4 x e^{-i(k\cdot x + \sigma \kappa \tau)} \varepsilon_s
\cdot [\stackrel{\rightarrow}{\partial_\tau} a(x,\tau) -
\stackrel{\leftarrow}{\partial_\tau} a(x,\tau)]
\nonumber \\
&=& \int d^4 x e^{-i(k\cdot x + \sigma \kappa \tau)} \varepsilon_s
\cdot [ \stackrel{\leftrightarrow}{\partial_\tau} a(x,\tau)],
\label{eqn:5.36}
\end{eqnarray}
so that
\begin{equation}
a(k,s) = \frac{-i\sigma g(s)}{(2\pi)^4} \int d^4 x e^{-i(k\cdot x +
\sigma \kappa \tau)} \varepsilon_s
\cdot [ \stackrel{\leftrightarrow}{\partial_\tau} a(x,\tau)]
\label{eqn:5.37}
\end{equation}
and
\begin{equation}
a^*(k,s) =
\frac{i\sigma g(s)}{(2\pi)^4} \int d^4 x e^{i(k\cdot x +
\sigma \kappa \tau)} \varepsilon_s
\cdot [ \stackrel{\leftrightarrow}{\partial_\tau} a(x,\tau)]
\label{eqn:5.38}
\end{equation}
where we have used the fact that $1/g(s) = g(s)$.

Expressions (\ref{eqn:5.37}) and (\ref{eqn:5.38}) permit us to
evaluate the commutators
\begin{eqnarray}
[a(k,s),a^*(k',s')] &=& \frac{g(s)g(s')}{(2\pi)^8}
\int d^4 x \int d^4 x' e^{-i(k\cdot x + \sigma \kappa \tau)}
e^{i(k'\cdot x' + \sigma \kappa' \tau')} \nonumber \\
& & \mbox{\qquad \qquad} \left[
\varepsilon_s (k) \cdot [ \stackrel{\leftrightarrow}{\partial_\tau}
a(x,\tau)],
\varepsilon_{s'} (k')
 \cdot [ \stackrel{\leftrightarrow}{\partial_{\tau'}}
a(x',\tau')] \right] ,
\label{eqn:5.39}
\end{eqnarray}
where
\begin{eqnarray}
\left[ \stackrel{\leftrightarrow}{\partial_\tau} a^\mu (x,\tau)
\right. &&\mbox{\hspace{-.4 in}}, \left. 
\stackrel{\leftrightarrow}{\partial_{\tau'}} a^\nu (x',\tau')
\right] = \left[
\left( \stackrel{\rightarrow}{\partial_\tau}
- \stackrel{\leftarrow}{\partial_\tau} \right)
a^\mu (x,\tau) ,
\left(\stackrel{\rightarrow}{\partial_{\tau'}}
- \stackrel{\leftarrow}{\partial_{\tau'}}
\right) a^\nu (x',\tau') \right] 
\nonumber \\
&=& \left[ \stackrel{\rightarrow}{\partial_\tau}a^\mu (x,\tau) ,
\stackrel{\rightarrow}{\partial_{\tau'}} a^\nu (x',\tau')  \right] 
+ \stackrel{\leftarrow}{\partial_\tau}
   \stackrel{\leftarrow}{\partial_{\tau'}}\left[ a^\mu (x,\tau),
a^\nu (x',\tau')\right]
\nonumber \\
&&- \stackrel{\leftarrow}{\partial_{\tau'}}
\left[ \stackrel{\rightarrow}{\partial_\tau}a^\mu (x,\tau) ,
 a^\nu (x',\tau')  \right] -
 \stackrel{\leftarrow}{\partial_\tau} \left[ a^\mu (x,\tau) 
, \stackrel{\rightarrow}{\partial_{\tau'}} a^\nu (x',\tau') \right] 
\label{eqn:5.40}
\end{eqnarray}
so
\begin{eqnarray}
[a(k,s),a^*(k',s')] &=& \frac{g(s)g(s')}{(2\pi)^8}
\varepsilon_{s\mu}(k) \varepsilon_{s'\nu}(k')
\int d^4 x \int d^4 x' e^{-i(k\cdot x + \sigma \kappa \tau)}
e^{i(k'\cdot x' + \sigma \kappa' \tau')} \nonumber \\
& & \mbox{\quad} \left\{  \left[ \dot a^\mu (x,\tau) ,
\dot a^\nu (x',\tau')  \right]
+ \kappa\kappa' \left[ a^\mu (x,\tau), a^\nu (x',\tau')\right]
\right. \nonumber \\ & & \mbox{\quad} \left.
- i\sigma\kappa'
\left[ \dot a^\mu (x,\tau) , a^\nu (x',\tau')  \right]
+i\sigma\kappa \left[ a^\mu (x,\tau) 
, \dot a^\nu (x',\tau') \right] 
\right\}.
\label{eqn:5.41}
\end{eqnarray}
Evaluating the commutators at equal-$\tau$ and using
(\ref{eqn:5.6})  and (\ref{eqn:5.8}) to establish
\begin{equation}
\left[ \dot a^\mu (x,\tau) , a^\nu (x',\tau) \right] =
\frac{i\sigma}{\lambda}  \delta^{\mu\nu}_{\perp} (x-x'),
\label{eqn:5.42}
\end{equation}
one obtains
\begin{eqnarray}
[a(k,s),a^*(k',s')] &=& \frac{g(s)g(s')}{(2\pi)^8}
\varepsilon_{s\mu}(k) \varepsilon_{s'\nu}(k')
\int d^4 x \int d^4 x' e^{-i(k\cdot x + \sigma \kappa \tau)}
e^{i(k'\cdot x' + \sigma \kappa' \tau')} \nonumber \\
& & \mbox{\qquad} \left\{  -i\sigma \kappa' \left(
\frac{i\sigma}{\lambda} \delta^{\mu\nu}_{\perp} (x-x') \right)
+ i\sigma \kappa \left( 
\frac{-i\sigma}{\lambda} \delta^{\mu\nu}_{\perp} (x-x') \right)
\right\}\nonumber \\
&=& \frac{g(s)g(s')}{(2\pi)^8} 
\varepsilon_{s\mu}(k) \varepsilon_{s'\mu}(k')
\frac{\kappa + \kappa'}{\lambda}
\int d^4 x e^{i[x \cdot (k'-k) + \sigma \tau (\kappa'- \kappa) ]}
\nonumber \\
&=& \frac{2\kappa}{\lambda}\frac{g(s)}{(2\pi)^4} \delta_{ss'}
\delta^4 (k-k') 
\label{eqn:5.43}
\end{eqnarray}
where we have used $\sigma^2 =1$, $g(s)^2=1$,
(\ref{eqn:5.17}), and (\ref{eqn:5.32}).

Given the commutation relations for the operators
$a(k,s)$ and $a^*(k,s)$, we may evaluate the photon
propagator as the vacuum expectation value of the $\tau$-ordered
product of the fields.  Thus,
\begin{eqnarray}
\langle 0| {\rm T} a_\mu (x,\tau)a_\nu (x',\tau')|0\rangle &=&
\theta(\tau - \tau')  
\langle 0|a_\mu(x,\tau) a_\nu(x',\tau') |0\rangle
\nonumber \\
& & \mbox{\qquad}  + \theta(\tau' - \tau)
\langle 0| a_\nu(x',\tau') a_\mu(x,\tau)|0\rangle
\label{eqn:5.44}
\end{eqnarray}
where we use (\ref{eqn:5.14}) to expand
\begin{eqnarray}
  \lefteqn{\langle 0|a^\mu(x,\tau) a^\nu(x',\tau') |0\rangle
  = \sum_{s,s'} \int \frac{d^4 k}{2\kappa} \frac{d^4 k'}{2\kappa'} 
  \varepsilon^\mu_s \varepsilon^\nu_{s'} }
\nonumber \\&&\mbox{\qquad}
  \langle 0| \left[
  a(k,s) e^{i(k\cdot x + \sigma \kappa \tau)} +
  a^*(k,s) e^{-i(k\cdot x + \sigma \kappa \tau)} 
  \right]
\nonumber \\&&\mbox{\qquad\qquad}
  \left[
  a(k',s') e^{i(k'\cdot x' + \sigma \kappa' \tau')} +
  a^*(k',s') e^{-i(k'\cdot x' + \sigma \kappa' \tau')} 
  \right] |0\rangle
\nonumber \\
  & & = \sum_{s,s'} \int \frac{d^4 k}{2\kappa}
  \frac{d^4 k'}{2\kappa'} 
  \varepsilon^\mu_s \varepsilon^\nu_{s'}
  e^{i(k\cdot x + \sigma \kappa \tau)}
  e^{-i(k'\cdot x' + \sigma \kappa' \tau')}
  \langle 0| a(k,s) a^*(k',s') |0\rangle
\nonumber \\
  & & = \sum_{s,s'} \int \frac{d^4 k}{2\kappa}
  \frac{d^4 k'}{2\kappa'} 
  \varepsilon^\mu_s \varepsilon^\nu_{s'}
  e^{i(k\cdot x + \sigma \kappa \tau)}
  e^{-i(k'\cdot x' + \sigma \kappa' \tau')}
  \frac{2\kappa}{\lambda}\frac{g(s)}{(2\pi)^4} \delta_{ss'}
  \delta^4 (x-x')
\nonumber \\
  & & = \sum_{s} \frac{g(s)}{(2\pi)^4\lambda} \int
  \frac{d^4 k}{2\kappa} 
  \varepsilon^\mu_s \varepsilon^\nu_s
  e^{i[k\cdot (x-x') + \sigma \kappa (\tau-\tau')]}
\label{eqn:5.45}
\end{eqnarray}
so that
\begin{eqnarray}
\langle 0| {\rm T} a^\mu (x,\tau)a^\nu (x',\tau')|0\rangle &=&
\sum_{s} \frac{g(s)}{(2\pi)^4\lambda} \int \frac{d^4 k}{2\kappa} 
\varepsilon^\mu_s \varepsilon^\nu_s e^{i[k\cdot (x-x')]}
\nonumber \\ & & \mbox{\qquad} \left[
\theta(\tau - \tau') e^{i\sigma \kappa (\tau-\tau')} +
\theta(\tau' - \tau) e^{i\sigma \kappa (\tau'-\tau)} 
\right].
\label{eqn:5.46}
\end{eqnarray}
We recognize the integral
\begin{equation}
\frac{1}{2\kappa} \left[
\theta(\tau - \tau') e^{i\sigma \kappa (\tau-\tau')}
\theta(\tau' - \tau) e^{i\sigma \kappa (\tau'-\tau)} 
\right] = - \frac{i}{2\pi} \int d K \frac{e^{i\sigma K (\tau-\tau')}}
{k^2 +\sigma K^2 -i\epsilon}
\label{eqn:5.47}
\end{equation}
which we may use to put the Green's function (the vacuum
expectation value of the $\tau$-ordered product of the fields)
in the form of the Feynman propagator for the five-dimensional
field:
\begin{eqnarray}
d^{\mu\nu}(x-x',\tau-\tau') &=&
i\langle 0| {\rm T} a^\mu (x,\tau)a^\nu (x',\tau')|0\rangle
\nonumber \\
&=& \int \frac{d^4 k d\kappa}{(2\pi)^5} \sum_{s} \frac{g(s)}{\lambda}
\varepsilon^\mu_s \varepsilon^\nu_s
\frac{e^{i[ k\cdot (x-x') + \sigma \kappa (\tau-\tau')]}}
{k^2 +\sigma \kappa^2 -i\epsilon}
\label{eqn:5.48}
\end{eqnarray}
In order to evaluate the sum over polarizations in (\ref{eqn:5.48}),
we must consider the cases \hbox{$\sigma=\pm 1$} separately.  For the case
that $\sigma=1$, we must satisfy $k^2=-\kappa^2 <0$, and we may
take
\begin{equation}
k = \lim_{\alpha \rightarrow 0} (\sqrt{\kappa^2 + \alpha^2},0,0,
\alpha) = (\kappa,0,0,0).
\label{eqn:5.49}
\end{equation}
Choosing
\begin{equation}
n=(1,0,0,0) \qquad \varepsilon_1 = (0,1,0,0) \qquad \varepsilon_2 = (0,0,1,0)
\label{eqn:5.50}
\end{equation}
we find from (\ref{eqn:5.31}) and (\ref{eqn:5.49}) that
\begin{equation}
\varepsilon_5 = (|\vec{k}|,k^0 \hat{k}) \frac{1}{\sqrt{-k^2}} =
\lim_{\alpha \rightarrow 0} \frac{1}{\sqrt{\kappa^2 + \alpha^2}}
(\alpha,0,0,\sqrt{\kappa^2 + \alpha^2}) = (0,0,0,1).
\label{eqn:5.51}
\end{equation}
The completeness relation may then be written in the form
\begin{equation}
g^{\mu\nu} = - \frac{1}{\kappa^2} k^\mu k^\nu +
(\varepsilon_1)^\mu (\varepsilon_1)^\nu + 
(\varepsilon_2)^\mu (\varepsilon_2)^\nu +
(\varepsilon_5)^\mu (\varepsilon_5)^\nu
\label{eqn:5.52}
\end{equation}
which we may rearrange to write
\begin{equation}
\sum_{s=1,2,5} g(s) (\varepsilon_s)^\mu (\varepsilon_s)^\nu =
\sum_{s=1,2,5} (\varepsilon_s)^\mu (\varepsilon_s)^\nu =
 g^{\mu\nu} + \frac{1}{\kappa^2} k^\mu k^\nu =
g^{\mu\nu} - \frac{1}{k^2} k^\mu k^\nu
\label{eqn:5.53}
\end{equation}
where we have used $g(1,2)=1$ and $g(5)=\sigma =1$.  

For the case of $\sigma = -1$, we have $k^2=\kappa^2 >0$ so we
may take
\begin{equation}
k = (0,0,0,\kappa).
\label{eqn:5.54}
\end{equation}
and using (\ref{eqn:5.31}) we find that
\begin{equation}
\varepsilon_5 = (|\vec{k}|,k^0 \hat{k}) \frac{1}{\sqrt{-k^2}} =
(1,0,0,0).
\label{eqn:5.55}
\end{equation}
In this case the completeness relation is
\begin{equation}
g^{\mu\nu} = - (\varepsilon_5)^\mu (\varepsilon_5)^\nu +
(\varepsilon_1)^\mu (\varepsilon_1)^\nu + 
(\varepsilon_2)^\mu (\varepsilon_2)^\nu +
\frac{1}{\kappa^2} k^\mu k^\nu
\label{eqn:5.56}
\end{equation}
and it may be rearranged as
\begin{equation}
\sum_{s=1,2,5} g(s) (\varepsilon_s)^\mu (\varepsilon_s)^\nu =
\sum_{s=1,2} (\varepsilon_s)^\mu (\varepsilon_s)^\nu
- (\varepsilon_5)^\mu (\varepsilon_5)^\nu =
 g^{\mu\nu} - \frac{1}{\kappa^2} k^\mu k^\nu =
g^{\mu\nu} - \frac{1}{k^2} k^\mu k^\nu
\label{eqn:5.57}
\end{equation}
where we have used $g(1,2)=1$ and $g(5)=\sigma =-1$.  Notice that
(\ref{eqn:5.57}) and (\ref{eqn:5.53}) are identical and
are consistent with the transversality requirements
on the polarization states.  Using this expression and observing
that the sum over polarization states gives the projection operator
$\P^{\mu\nu}(k)$ defined in (\ref{eqn:5.2}), equation
(\ref{eqn:5.7}) may be verified explicitly.

To evaluate the Hamiltonian in the momentum representation,
\begin{eqnarray}
\frac{\lambda}{4} f_{\mu\nu}f^{\mu\nu} &=& \frac{\lambda}{4}
(\partial_\mu a_\nu - )
(\partial^\mu a^\nu -\partial^\nu a^\mu)
\nonumber \\
&=& \frac{\lambda}{2}[ (\partial_\mu a_\nu)(\partial^\mu a^\nu)
- (\partial_\nu a_\mu)(\partial^\mu a^\nu) ]
\nonumber \\
&=& \frac{\lambda}{2} [ \partial^\mu (a^\nu \partial_\mu a_\nu) -
a_\nu \Box a^\nu - \partial^\mu (a^\nu\partial_\nu a_\mu) -
a^\nu\partial_\nu (\partial^\mu a_\mu) ]
\nonumber \\
&=& - \frac{\lambda}{2} a_\nu \Box a^\nu + {\rm divergence}
\label{eqn:5.58}
\end{eqnarray}
and
\begin{eqnarray}
- \frac{\lambda\sigma}{2} \epsilon^\mu \epsilon_\mu &=&
- \frac{\lambda\sigma}{2} (\sigma \partial_\tau a_\mu)
(\sigma \partial_\tau a^\mu)
\nonumber \\
&=& - \frac{\lambda\sigma}{2}[\partial_\tau
(a^\mu \partial_\tau a_\mu) -a_\mu \partial_\tau^2  a^\mu ]
\nonumber \\
&=& \frac{\lambda\sigma}{2} a_\mu \partial_\tau^2  a^\mu .
\label{eqn:5.59}
\end{eqnarray}
So the Hamiltonian becomes
\begin{equation}
{\rm K} = \frac{\lambda}{2} \int d^4 x a_\mu
[ - \Box + \sigma \partial_\tau^2 ] a^\mu =
\lambda\sigma \int d^4 x \ \ a_\mu \partial_\tau^2 a^\mu
\label{eqn:5.60}
\end{equation}
where we have used the wave equation (\ref{eqn:5.13}).  Using the
momentum expansion (\ref{eqn:5.14}) and the commutation relations
(\ref{eqn:5.43}) we find for the normal ordered Hamiltonian
\begin{eqnarray}
\lefteqn{{\rm :K:} =\lambda\sigma \int d^4 x
\sum_{s,s'} \int \frac{d^4 k d^4 k'}{4\kappa \kappa'}
(\varepsilon_s)^\mu (\varepsilon_{s'})_\mu (i\sigma\kappa')^2} \\
\nonumber \\
& &\mbox{\qquad}
:\left\{ a(k,s) a(k',s') e^{i[(k+k')\cdot x +\sigma
(\kappa+\kappa')\tau]} + a^*(k,s) a^*(k',s')
e^{-i[(k+k')\cdot x +\sigma (\kappa+\kappa')\tau]}
\right. \nonumber \\ & & \mbox{\qquad\qquad} \left.
+ a^*(k,s) a(k',s')
e^{-i[(k-k')\cdot x +\sigma (\kappa-\kappa')\tau]}
+ a^*(k',s') a(k,s)
e^{i[(k-k')\cdot x +\sigma (\kappa-\kappa')\tau]} \right\}:
\nonumber \\
& &= -\lambda\sigma (2\pi)^4 \sum_{s,s'}
\int \frac{d^4 k}{4\kappa^2} \kappa^2
(\varepsilon_s)^\mu (\varepsilon_{s'})_\mu
\nonumber \\
& & \mbox{\qquad}
:\left\{ a(k,s) a(-k,s') e^{2i\sigma\kappa\tau} +
a^*(k,s) a^*(-k,s')e^{-2i\sigma\kappa\tau}
\right. \nonumber \\ & & \mbox{\qquad\qquad} \left.
+ a^*(k,s) a(k,s') + a^*(k,s') a(k,s)\right\}:
\label{eqn:5.61}
\end{eqnarray}
By replacing $k\rightarrow-k$ in (\ref{eqn:5.14}), we see that we
may identify
\begin{equation}
a(-k,s)e^{i\sigma\kappa\tau} =
a^*(k,s) e^{-i\sigma\kappa\tau}
\qquad a^*(-k,s) e^{-i\sigma\kappa\tau} =
a(k,s)e^{i\sigma\kappa\tau}.
\label{eqn:5.62}
\end{equation}
Using (\ref{eqn:5.62}) in (\ref{eqn:5.61}), we obtain
\begin{eqnarray}
:{\rm K}: &=& -\lambda\sigma (2\pi)^4 \sum_{s,s'}
\int \frac{d^4 k}{4\kappa^2} \kappa^2
(\varepsilon_s)^\mu (\varepsilon_{s'})_\mu 2 :\left\{
a^*(k,s) a(k,s') + a^*(k,s') a(k,s)\right\}:
\nonumber \\
&=& -\lambda\sigma (2\pi)^4 \sum_{s,s'}
\int \frac{d^4 k}{4\kappa^2} 2 \kappa^2
[g(s)\delta_{ss'} ] \left\{
a^*(k,s) a(k,s') + a^*(k,s') a(k,s)\right\}
\nonumber \\
&=& -\lambda\sigma (2\pi)^4 \sum_{s}g(s)
\int d^4 k \ a^*(k,s) a(k,s)
\label{eqn:5.63}
\end{eqnarray}
where we have used (\ref{eqn:5.32}).

%
\section{Perturbation Theory}
\setcounter{equation}{0}

We define $|\psi\rangle  $ to be a $\tau$-independent Heisenberg state and 
\begin{equation}
|\psi (\tau) \rangle   = U(\tau) |\psi(0)\rangle
=  U(\tau) |\psi\rangle
\label{eqn:6.1}
\end{equation}
to be the corresponding Schr\"odinger state.  We require that for
all $\tau$,
\begin{equation}
\langle  \psi (\tau)|\psi (\tau)\rangle   =
\langle  \psi |\overline{U} U | \psi \rangle  =
\langle  \psi |  \psi \rangle ,
\label{eqn:6.2}
\end{equation}
so that
\begin{equation}
\overline{U} (\tau) U (\tau) = 1
\label{eqn:6.3}
\end{equation}
Since $U$ is unitary and $U(0)=1$, it may be expressed in terms of
a Hermitian generator $K(\tau)$, such that
\begin{equation}
U(\tau) = e^{-iK(\tau)}  \qquad \qquad \left. i\partial_\tau U(\tau)
\right|_{\tau=0} = K'(0).
\label{eqn:6.4}
\end{equation}
But also,
\begin{equation}
U(\tau +\tau') = U(\tau) U(\tau')
\label{eqn:6.5}
\end{equation}
so that
\begin{eqnarray}
i\partial_\tau U(\tau) &=& \lim_{\tau' \rightarrow 0}
i \frac{U(\tau + \tau') - U(\tau)}{\tau'}
\nonumber \\
&=& \lim_{\tau' \rightarrow 0} i
\frac{U(\tau')U(\tau) - U(0)U(\tau)}{\tau'}
\nonumber \\
&=& \lim_{\tau' \rightarrow 0} i
\frac{U(\tau') - U(0)}{\tau'} \ U(\tau)
\nonumber \\
&=& \left. i\partial_\tau U(\tau) \right|_{\tau =0} \ U(\tau)
\nonumber \\
&=& K'(0)  \ U(\tau).
\label{eqn:6.6}
\end{eqnarray}
Therefore, we find
\begin{equation}
K(\tau) = {\rm K} \tau
\label{eqn:6.7}
\end{equation}
which says that $|\psi (\tau)\rangle  $ satisfies the Schr\"odinger
equation.  For a $\tau$-independent operator $A_S$ in the
Schr\"odinger picture,
\begin{equation}
A(\tau) = \langle  \psi (\tau)|A_S | \psi (\tau)\rangle   =
\langle  \psi |\overline{U}(\tau) A_S U(\tau) | \psi \rangle
= \langle  \psi |A_H(\tau) | \psi \rangle   
\label{eqn:6.8}
\end{equation}
where the Heisenberg operator $A_H(\tau)$ satisfies
\begin{equation}
i\partial_\tau A_H(\tau) = i\partial_\tau [e^{i{\rm K}\tau}
A_S e^{-i{\rm K}\tau} ] = [ A_H(\tau), {\rm K}] 
\label{eqn:6.9}
\end{equation}

Suppose that
\begin{equation}
{\rm K}={\rm K}_0 + {\rm K}'
\label{eqn:6.10}
\end{equation}
where ${\rm K}_0$ is the Hamiltonian of a free event.  The
field operator $\phi_H$ for the interacting event must satisfy
(\ref{eqn:6.9}) and must have a corresponding Schr\"odinger
operator such that
\begin{equation}
\phi_H (x,\tau) = e^{i{\rm K}\tau} \phi_S (x) e^{-i{\rm K}\tau} 
\label{eqn:6.11}
\end{equation}
The interaction picture operator is defined by \cite{scattering}
\begin{equation}
\phi_I (x,\tau) = e^{i{\rm K}_0\tau} \phi_S (x) e^{-i{\rm K}_0\tau}
= e^{i{\rm K}_0\tau} e^{-i{\rm K}\tau} \phi_H (x,\tau)
e^{i{\rm K}\tau} e^{-i{\rm K}_0\tau}
\label{eqn:6.12}
\end{equation}
so that in the limit that ${\rm K}' \rightarrow 0$, $\phi_I
\rightarrow \phi_H$.  Since $\phi_S (x)$ is $\tau$-independent,
$\phi_I$ satisfies
\begin{equation}
i\partial_\tau  \phi_I (x,\tau)  = [ \phi_I (x,\tau) , {\rm K}_0] 
\label{eqn:6.13}
\end{equation}
while the states of the interaction picture are
\begin{equation}
|\phi (\tau)\rangle  _I = e^{i{\rm K}_0\tau} |\phi (\tau)\rangle  _S =
e^{i{\rm K}_0\tau} e^{-i{\rm K}\tau} |\phi \rangle  _H
\label{eqn:6.14}
\end{equation}
and go over to the Heisenberg states in the absence of
interaction.  From (\ref{eqn:6.14}) we see that
\begin{eqnarray}
i\partial_\tau |\phi (\tau)\rangle  _I &=& e^{i{\rm K}_0\tau}
{\rm K} e^{-i{\rm K}\tau} |\phi (\tau)\rangle  _H -
e^{i{\rm K}_0\tau} {\rm K}_0  e^{-i{\rm K}\tau} |\phi (\tau)\rangle  _H
\nonumber \\
&=&  e^{i{\rm K}_0\tau} [{\rm K}- {\rm K}_0 ]
e^{-i{\rm K}\tau} |\phi (\tau)\rangle  _H
\nonumber \\
&=& e^{i{\rm K}_0\tau} {\rm K}' e^{-i{\rm K}_0\tau}
e^{i{\rm K}_0\tau} e^{-i{\rm K}\tau} |\phi (\tau)\rangle  _H
\nonumber \\
&=& e^{i{\rm K}_0\tau} {\rm K}' e^{-i{\rm K}_0\tau}
|\phi (\tau)\rangle  _I
\nonumber \\
&=& {\rm K}_I (\tau) |\phi (\tau)\rangle  _I.
\label{eqn:6.15}
\end{eqnarray}
Equation (\ref{eqn:6.15}) has the formal solution
\begin{equation}
|\phi (\tau)\rangle  _I = V(\tau,\tau') |\phi (\tau')\rangle  _I =
{\rm T} e^{-i\int_{\tau'}^\tau d\tau'' {\rm K}_I (\tau'')}
|\phi (\tau')\rangle  _I
\label{eqn:6.16}
\end{equation}
in which the symbol T refers to $\tau$-ordering.  The S-matrix is
given by
\begin{equation}
S= \lim_{\stackrel{\tau' \rightarrow -\infty}{\tau \rightarrow \infty }}
{\rm T} e^{-i\int_{\tau'}^\tau d\tau'' {\rm K}_I (\tau'')} =
{\rm T} e^{-i\int d^4 x d\tau \K_I }
\label{eqn:6.17}
\end{equation}
Scattering amplitudes may be computed from the interaction
picture states as
\begin{equation}
W_{i\rightarrow f} = \langle  f|S|i \rangle   .
\label{eqn:6.18}
\end{equation}
Since the initial and final states are defined when there is no
interaction, the asymptotic interaction picture states are
identical with the asymptotic Heisenberg states.  Moreover, since
the Heisenberg picture Hamiltonian is formed of bilinear
combinations of the fields without $\tau$-derivatives, the
interacting Hamiltonian in (\ref{eqn:6.17}) 
has the same form as the Heisenberg picture Hamiltonian.

{\bf Reduction Formulas}

In order to use Wick's theorem for perturbation expansions of the
S-matrix, \cite{I-Z}, we must adapt the
LSZ reduction formulas \cite{I-Z} to the present theory.
Consider the amplitude
\begin{equation}
\langle k' \  {\rm out} | k \ {\rm in} \rangle =
\langle k' \ {\rm out} |
b^*_{in} (k) | 0 \rangle 
\label{eqn:6.19}
\end{equation}
  From (\ref{eqn:4.6}) we find that
\begin{equation}
b^*(k) = \int d^4 x e^{i(k\cdot x - \kappa \tau)}
 \ \psi^* (x,\tau) \qquad
b(k) = \int d^4 x e^{i(k\cdot x - \kappa \tau)} \ \psi (x,\tau)
\label{eqn:6.20}
\end{equation}
which we may insert in (\ref{eqn:6.19}) to obtain
\begin{equation}
\langle k' \ {\rm out} | k \ {\rm in} \rangle = \int d^4 x
\langle k' \ {\rm out} |
\psi^*_{\rm in} (x,\tau) | 0 \rangle e^{i(k\cdot x - \kappa \tau)}
\label{eqn:6.21}
\end{equation}
which is valid for arbitrary $\tau_{in}$.  Now, consider the
integral
\begin{eqnarray}
\int_{\tau_i}^{\tau_f} d\tau & & \!\!\!\!\!\!\!\!\!\!
\int d^4 x \ \partial_\tau \left\{
 \langle k' {\rm out} |
\psi^* (x,\tau) | 0 \rangle
e^{i(k\cdot x - \kappa \tau)}
\right\} =
\nonumber \\
&=& \int d^4 x e^{i(k\cdot x - \kappa \tau)}
\langle k' {\rm out} | \psi^*(x,\tau_f) - \psi^*(x,\tau_i)
| 0 \rangle \nonumber \\
&=&\langle k' \ {\rm out} | b^*_{out} (k) | 0 \rangle
- \int d^4 x e^{i(k\cdot x - \kappa \tau)} \langle k' \ {\rm out} |
\psi^*(x,\tau_i) | 0 \rangle,
\label{eqn:6.22}
\end{eqnarray}
which we may rearrange as
\begin{eqnarray}
\lefteqn{\langle k' \ {\rm out} | k \ {\rm in} \rangle =
-  \int d^4 x d\tau \partial_\tau
\left\{ e^{i(k\cdot x - \kappa \tau)} \langle k' \ {\rm out} |
\psi^* (x,\tau) | 0 \rangle \right\}
+\langle k' \ {\rm out} | b^*_{out} (k) | 0 \rangle }
\nonumber \\
& & = -  \int d^4 x d\tau e^{i(k\cdot x - \kappa \tau)}
\left\{ -i\kappa + \partial_\tau \right\}
\langle k' \ {\rm out} | \psi^* (x,\tau) | 0 \rangle 
+ \langle k' {\rm out} | b^*_{out} (k) | 0 \rangle 
\nonumber \\
& & = -  \int d^4 x d\tau e^{i(k\cdot x - \kappa \tau)}
\left\{ i\frac{k^2}{2M}  + \partial_\tau \right\}
\langle k' \ {\rm out} | \psi^* (x,\tau) | 0 \rangle 
+ \langle k' \ {\rm out} | b^*_{out} (k) | 0 \rangle 
\nonumber \\
& & = -  \int d^4 x d\tau e^{i(k\cdot x - \kappa \tau)}
\left\{ \frac{i}{2M} \Box  + \partial_\tau \right\}
\langle k' \ {\rm out} | \psi^* (x,\tau) | 0 \rangle 
+ \langle k' \ {\rm out} | b^*_{out} (k) | 0 \rangle 
\nonumber \\
& & = i \int d^4 x d\tau e^{i(k\cdot x - \kappa \tau)}
\left\{ i\partial_\tau -\frac{1}{2M} \Box \right\}
\langle k' \ {\rm out} | \psi^* (x,\tau) | 0 \rangle
+ \langle k' \ {\rm out} | b^*_{out} (k) | 0 \rangle
\nonumber \\
\label{eqn:6.23}
\end{eqnarray}
where we performed two integrations by parts in the second to
last step.  The last term is a ``disconnected term''; it corresponds to
the non-scattering path and makes no contribution to the
scattering amplitude.  Now consider the integral
\begin{eqnarray}
\int_{\tau_i}^{\tau_f} d\tau'
\int d^4 x' \ \partial_{\tau'} 
&& \!\!\!\!\!\!\!\!\!\!
 \left[ \langle 0 | {\rm T} \psi (x',\tau')
\psi^* (x,\tau) | 0 \rangle 
e^{-i(k'\cdot x' - \kappa' \tau')} \right] =
\nonumber \\
 & & =\int d^4 x' e^{-i(k'\cdot x' - \kappa' \tau')}
 \langle 0 | {\rm T} \psi (x',\tau')
\psi^* (x,\tau) | 0 \rangle \left. \right|_{\tau_i}^{\tau_f}
\mbox{\hskip 2 true cm ~} 
\nonumber \\
& & = \langle 0 | b_{out} (k') \psi^* (x,\tau) | 0 \rangle
- \langle 0 | \psi^* (x,\tau) b_{in} (k') | 0 \rangle
\label{eqn:6.24}
\end{eqnarray}
in which the second term in (\ref{eqn:6.24}) vanishes.  
For the second term in (\ref{eqn:6.23}), we use (\ref{eqn:6.24})
to similarly expand as
\begin{eqnarray}
\langle k' \ {\rm out} | \psi^* (x,\tau) | 0 \rangle &=&
\langle 0 | b_{out}(k') \psi^* (x,\tau) | 0 \rangle
\nonumber \\
&=& \int d^4 x' e^{i(k'\cdot x' - \kappa' \tau')} \langle 0 |
\psi_{\rm out} (x',\tau') \psi^* (x,\tau) | 0 \rangle
\nonumber \\
&=& \int d^4 x' d\tau' \partial_{\tau'}
\left\{ e^{-i(k'\cdot x' - \kappa' \tau')}
\langle 0 |{\rm T} \psi(x',\tau') \psi^* (x,\tau) | 0 \rangle
\right\} +
\nonumber \\
& & \mbox{\qquad} \langle 0 | \psi^* (x,\tau) b_{in} (k') | 0 \rangle
\label{eqn:6.25}
\end{eqnarray}
and again the second term vanishes.  Finally, we arrive at
\begin{eqnarray}
\langle k' \ {\rm out} | k \ {\rm in} \rangle &=& i^2
\int d^4 x d\tau d^4 x' d\tau'
e^{i(k\cdot x - \kappa \tau)} e^{-i(k'\cdot x' - \kappa' \tau')}
\nonumber \\
& & \mbox{\qquad}
[i\partial_{\tau'} + \frac{1}{2M} \Box']
[i\partial_{\tau} - \frac{1}{2M} \Box]
\langle 0 |{\rm T} \psi(x',\tau') \psi^* (x,\tau) | 0 \rangle
\label{eqn:6.26}
\end{eqnarray}
By continuing this process, we can write
\begin{eqnarray}
\lefteqn{\langle k'_1 \cdots k'_n \ {\rm out} | k_1 \cdots k_m \
{\rm in} \rangle =
i^{n+m} \int d^5 x_1 \cdots d^5 x_m d^5 x'_1 \cdots d^5 x'_n
} \nonumber \\
& & \mbox{\qquad} e^{i(k_1\cdot x_1 +\cdots + k_m\cdot x_m
- \kappa_1 \tau_1 - \cdots - \kappa_m \tau_m)}
e^{-i(k'_1\cdot x'_1 +\cdots + k'_n\cdot x'_n
- \kappa'_1 \tau'_1 - \cdots - \kappa'_n \tau'_n)}
\nonumber \\
& & \mbox{\qquad}
[i\partial_{\tau'_1} + \frac{1}{2M} \Box'_1] \cdots 
[i\partial_{\tau'_n} + \frac{1}{2M} \Box'_n]
[i\partial_{\tau_1} - \frac{1}{2M} \Box_1]  \cdots
[i\partial_{\tau_m} - \frac{1}{2M} \Box_m]
\nonumber \\
& & \mbox{\qquad}
\langle 0 |{\rm T} \psi(x'_1,\tau'_1) \cdots \psi(x'_n,\tau'_n)
\psi^* (x_1,\tau_1) \cdots
\psi^* (x_m,\tau_m)| 0 \rangle
\label{eqn:6.27}
\end{eqnarray}
where we denote $d^5 x = d^4 xd\tau$.

For the gauge field, the reduction formula follows closely the
development for the usual Maxwell case.  We begin with
\begin{eqnarray}
\langle \beta , k,s \ {\rm out} | \alpha \ {\rm in}\rangle &=&
\langle \beta \ {\rm out} | a(k,s) | \alpha \ {\rm in}\rangle
\nonumber \\
&=& \frac{-i\sigma g(s)}{(2\pi)^4} \int d^4 x e^{-i(k\cdot x +
\sigma \kappa \tau)} \langle \beta \ {\rm out} | \varepsilon_s \cdot
\stackrel{\leftrightarrow}{\partial}_\tau a(x,\tau)| \alpha \ {\rm in}\rangle
\label{eqn:6.28}
\end{eqnarray}
where $\alpha$ and $\beta$ are the other quantum numbers for the
states.  We may again use
\begin{equation}
\int_{\tau_i}^{\tau_f} d\tau \int d^4 x \partial_\tau g(x,\tau)
= \int d^4 x [ g(x,\tau_f) - g(x,\tau_i) ]
\label{eqn:6.29}
\end{equation}
to write
\begin{eqnarray}
\lefteqn{\int_{\tau_i}^{\tau_f}  d\tau \int d^4 x \partial_\tau
\frac{-i\sigma g(s)}{(2\pi)^4}e^{-i(k\cdot x +
\sigma \kappa \tau)} \langle \beta \ {\rm out} | \varepsilon_s \cdot
\stackrel{\leftrightarrow}{\partial}_\tau a(x,\tau)| \alpha \ 
{\rm in}\rangle =} \nonumber \\
& & = \frac{-i\sigma g(s)}{(2\pi)^4} \int d^4 x e^{-i(k\cdot x +
\sigma \kappa \tau)} \left\{ \langle \beta \ {\rm out} | \varepsilon_s \cdot
\stackrel{\leftrightarrow}{\partial}_\tau a(x,\tau_f)|
\alpha \ {\rm in}\rangle 
\right. \nonumber \\ & & \mbox{\qquad\qquad} \left.
\!-\! \langle \beta \ {\rm out} | \varepsilon_s \cdot
\stackrel{\leftrightarrow}{\partial}_\tau a(x,\tau_i)|
\alpha \ {\rm in}\rangle \right\}
\nonumber \\
& & = \frac{-i\sigma g(s)}{(2\pi)^4} \int d^4 x e^{-i(k\cdot x +
\sigma \kappa \tau)} \langle \beta \ {\rm out} | \varepsilon_s \cdot
\stackrel{\leftrightarrow}{\partial}_\tau a(x,\tau_f)| \alpha \
{\rm in}\rangle
\label{eqn:6.30}
\end{eqnarray}
where we have used the fact that $a(x,\tau_i)|0 \ {\rm in}\rangle =0$.
By expanding $\stackrel{\leftrightarrow}{\partial}_\tau$, using
the wave equation (\ref{eqn:5.13}), and performing
two integrations by parts, we arrive at
\begin{equation}
\langle \beta , k,s \ {\rm out} | \alpha \ {\rm in}\rangle =
\frac{-ig(s)}{(2\pi)^4} \int d^5 x e^{-i(k\cdot x +
\sigma \kappa \tau)} [\Box + \sigma \partial_\tau^2 ]
 \langle \beta \ {\rm out} | \varepsilon_s \cdot
 a(x,\tau)| \alpha \ {\rm in}\rangle
\label{eqn:6.31}
\end{equation}
A second application of this procedure enables us to write
\begin{eqnarray}
\langle \beta , k,s \ {\rm out} &&
\!\!\!\!\!\!\!\!\!\! | \alpha , k',s' \
{\rm in}\rangle =
\frac{(-i)^2 g(s)g(s')}{(2\pi)^8} \int d^5 x d^5 x' e^{-i(k\cdot x +
\sigma \kappa \tau)} e^{i(k'\cdot x' + \sigma \kappa' \tau')}
\nonumber \\ && \mbox{\qquad}
[\Box + \sigma \partial_\tau^2 ] [\Box' + \sigma \partial_{\tau'}^2 ]
\langle \beta \ {\rm out} | \varepsilon_s \cdot  a(x,\tau)
\varepsilon_{s'} \cdot a(x',\tau') | \alpha \ {\rm in}\rangle
\label{eqn:6.32}
\end{eqnarray}

{\bf Feynman Rules}

In order to write the Feynman rules for the interacting off-shell
theory, may use the general expression for Green's functions,
which may be derived from the path integral \cite{ryder} or by
operator methods \cite{I-Z},
\begin{equation}
G^{(n)} (x_1,\tau_1 , \cdots , x_n,\tau_n ) = \langle 0| {\rm T}
\phi (x_1,\tau_1) , \cdots , \phi (x_n,\tau_n)
e^{i\int d^4 y d\tau \L_{\rm int} } | 0 \rangle
\label{eqn:6.33}
\end{equation}
in which vacuum-vacuum diagrams are excluded from (\ref{eqn:6.33}),
and $\phi$ represents the non-interacting free fields of the theory,
which permits us to use Wick's theorem and the free propagators
(\ref{eqn:4.16}) and (\ref{eqn:5.48}).  In (\ref{eqn:6.33}), we
adopt the conventional notation for Green's functions in field
theory; in relation to the non-interaction propagators calculated
above, these are
\begin{eqnarray}
G^{(2)} (x,\tau) &=&  \langle 0| {\rm T}
\psi (x,\tau) \psi^*(0) | 0 \rangle _{\rm tree} =
-i G(x,\tau)
\nonumber \\
d^{(2)}_{\mu\nu} (x,\tau) &=& \langle 0| {\rm T}
a_\mu (x,\tau) a_\nu (0) | 0 \rangle _{\rm tree} =
-i d_{\mu\nu} (x,\tau)
\label{eqn:6.green-defs}
\end{eqnarray}
where $G(x,\tau)$ and $d_{\mu\nu} (x,\tau)$ are defined in
(\ref{eqn:4.12}) and (\ref{eqn:5.48}).

Using (\ref{eqn:2.37}) for the interacting Hamiltonian, we have
\begin{equation}
\L_{\rm int} = -\K_{\rm interaction} = - \frac{ie_0}{2M} a_\mu
(\psi^* \partial^\mu \psi - (\partial^\mu \psi^*) \psi )
- \frac{e_0^2}{2M} a_\mu a^\mu |\psi|^2.
\label{eqn:6.34}
\end{equation}
The two terms in (\ref{eqn:6.34}) correspond to the two basic
diagrams
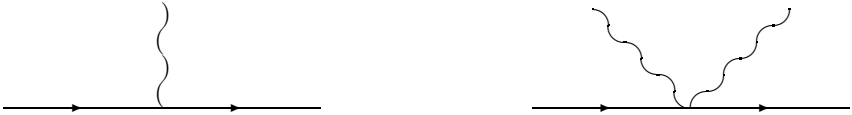
\newline
\begin{picture}(18000,12000)(0,12000)

  \drawline\photon[\N\REG](10000,16000)[4]
  \drawline\fermion[\W\REG](\photonfrontx,\photonfronty)[6000]
\drawarrow[\E\ATTIP](\pmidx,\pmidy)
  \drawline\fermion[\E\REG](\photonfrontx,\photonfronty)[6000]
\drawarrow[\E\ATTIP](\pmidx,\pmidy)

  \drawline\photon[\NE\REG](30000,16000)[6]
  \drawline\photon[\NW\REG](30000,16000)[6]
  \drawline\fermion[\W\REG](\photonfrontx,\photonfronty)[6000]
\drawarrow[\E\ATTIP](\pmidx,\pmidy)
  \drawline\fermion[\E\REG](\photonfrontx,\photonfronty)[6000]
\drawarrow[\E\ATTIP](\pmidx,\pmidy)

\end{picture}
\newline
and we may calculate the vertex factors separately.  For the
first diagram, we have
\begin{equation}
G^{(3)}_\mu (x_1,\tau_1,x_2,\tau_2,x_3,\tau_3) = \langle 0| {\rm T}
\psi^*(x_1,\tau_1) \psi (x_2,\tau_2) a_\mu (x_3,\tau_3)
e^{i\int d^4 y d\tau e_0 a_\mu (y,\tau)
j^\mu (y,\tau) } | 0 \rangle _{\rm tree}
\label{eqn:6.35}
\end{equation}
where only the tree-level diagram is considered in (\ref{eqn:6.35})
and $j^\mu$ is the vector part of the free matter field current
(\ref{eqn:1.23}),
\begin{equation}
j^\mu = \frac{-i}{2M} (\psi^* \partial^\mu \psi -
(\partial^\mu \psi^*) \psi )
\label{eqn:6.36}
\end{equation}
Expanding the exponential in (\ref{eqn:6.35}), we find that the
tree-level term is given by
\begin{eqnarray}
\lefteqn{G^{(3)}_\mu (x_1,\tau_1,x_2,\tau_2,x_3,\tau_3) =} \nonumber \\
& &=\langle 0| {\rm T}
\psi^*(x_1,\tau_1) \psi (x_2,\tau_2) a_\mu (x_3,\tau_3)
\left\{ie_0\int d^4 y d\tau a_\mu (y,\tau)
j^\mu (y,\tau) \right\} | 0 \rangle
\nonumber \\
& & = \frac{e_0}{2M}  \langle 0| {\rm T}
\psi^*(x_1,\tau_1) \psi (x_2,\tau_2) a_\mu (x_3,\tau_3)
\int d^4 y d\tau a^\nu (y,\tau) \psi^* (y,\tau)
\stackrel{\leftrightarrow}{\partial}_{y^\nu} \psi (y,\tau)| 0 \rangle
\nonumber \\
& & = \frac{e_0}{2M}\int d^4 y d\tau 
\langle 0| {\rm T} a_\mu (x_3,\tau_3) a^\nu (y,\tau) | 0 \rangle
\langle 0| {\rm T} \psi^*(x_1,\tau_1) \psi (x_2,\tau_2)
\psi^* (y,\tau)
\stackrel{\leftrightarrow}{\partial}_{y^\nu} \psi (y,\tau)| 0 \rangle
\nonumber \\
& & = \frac{e_0}{2M}\int d^4 y d\tau
\langle 0| {\rm T} a_\mu (x_3,\tau_3) a^\nu (y,\tau) | 0 \rangle
\nonumber \\ & &\mbox{\qquad\qquad\qquad} 
[\langle 0|{\rm T}\psi(x_2,\tau_2)\psi^*(y,\tau) | 0 \rangle
\stackrel{\leftrightarrow}{\partial}_{y^\nu}
\langle 0|{\rm T}\psi^*(x_1,\tau_1)\psi (y,\tau)| 0
\rangle]
\nonumber \\
& & = \frac{e_0}{2M}\int d^4 y d\tau [-id_\mu^\nu (x_3-y,\tau-\tau_3)]
\nonumber \\ & &\mbox{\qquad\qquad\qquad} 
[(-iG(x_2-y,\tau_2 - \tau))
\stackrel{\leftrightarrow}{\partial}_{y^\nu}
(-iG(y-x_1,\tau - \tau_1))]
\label{eqn:6.37}
\end{eqnarray}
We now use the Fourier expansions for the free propagators to
write
\begin{eqnarray}
\lefteqn{G^{(3)}_\mu (x_1,\tau_1,x_2,\tau_2,x_3,\tau_3) =} \nonumber \\
& & = \frac{ie_0}{2M}\int d^4 y d\tau
\frac{d^5 k d^5 p d^5 p'}{(2\pi)^{15} \lambda}
\P^\nu_\mu (k) 
\frac{e^{i[ k\cdot (x_3-y) + \sigma \kappa (\tau_3-\tau)]}}
{k^2 +\sigma \kappa^2 -i\epsilon}
\nonumber \\ & & \mbox{\qquad\qquad\qquad}
\left[ \frac{e^{i(p'\cdot (x_2- y)- P'(\tau_2-\tau))}}
{\frac{1}{2M} (p')^2 -P' -i\epsilon}
\stackrel{\leftrightarrow}{\partial}_{y^\nu}
\frac{e^{i(p\cdot (y-x_1)- P(\tau-\tau_1))}}
{\frac{1}{2M} p^2 -P -i\epsilon} \right]
\nonumber \\
&  & = \frac{ie_0}{2M} \int d^4 y d\tau
\frac{d^5 k d^5 p d^5 p'}{(2\pi)^{15} \lambda}
\P^\nu_\mu (k) \ i (p+p')^\nu \
e^{iy\cdot [p-p'-k] -i\tau [P-P'+\sigma \kappa]}
\nonumber \\ & & \mbox{\qquad\qquad\qquad}
\frac{e^{i( k\cdot x_3 + \sigma \kappa \tau_3)}}
{k^2 +\sigma \kappa^2 -i\epsilon}
\frac{e^{i(p'\cdot x_2)- P'\tau_2)}}
{\frac{1}{2M} (p')^2 -P' -i\epsilon}
\frac{e^{-i(p\cdot x_1)- P\tau_1)}}
{\frac{1}{2M} p^2 -P -i\epsilon}
\nonumber \\
&  & = \frac{-e_0}{2M}\int \frac{d^5 k d^5 p d^5 p'}{(2\pi)^{15} \lambda}
\P^\nu_\mu (k) \ (p+p')^\nu \
(2\pi)^5 \delta^4(p-p'-k)\delta(P-P'+\sigma \kappa)
\nonumber \\ & & \mbox{\qquad\qquad\qquad}
\frac{e^{i( k\cdot x_3 + \sigma \kappa \tau_3)}}
{k^2 +\sigma \kappa^2 -i\epsilon}
%
\frac{e^{i(p'\cdot x_2)- P'\tau_2)}}
{\frac{1}{2M} (p')^2 -P' -i\epsilon}
\frac{e^{-i(p\cdot x_1)- P\tau_1)}}
{\frac{1}{2M} p^2 -P -i\epsilon}
\label{eqn:6.38}
\end{eqnarray}
Transforming to the momentum space Green's function, with diagram
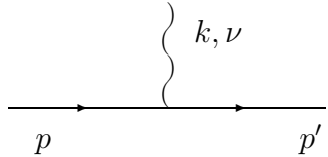
\newline
\begin{picture}(18000,13000)(0,10000)

  \drawline\photon[\N\REG](20000,16000)[4]
  \global\advance\pmidy by   500
  \global\advance\pmidx by  1000
  \put(\pmidx,\pmidy){$k,\nu$}
  \drawline\fermion[\W\REG](\photonfrontx,\photonfronty)[6000]
\drawarrow[\E\ATTIP](\pmidx,\pmidy)
  \global\advance\pmidy by -1500
  \global\advance\pmidx by -2000
  \put(\pmidx,\pmidy){$p$}
  \drawline\fermion[\E\REG](\photonfrontx,\photonfronty)[6000]
\drawarrow[\E\ATTIP](\pmidx,\pmidy)
  \global\advance\pmidy by -1500
  \global\advance\pmidx by  2000
  \put(\pmidx,\pmidy){$p'$}

\end{picture}
\newline
we find
\begin{eqnarray}
G^{(3)}_\mu (p,p',k) &=& \frac{e_0}{2M} \P^\nu_\mu (k) \ i (p+p')_\nu \
(2\pi)^5 \delta^4(p-p'-k)\delta(P-P'+\sigma \kappa)
\nonumber \\ & & \mbox{\qquad}
\frac{1}{\lambda} \frac{-i}
{k^2 +\sigma \kappa^2 -i\epsilon}
\frac{-i}
{\frac{1}{2M} (p')^2 -P' -i\epsilon}
\frac{-i}
{\frac{1}{2M} p^2 -P -i\epsilon}
\label{eqn:6.39}
\end{eqnarray}
which we identify as the product of the three propagators and the
vertex factor for the interaction
\begin{equation}
\frac{e_0}{2M} i (p+p')^\nu \
(2\pi)^5 \delta^4(p-p'-k)\delta(P-P'+\sigma \kappa)
\label{eqn:6.40}
\end{equation}
in which
\begin{equation}
\kappa=\sqrt{-\sigma k^2} \qquad P=\frac{p^2}{2M} 
\qquad P'=\frac{p'^2}{2M}
\label{eqn:6.41}
\end{equation}

For the second diagram, we calculate
\begin{eqnarray}
  \lefteqn{G^{(4)}_{\mu\nu}
  (x_1,\tau_1,x_2,\tau_2,x_3,\tau_3,x_4,\tau_4) =} 
\nonumber \\& &
  = \langle 0| {\rm T}\psi^*(x_1,\tau_1) 
  \psi (x_2,\tau_2) a_\mu (x_3,\tau_3) a_\nu (x_4,\tau_4)
  e^{-i\int d^4 y d\tau \frac{e_0^2}{2M} a_\rho a^\rho |\psi|^2}
  | 0 \rangle _{\rm tree}
\nonumber \\& &
  = \frac{-ie_0^2}{2M} \int d^4 y d\tau \langle 0| {\rm T}
  \psi^*(x_1,\tau_1) \psi (x_2,\tau_2) a_\mu (x_3,\tau_3)
  a_\nu (x_4,\tau_4)
\nonumber \\ & & \mbox{\qquad}
  a_\rho (y,\tau) a^\rho (y,\tau) \psi^*(y,\tau) \psi (y,\tau)
  | 0 \rangle
\nonumber \\& &
  = \frac{-ie_0^2}{2M} \int d^4 y d\tau 
  \langle 0| {\rm T} \psi^*(x_1,\tau_1) \psi (x_2,\tau_2)
  \psi^*(y,\tau) \psi (y,\tau) | 0 \rangle
\nonumber \\ & & \mbox{\qquad\qquad\qquad}
  \langle 0| {\rm T} a_\mu (x_3,\tau_3) a_\nu (x_4,\tau_4)
  a_\rho (y,\tau) a^\rho (y,\tau) | 0 \rangle
\nonumber \\& &
  = \frac{-ie_0^2}{2M} \int d^4 y d\tau 
  \langle 0| {\rm T} \psi^*(x_1,\tau_1) \psi (y,\tau) | 0 \rangle
  \langle 0| {\rm T} \psi (x_2,\tau_2) \psi^*(y,\tau) | 0 \rangle
\nonumber \\ & & \mbox{\qquad\qquad\qquad}
  2 \Biggl\{
  \langle 0| {\rm T} a_\mu (x_3,\tau_3) a_\rho (y,\tau) | 0 \rangle
  \langle 0| {\rm T} a_\nu (x_4,\tau_4)  a^\rho (y,\tau) | 0 \rangle
  \Biggr\}
\nonumber \\& &
  = \frac{-ie_0^2}{M} \int d^4 y d\tau
  \frac{d^5 k d^5 k' d^5 p d^5 p'}{(2\pi)^{20} \lambda^2}
  \P^\rho_\mu (k)
  \frac{e^{i[ k\cdot (y-x_3) + \sigma \kappa (\tau-\tau_3)]}}
  {k^2 +\sigma \kappa^2 -i\epsilon} \P_{\rho\nu} (k') \times
\nonumber \\ & & \mbox{\qquad\qquad} \times
  \frac{e^{i[ k'\cdot (x_4-y) + \sigma \kappa' (\tau_4-\tau)]}}
  {(k')^2 +\sigma (\kappa')^2 -i\epsilon}
  \frac{e^{i[p'\cdot (x_2- y)- P'(\tau_2-\tau)]}}
  {\frac{1}{2M} (p')^2 -P' -i\epsilon}
  \frac{e^{i[p\cdot (y-x_1)- P(\tau-\tau_1)]}}
  {\frac{1}{2M} p^2 -P -i\epsilon}
\nonumber \\& &
  = \frac{-ie_0^2}{M\lambda^2} \int
  \frac{d^5 k d^5 k' d^5 p d^5 p'}{(2\pi)^{20}}(2\pi)^5
  \delta^4(k-k'- p'+ p)
  \delta(\sigma \kappa - \sigma \kappa'+ P' - P)
  \P^\rho_\mu (k)
\nonumber \\ & & \mbox{\qquad\qquad}
  \P_{\rho\nu} (k') e^{-i[ k\cdot x_3 + \sigma \kappa \tau_3]}
  e^{i[ k'\cdot x_4 + \sigma \kappa' \tau_4]}
  e^{i[p'\cdot x_2- P'\tau_2]}
  e^{-i[p\cdot x_1- P \tau_1]}
\nonumber \\ & & \mbox{\qquad\qquad}
  \frac{-i}{k^2 +\sigma \kappa^2 -i\epsilon}
  \frac{-i}{(k')^2 +\sigma (\kappa')^2 -i\epsilon}
  \frac{-i}{\frac{1}{2M} (p')^2 -P' -i\epsilon}
  \frac{-i}{\frac{1}{2M} p^2 -P -i\epsilon}
\nonumber \\
\label{eqn:6.42}
\end{eqnarray}
Notice that a factor of 2 appears in the fourth line,
because there are two ways to contract the photon operators.
Transforming to the momentum space Green's function, with diagram
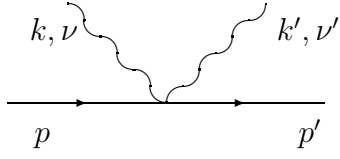
\newline
\begin{picture}(18000,13000)(0,10000)

  \drawline\photon[\NE\REG](20000,16000)[6]
  \global\advance\pmidy by   500
  \global\advance\pmidx by  2300
  \put(\pmidx,\pmidy){$k',\nu'$}
  \drawline\photon[\NW\REG](20000,16000)[6]
  \global\advance\pmidy by   500
  \global\advance\pmidx by -3300
  \put(\pmidx,\pmidy){$k,\nu$}
  \drawline\fermion[\W\REG](\photonfrontx,\photonfronty)[6000]
\drawarrow[\E\ATTIP](\pmidx,\pmidy)
  \global\advance\pmidy by -1500
  \global\advance\pmidx by -2000
  \put(\pmidx,\pmidy){$p$}
  \drawline\fermion[\E\REG](\photonfrontx,\photonfronty)[6000]
\drawarrow[\E\ATTIP](\pmidx,\pmidy)
  \global\advance\pmidy by -1500
  \global\advance\pmidx by  2000
  \put(\pmidx,\pmidy){$p'$}

\end{picture}
\newline
we find
\begin{eqnarray}
\lefteqn{G^{(4)}_{\mu\nu} (p,p',k,k') =} \nonumber \\
& & =\frac{-ie_0^2}{M\lambda^2} (2\pi)^5
\delta^4(k-k'- p'+ p)\delta(\sigma \kappa - \sigma \kappa'+ P' - P)
\P^\rho_\mu (k) \P_{\rho\nu} (k')
\nonumber \\ & & \mbox{\qquad}
\frac{-i}{k^2 +\sigma \kappa^2 -i\epsilon}
\frac{-i}{(k')^2 +\sigma (\kappa')^2 -i\epsilon}
\frac{-i}{\frac{1}{2M} (p')^2 -P' -i\epsilon}
\frac{-i}{\frac{1}{2M} p^2 -P -i\epsilon}
\label{eqn:6.43}
\end{eqnarray}
which we recognize as the product of the four propagators and the
vertex factor
\begin{equation}
\frac{-ie_0^2}{M} (2\pi)^5 g_{\mu\nu}
\delta^4(k-k'- p'+ p)\delta(\sigma \kappa - \sigma \kappa'+ P' - P)
\label{eqn:6.44}
\end{equation}

We may summarize the Feynman rules for the momentum space
Green's functions as follows:
\begin{enumerate}
\item  For each matter field propagator, draw a directed line
associated with the factor
$$\frac{1}{(2\pi)^5} \frac{-i}{\frac{1}{2M} p^2 -P -i\epsilon}$$
\item  For each photon propagator, draw a photon line associated
with the factor $$\frac{1}{\lambda} \P^{\mu\nu}
\frac{-i}{k^2 +\sigma \kappa^2 -i\epsilon}$$
\item  For the three-particle interaction,
write the vertex factor
$$\frac{e_0}{2M} i (p+p')^\nu \
(2\pi)^5 \delta^4(p-p'-k)\delta(P-P'+\sigma \kappa)$$
\item  For the four-particle interaction,
write the vertex factor
$$\mbox{\qquad}\frac{-ie_0^2}{M} (2\pi)^5 g_{\mu\nu}
\delta^4(k-k'- p'+ p)\delta(\sigma \kappa - \sigma \kappa'+ P'
- P)$$
\end{enumerate}

To obtain the Feynman rules for the S-matrix elements, we use the
reduction formulas together with the free propagators for the
incoming and outgoing particles.  From (\ref{eqn:6.27}) for the
matter fields, we see that inserting a propagator will precisely
cancel the derivative operator, so that in calculating S-matrix
elements, the incoming and outgoing propagators are replaced by
1.

Using (\ref{eqn:6.31}) for the photons, we see that the
derivative operator will cancel the factor
$-i/(k^2 +\sigma \kappa^2 -i\epsilon)$, but not the factor
$\frac{1}{\lambda}$.  So in the rules for the S-matrix elements
we replace the incoming and outgoing photon propagators with the
factor
\begin{equation}
\pm \frac{ig(s)}{(2\pi)^4 \lambda} \varepsilon^\mu (k,s)
\label{eqn:6.45}
\end{equation}
where the incoming photon takes the sign $-$ and the outgoing
photon takes the sign $+$.

%
\section{Scattering Cross-Sections}
\setcounter{equation}{0}

In this section, we calculate the relationship of the scattering
cross-section to the transition amplitudes calculated from the
S-matrix expansions discussed in the previous section.  In
particular, we specialize the five-dimensional quantum theory to
the case of scattering.  The development here generally follows the
presentation given in \cite{I-Z}, with elements taken from
\cite{goldberger} and \cite{novozhilov}.

The initial state will be represented by constructing wave packets
of the form
\begin{equation}
| \Psi_p \rangle =\int \frac{d^4 q }{(2\pi)^4}\tilde\psi_p (q)
| q \rangle =\int \frac{d^4 q }{(2\pi)^4}\tilde\psi (q-p) | q \rangle
\label{eqn:7.1}
\end{equation}
in which we take $\tilde\psi_p (q) = \tilde\psi (q-p)$ to be a
narrow distribution centered around the momentum $p$.  These wave
packets correspond to linear superpositions of solutions to the
quantum mechanical eigenvalue equation
\begin{equation}
K \psi_q = \kappa_q \psi_q 
\label{eqn:7.2}
\end{equation}
where
\begin{equation}
K = \frac{p^2}{2M} + V \qquad {\rm and} \qquad
\kappa_q = \frac{q^2}{2M} +\delta\kappa_q .
\label{eqn:7.3}
\end{equation}
The wavefunctions may be written in the form
\begin{equation}
\Psi_p (x) = \int \frac{d^4 q }{(2\pi)^4}\tilde\psi (q-p)
e^{iq\cdot x}  =
e^{ip\cdot x} \int \frac{d^4 \rho}{(2\pi)^4}  \tilde\psi (\rho)
e^{i\rho\cdot x} = e^{ip\cdot x} G(x) 
\label{eqn:7.4}
\end{equation}
in which $G(x)$ is a slowly varying, localized function of $x$.
The $\tau$-dependent wavefunctions are
\begin{equation}
\Psi(x,\tau) = e^{-iK\tau} \Psi_\rho (x) =
\int \frac{d^4q}{(2\pi)^4} \tilde\psi (q-p) 
e^{i(q\cdot x - \kappa_q \tau)}.
\label{eqn:7.5}
\end{equation}
Since the momentum distribution is narrowly centered around $p$,
the wavepacket in (\ref{eqn:7.5}), may be expanded in
$q=p+\rho$.  To do this note that
\begin{equation}
\kappa_q=\kappa_p+\rho^\mu \partial_{p^\mu} \kappa_p
+ \frac{1}{2} \rho^\mu \rho^\nu \partial_{p^\mu} \partial_{p^\nu}
\kappa_p + \cdots = \kappa_p+\rho \cdot u + o(\rho^2)
\label{eqn:7.6}
\end{equation}
where $u^\sigma = \partial \kappa_p / \partial p_\sigma$ is the group
4-velocity.  Then, to first order in $\rho$,
\begin{eqnarray}
\Psi (x,\tau) &=& \int \frac{d^4 \rho}{(2\pi)^4}
e^{i[(p+\rho)\cdot x-(\kappa_p + \rho\cdot u)\tau]} \tilde\psi(\rho)
\nonumber \\
&=& e^{i[p\cdot x-\kappa_p \tau]} \int \frac{d^4 \rho}{(2\pi)^4}
\tilde\psi(\rho) e^{i\rho\cdot [x-u \tau]}
\nonumber \\
&=& e^{i[p\cdot x-\kappa_p \tau]} G(x-u \tau)
\label{eqn:7.7}
\end{eqnarray}
and the slowly varying function $G(x-u \tau)$ describes the
Ehrenfest motion of the wavepacket.  The event density
corresponding to the wavefunction in (\ref{eqn:7.7}) is given
by the integral of the current
\begin{equation}
j^4 (x,\tau) = \overline{\Psi}_p \Psi_p = \left|G(x-u \tau)\right|^2.
\label{eqn:7.8}
\end{equation}

In scattering, the initial state is a product of
the target state and the beam state, and is represented by a
wavepacket of the form
\begin{equation}
|i\rangle = | \Psi_{\rm target} \rangle \ | \Psi_{\rm beam} \rangle
= \int \frac{d^4q_{\rm T}}{(2\pi)^4}
\frac{d^4q_{\rm B}}{(2\pi)^4}  \tilde\psi_{\rm T}
(q_{\rm T}-p_{\rm T})  \tilde\psi_{\rm B} (q_{\rm B}-p_{\rm B})
| q_{\rm T} \ q_{\rm B} \rangle .
\label{eqn:7.9}
\end{equation}
Writing the transition matrix T in terms of the scattering matrix
S \cite{scattering},
\begin{eqnarray}
{\rm S} &=& 1 + i{\rm T} 
\nonumber \\
\langle f| {\rm S} | i \rangle &=& \delta_{if} +
i \langle f| {\rm T} | i \rangle
\nonumber \\
&=& \delta_{if} + i (2\pi)^5 \delta(\kappa_f - \kappa_i)
\delta^4 (p_f - p_i) \langle f| \T | i \rangle
\label{eqn:7.10}
\end{eqnarray}
we find for the initial state defined in (\ref{eqn:7.9}),
\begin{eqnarray}
\langle f| \T | i \rangle &=& \int \frac{d^4q_{\rm T}}{(2\pi)^4}
\frac{d^4q_{\rm B}}{(2\pi)^4} 
\tilde\psi_{\rm T}(q_{\rm T}-p_{\rm T})
\tilde\psi_{\rm B} (q_{\rm B}-p_{\rm B})
\langle f | \T | q_{\rm T} q_{\rm B} \rangle .
\label{eqn:7.11}
\end{eqnarray}
The transition probability becomes
\begin{eqnarray}
|\langle f| {\rm T} | i \rangle|^2 &=& \int
\frac{d^4q_{\rm T}}{(2\pi)^4}
\frac{d^4q'_{\rm T} }{(2\pi)^4}
\frac{d^4q_{\rm B}}{(2\pi)^4}
\frac{d^4q'_{\rm B}}{(2\pi)^4} 
\tilde\psi_{\rm T}(q_{\rm T}-p_{\rm T})
\tilde\psi^*_{\rm T}(q'_{\rm T}-p_{\rm T})
\tilde\psi_{\rm B} (q_{\rm B}-p_{\rm B})
\tilde\psi^*_{\rm B} (q'_{\rm B}-p_{\rm B})
\nonumber \\ && 
\langle f | \T | q_{\rm T} q_{\rm B} \rangle
\langle f | \T | q'_{\rm T} q'_{\rm B} \rangle ^*
\delta(\kappa_f - \kappa_i) \delta^4 (p_f - p_i)
\delta(\kappa_f - \kappa'_i) \delta^4 (p_f - p'_i) .
\nonumber \\
\label{eqn:7.12}
\end{eqnarray}
where
\begin{eqnarray}
p_i &=& q_{\rm T} + q_{\rm B} = p_{\rm T} + p_{\rm B} +
\rho_{\rm T} + \rho_{\rm B}
\label{eqn:7.13}\\
p'_i &=& q'_{\rm T} + q'_{\rm B} = p_{\rm T} + p_{\rm B} +
\rho'_{\rm T} + \rho'_{\rm B}
\label{eqn:7.14}\\
\kappa_i &=& \kappa_{\rm T} + \kappa_{\rm B} =
\frac{q_{\rm T}^2}{2M_{\rm T}} + \frac{q_{\rm B}^2}{2M_{\rm B}} .
\label{eqn:7.15}
\end{eqnarray}
We assume that the interaction occurs close to the central
momenta, so that
\begin{equation}
\langle f | \T | q_{\rm T} q_{\rm B} \rangle
\langle f | \T | q'_{\rm T} q'_{\rm B} \rangle ^*
\simeq |\langle f | \T | p_{\rm T} p_{\rm B}
\rangle|^2 = |\T_{fi}|^2 .
\label{eqn:7.16}
\end{equation}
We may re-write the $\delta$-functions, replacing $q=p+\rho$ for each
momentum:
\begin{eqnarray}
\delta^4(p_f -p_i) \delta^4(p_f -p'_i) &=& \delta^4  (
p_f - p_{\rm T} - \rho_{\rm T} - p_{\rm B} - \rho_{\rm B})
\nonumber \\ && \mbox{\qquad}
\delta^4  ( p_f - p_{\rm T} - \rho'_{\rm T} -
p_{\rm B} - \rho'_{\rm B})
\nonumber \\
&=& \delta^4  ( p_f  - p_{\rm T} - p_{\rm B} 
-( \rho_{\rm T} + \rho_{\rm B} )  )
\nonumber \\ && \mbox{\qquad}
\delta^4  ( p_f - p_{\rm T} - p_{\rm B} 
- ( \rho'_{\rm T} + \rho'_{\rm B} )  )
\nonumber \\
&=& \delta^4  ( p_f - p_{\rm T} - p_{\rm B} -
( \rho_{\rm T} + \rho_{\rm B} ) )
\nonumber \\ && \mbox{\qquad}
\delta^4  ( \rho_{\rm T} + \rho_{\rm B} 
- \rho'_{\rm T} - \rho'_{\rm B} )
\nonumber \\
&\simeq& \delta^4  ( p_f - p_{\rm T} - p_{\rm B}  )
\nonumber \\ && \mbox{\qquad}
\delta^4
( \rho_{\rm T}- \rho'_{\rm T} + \rho_{\rm B} -  \rho'_{\rm B}).
\label{eqn:7.17}
\end{eqnarray}
Similarly,
\begin{equation}
\delta (\kappa_f -\kappa_i) \delta (\kappa_f -\kappa'_i)
\simeq \delta ( \kappa_{p_f} - \kappa_{p_{\rm T}}
- \kappa_{p_{\rm B}} )
\delta  (
\kappa_{\rho_{\rm T}} + \kappa_{\rho_{\rm B}}
-\kappa_{\rho'_{\rm T}} - \kappa_{\rho'_{\rm B}}).
\label{eqn:7.18}
\end{equation}
so that, since $d^4q = d^4\rho$, (\ref{eqn:7.12}) becomes
\begin{eqnarray}
\lefteqn{|\langle f| {\rm T} | i \rangle|^2 =
(2\pi)^{10}\delta ( \kappa_{p_f} - \kappa_{p_{\rm T}}
- \kappa_{p_{\rm B}} )
\delta^4  ( p_f - p_{\rm T} - p_{\rm B}  )
|\T_{fi}|^2 } \nonumber \\
&& 
\int
\frac{d^4\rho_{\rm T}}{(2\pi)^4} 
\frac{d^4\rho'_{\rm T}}{(2\pi)^4} 
\frac{d^4\rho_{\rm B}}{(2\pi)^4} 
\frac{d^4\rho'_{\rm B}}{(2\pi)^4} 
\tilde\psi_{\rm T}(\rho_{\rm T})
\tilde\psi^*_{\rm T}(\rho'_{\rm T})
\tilde\psi_{\rm B} (\rho_{\rm B})
\tilde\psi^*_{\rm B} (\rho'_{\rm B})
\delta^4 (\rho_{\rm T}- \rho'_{\rm T} + \rho_{\rm B}- \rho'_{\rm B} )
\nonumber \\ && \mbox{\qquad}
\delta (\kappa_{\rho_{\rm T}} + \kappa_{\rho_{\rm B}}
-\kappa_{\rho'_{\rm T}} - \kappa_{\rho'_{\rm B}} ) .
\label{eqn:7.19}
\end{eqnarray}
Notice that since $\kappa_\rho \simeq \rho\cdot u $,
\begin{eqnarray}
\kappa_{\rho_{\rm T}} - \kappa_{\rho'_{\rm T}}
+ \kappa_{\rho_{\rm B}} - \kappa_{\rho'_{\rm B}} &\simeq&
\rho_{\rm T} \cdot u_{\rm T} - \rho'_{\rm T} \cdot u_{\rm T}
+ \rho_{\rm B}\cdot u_{\rm B} - \rho'_{\rm B} \cdot u_{\rm B} 
\nonumber \\
&=& (\rho_{\rm T} - \rho'_{\rm T}) \cdot u_{\rm T}
+ (\rho_{\rm B}- \rho'_{\rm B}) \cdot u_{\rm B}
\nonumber \\
&=& (u_{\rm B} - u_{\rm T}) \cdot (\rho_{\rm B}- \rho'_{\rm B})
+ u_{\rm T} \cdot ( \rho_{\rm B}- \rho'_{\rm B} +
\rho_{\rm T} - \rho'_{\rm T})
\nonumber \\
&=& (u_{\rm B} - u_{\rm T}) \cdot (\rho_{\rm B}- \rho'_{\rm B})
\label{eqn:7.20}
\end{eqnarray}
where we have used (\ref{eqn:7.17}) in the last line of
(\ref{eqn:7.20}).  To continue, we write integral representations
for the $\delta$-functions.  Thus,
\begin{equation}
\delta^4 (\rho_{\rm T}- \rho'_{\rm T} + \rho_{\rm B}- \rho'_{\rm B})
= \frac{1}{(2\pi)^4} \int d^4 x
e^{ix\cdot (\rho_{\rm T}- \rho'_{\rm T} +
\rho_{\rm B}- \rho'_{\rm B})}
\label{eqn:7.21}
\end{equation}
and
\begin{equation}
\delta((u_{\rm B} - u_{\rm T}) \cdot (\rho_{\rm B}- \rho'_{\rm B}))
= \int \frac{d\alpha}{2\pi} e^{i\alpha
(u_{\rm B} - u_{\rm T}) \cdot (\rho_{\rm B}- \rho'_{\rm B})} .
\label{eqn:7.22}
\end{equation}
Inserting (\ref{eqn:7.20}), (\ref{eqn:7.21}) and (\ref{eqn:7.22}) into
(\ref{eqn:7.19}), we have
\begin{eqnarray}
\lefteqn{|\langle f| {\rm T} | i \rangle|^2 =
(2\pi)^{10}\delta ( \kappa_{p_f} - \kappa_{p_{\rm T}}
- \kappa_{p_{\rm B}} )
\delta^4  ( p_f - p_{\rm T} - p_{\rm B}  )
|\T_{fi}|^2 } \nonumber \\
&& \mbox{\qquad}
\frac{1}{(2\pi)^5} \int d^4x d\alpha
\frac{d^4\rho_{\rm T}}{(2\pi)^4} 
\frac{d^4\rho'_{\rm T}}{(2\pi)^4} 
\frac{d^4\rho_{\rm B}}{(2\pi)^4} 
\frac{d^4\rho'_{\rm B}}{(2\pi)^4} 
\tilde\psi_{\rm T}(\rho_{\rm T})
\tilde\psi^*_{\rm T}(\rho'_{\rm T})
\tilde\psi_{\rm B} (\rho_{\rm B})
\tilde\psi^*_{\rm B} (\rho'_{\rm B})
\nonumber \\
&& \mbox{\qquad}
e^{ix\cdot (\rho_{\rm T}- \rho'_{\rm T} +
\rho_{\rm B}- \rho'_{\rm B})}
e^{i\alpha(u_{\rm B} - u_{\rm T}) \cdot (\rho_{\rm B}- \rho'_{\rm B})}
\nonumber \\
&& = (2\pi)^5 \delta ( \kappa_{p_f} - \kappa_{p_{\rm T}} - 
\kappa_{p_{\rm B}} )
\delta^4  ( p_f - p_{\rm T} - p_{\rm B}  )
|\T_{fi}|^2
\nonumber \\ && \mbox{\qquad}
\int d^4x d\alpha \left|G_{\rm T} (x)\right|^2
\left| G_{\rm B}\Bigl(x+\alpha(u_{\rm B} - u_{\rm T})\Bigr)
\right| ^2 .
\label{eqn:7.23}
\end{eqnarray}
The transition probability per unit spacetime volume is then
\begin{eqnarray}
\frac{d}{dVdT} |\langle f| {\rm T} | i \rangle|^2 &=&
(2\pi)^5 \delta ( \kappa_{p_f} - \kappa_{p_{\rm T}}
- \kappa_{p_{\rm B}} )
\delta^4  ( p_f - p_{\rm T} - p_{\rm B}  )
|\T_{fi}|^2  \nonumber \\ && \mbox{\qquad}
\int d\alpha \left|G_{\rm T} (x)\right|^2
\left| G_{\rm B}\Bigl(x+\alpha(u_{\rm B} - u_{\rm T})\Bigr)
\right| ^2
\nonumber \\
&=& (2\pi)^5 \delta ( \kappa_{p_f} - \kappa_{p_{\rm T}}
- \kappa_{p_{\rm B}} )
\delta^4  ( p_f - p_{\rm T} - p_{\rm B}  )
|\T_{fi}|^2 \left|G_{\rm T} (x)\right|^2
\nonumber \\ && \mbox{\qquad} \int d\alpha 
\left| G_{\rm B}\Bigl(\alpha(u_{\rm B} - u_{\rm T})\Bigr)
\right| ^2 .
\label{eqn:7.24}
\end{eqnarray}
We take the target-beam axis to be along
the $z$-axis.  Then, the relative group 4-velocity becomes
\begin{equation}
u_{\rm B} - u_{\rm T} = (u^0_{\rm B} - u^0_{\rm T},0, 0,
u^3_{\rm B} - u^3_{\rm T})
\label{eqn:7.25}
\end{equation}
and so
\begin{equation}
\int d\alpha
\left| G_{\rm B}\Bigl(\alpha(u_{\rm B} - u_{\rm T})\Bigr)
\right| ^2
= \int d\alpha
\left| G_{\rm B}\Bigl(\alpha(u^0_{\rm B} - u^0_{\rm T}),0, 0
,\alpha(u^3_{\rm B} - u^3_{\rm T})\Bigr) \right| ^2 .
\label{eqn:7.26}
\end{equation}
Making the change of variable $\xi = \alpha(u^3_{\rm B} -
u^3_{\rm T})$, this becomes
\begin{equation}
\int d\alpha
\left|G_{\rm B}\Bigl(\alpha(u_{\rm B} - u_{\rm T})\Bigr)
\right|^2
= \frac{1}{(u^3_{\rm B} - u^3_{\rm T})}
\int d\xi \left|G_{\rm B}\Bigl(\frac{\xi}{v}, 0, 0, \xi\Bigr)
\right|^2
\label{eqn:7.27}
\end{equation}
where
\begin{equation}
v=\frac{u^3_{\rm B} - u^3_{\rm T}}{u^0_{\rm B} - u^0_{\rm T}}
\label{eqn:7.28}
\end{equation}
corresponds to the relative speed ($\sim dx / dt$) of the target
and beam events.  By (\ref{eqn:7.8}), and since $G(x-u\tau)$
satisfies
\begin{equation}
\partial_\tau G(x-u\tau) = -u^\mu \partial_{x^\mu} G(x-u\tau) .
\label{eqn:7.29}
\end{equation}
we may take $j^\mu$ to be
\begin{equation}
j^\mu = u^\mu G(x-u\tau)
\label{eqn:7.30}
\end{equation}

In order to understand the integral over $\xi$ in
(\ref{eqn:7.27}), we notice that for
\mbox{$u^3 = u^3_{\rm B} - u^3_{\rm T} >0$}, the relative motion will
cross from $x^3<0$ to $x^3>0$ during the scattering.
We define by ${\rm N}^+ (\tau)$
the proportion of all events for which $x^3>0$, given by
\begin{equation}
{\rm N}^+ (\tau) = \int_{0}^{\infty} dx^3 
\int dt d^2 x_\perp j^4 (x,\tau)
\label{eqn:7.31}
\end{equation}
where $x_\perp$ refers to the 1-2 plane.  The rate at which
events cross $x^3=0$ is then given by
\begin{eqnarray}
\frac{d}{d\tau} {\rm N}^+ (\tau) &=&
\int_{0}^{\infty} dx^3 
\int dt d^2 x_\perp \partial_\tau \ j^4 (x,\tau)
\nonumber \\
&=& \int_{0}^{\infty} dx^3 \int_{-\infty}^{\infty}
dt \int d^2 x_\perp [-u^\mu \partial_\mu \ j^4 (x,\tau) ]
\nonumber \\
&=& \int_{0}^{\infty} dx^3 
\int dt d^2 x_\perp [-u^3\partial_3 \ j^4 (x,\tau) ]
\label{eqn:7.32}
\end{eqnarray}
where in the last line, we used $j^4(x,\tau) \rightarrow 0$
as $x^\mu \rightarrow \pm\infty$.
Evaluating the integral on $dx^3$, we find
\begin{equation}
\frac{d}{d\tau} {\rm N}^+ (\tau) = -u^3 \int dt d^2 x_\perp
j^4 (t,x_\perp,x^3,\tau)|^{x^3=\infty}_{x^3 = 0}
= u^3 \int dt d^2 x_\perp j^4 (t,x_\perp,0,\tau).
\label{eqn:7.33}
\end{equation}
In terms of (\ref{eqn:7.32}) and (\ref{eqn:7.8}),
the total number of events
which cross $x^3=0$ for all $\tau$ is given by
\begin{eqnarray}
{\rm N}^+ (\infty) - {\rm N}^+ (-\infty) &=&
\int d\tau \frac{d}{d\tau} {\rm N}^+ (\tau)
\nonumber \\
&=& \int d\tau u^3 \int dt d^2 x_\perp j^4 (t,x_\perp,0,\tau)
\nonumber \\
&=& \int d\tau u^3 \int dt d^2 x_\perp 
\left|G\Bigl((t,x_\perp,0)-(u^0,0,0,u^3)\tau\Bigr)\right|^2
\nonumber \\
&=& \int d\tau u^3 \int dt d^2 x_\perp 
\left|G\Bigl(t-u^0\tau,x_\perp,-u^3\tau\Bigr)\right|^2.
\label{eqn:7.34}
\end{eqnarray}
Making the change of variables $\xi = u^3\tau$ puts
(\ref{eqn:7.34}) into the form
\begin{eqnarray}
{\rm N}^+ (\infty) - {\rm N}^+ (-\infty) &=& \int d\xi
\int dt d^2 x_\perp 
\left|G\Bigl(t-\frac{u^0}{u^3} \xi,x_\perp,-\xi\Bigr)\right|^2
\nonumber \\
&=& \int d\xi \int dt d^2 x_\perp 
\left|G\Bigl(t+\frac{1}{v} \xi,x_\perp,\xi\Bigr)\right|^2.
\label{eqn:7.35}
\end{eqnarray}
in which $v$ is given by (\ref{eqn:7.28}).
The total number of events which cross $x^3=0$, per unit
area (in the 1-2 plane) per unit time (which defines
the 3-flux $F^{(3)}$ of the beam) is just,
\begin{equation}
F^{(3)} = \int d\xi \left|G\Bigl(\frac{1}{v} \xi,0,0,\xi\Bigr)
\right|^2.
\label{eqn:7.36}
\end{equation}
Comparison with (\ref{eqn:7.27}) shows that
\begin{equation}
\int d\alpha
\left|G_{\rm B}\Bigl(\alpha(u_{\rm B} - u_{\rm T})\Bigr)
\right|^2
= \frac{1}{\left|\vec{u}_{\rm B} - \vec{u}_{\rm T}\right|} F^{(3)} 
\label{eqn:7.37}
\end{equation}

Therefore, we may re-write (\ref{eqn:7.24}) as
\begin{eqnarray}
\frac{d}{dVdT} |\langle f| {\rm T} | i \rangle|^2 &=&
(2\pi)^5 \delta ( \kappa_{p_f} - \kappa_{p_{\rm T}}
- \kappa_{p_{\rm B}} )
\delta^4  ( p_f - p_{\rm T} - p_{\rm B}  )
|\T_{fi}|^2 \left|G_{\rm T} (x)\right|^2
\nonumber \\ && \mbox{\qquad}
\frac{1}{\left|\vec{u}_{\rm B} - \vec{u}_{\rm T}\right|} F^{(3)} .
\label{eqn:7.38}
\end{eqnarray}
The scattering 3-cross-section is defined through
\begin{equation}
\frac{d}{dVdT} |\langle f| {\rm T} | i \rangle|^2 = d\sigma^{(3)}
\ \times \ {\rm beam \ flux} \ \times \ {\rm target \ density} 
\label{eqn:7.39}
\end{equation}
and so, we find that
\begin{equation}
d\sigma^{(3)} = (2\pi)^5 \delta ( \kappa_{p_f} - \kappa_{p_{\rm T}}
- \kappa_{p_{\rm B}} )
\delta^4  ( p_f - p_{\rm T} - p_{\rm B}  )
\frac{|\langle f | \T | p_{\rm T} p_{\rm B}
\rangle|^2 }{\left|\vec{u}_{\rm B} - \vec{u}_{\rm T}\right|} .
\label{eqn:7.40}
\end{equation}
\newpage
{\bf The Process ${\rm B} + {\rm T} \rightarrow 1 + 2$}

We will find it convenient to treat the case of two particle
final states in relative coordinates, and we make the following
definitions
\begin{eqnarray}
P = \frac{1}{2} (p_{\rm T} + p_{\rm B}) \qquad\qquad
p= p_{\rm T} - p_{\rm B}
\label{eqn:7.41}
\\
P' = \frac{1}{2} (p_1 + p_2) \qquad\qquad
p'= p_1 - p_2
\label{eqn:7.42}
\end{eqnarray}
which have the inverse relations
\begin{eqnarray}
p_{\rm T} = P + \frac{1}{2} p \qquad\qquad
p_{\rm B} = P - \frac{1}{2} p 
\label{eqn:7.43}
\\
p_1 = P' + \frac{1}{2} p' \qquad\qquad
p_2 = P' - \frac{1}{2} p'
\label{eqn:7.44}
\end{eqnarray}
In the center of mass system, we have that
\begin{equation}
\vec{p}_{\rm B} + \vec{p}_{\rm T} = \vec{p}_1 + \vec{p}_2 =0
\label{eqn:7.45}
\end{equation}
so that
\begin{equation}
P=\frac{1}{2} \left[ \begin{array}{c}
           E(\vec{p}_{\rm B}) + E(\vec{p}_{\rm T}) \\
           \vec{p}_{\rm B} + \vec{p}_{\rm T}
\end{array} \right] =
\left[ \begin{array}{c}
           \frac{1}{2} [E(\vec{p}_{\rm B}) + E(\vec{p}_{\rm T})] \\
           \vec{0}
\end{array} \right] =
\left[ \begin{array}{c}
           \frac{1}{2} \sqrt{s} \\
           \vec{0}
\end{array} \right] 
\label{eqn:7.46}
\end{equation}
where
\begin{equation}
s=-(p_{\rm B}+ p_{\rm T})^2
\label{eqn:7.47}
\end{equation}
is the usual Mandelstam parameter in this metric (and so
$P^2 = -s/4$), and
\begin{equation}
E(\vec{p}) = \sqrt{(\vec{p})^2 + m^2} .
\label{eqn:7.48}
\end{equation}
Similarly,
\begin{equation}
p' = p_1 - p_2 = \left[ \begin{array}{c}
E(p_1)-E(p_2)  \\ 2\vec{p}_1
\end{array} \right]
\label{eqn:7.49}
\end{equation}
and
\begin{eqnarray}
p_1^2 = P^2 + \frac{1}{4} (p')^2 + P\cdot p' = -\frac{1}{4}s
+\frac{1}{4} (p')^2 - \frac{1}{2} \sqrt{s} \ [E(p_1)-E(p_2)]
\nonumber \\
p_2^2 = P^2 + \frac{1}{4} (p')^2 - P\cdot p' = -\frac{1}{4}s
+\frac{1}{4} (p')^2 + \frac{1}{2} \sqrt{s} \ [E(p_1)-E(p_2)].
\label{eqn:7.50}
\end{eqnarray}
Since the relative momentum is spacelike, we may introduce the
parameterization
\begin{equation}
p'= \rho \left[ \begin{array}{c}
\sinh \beta  \\ \cosh \beta \hat{n}
\end{array} \right]
\label{eqn:7.51}
\end{equation}
so that (\ref{eqn:7.50}) becomes
\begin{equation}
p_1^2 = -\frac{1}{4}( s - \rho^2 + 2\sqrt{s}\rho \sinh \beta )
\qquad
p_2^2 = -\frac{1}{4}( s - \rho^2 - 2\sqrt{s}\rho \sinh \beta ) .
\label{eqn:7.52}
\end{equation}
We may see the utility of this approach
in the conventional description of relativistic scattering,
where the cross-section has the O(3,1) invariant measure
\begin{equation}
dR_2 = \delta^4  ( p_1 + p_2 - p_{\rm T} - p_{\rm B}  )
\delta (p_1^2 + m_1^2) \delta (p_2^2 + m_2^2)
\frac{d^4p_1}{(2\pi)^4}  \frac{d^4p_2}{(2\pi)^4} 
\label{eqn:7.53}
\end{equation}
which in relative coordinates becomes
\begin{eqnarray}
dR_2 &=& \frac{1}{2(2\pi)^8} \delta^4 (P-P')
\delta (p_1^2 + m_1^2) \delta (p_2^2 + m_2^2) d^4P' d^4p'
\nonumber \\
&=& \frac{1}{2(2\pi)^8} \delta (p_1^2 + m_1^2) \delta (p_2^2 + m_2^2) d^4p'
\nonumber \\
&=& \frac{1}{2(2\pi)^8} \cdot 4 \cdot 4 \cdot
\delta ( s - \rho^2 + 2\sqrt{s}\rho \sinh \beta -4m_1^2)
\delta ( s - \rho^2 - 2\sqrt{s}\rho \sinh \beta -4m_2^2)
\nonumber \\ && \mbox{\qquad}
\rho^3 \cosh^2 \beta d\rho d\beta d\Omega
\nonumber \\
&=& \frac{4}{(2\pi)^8}  \rho^3 \cosh^2 \beta d\rho d\beta d\Omega
\delta ( \rho^2 - s + (m_1^2 + m_2^2 ))
\delta ( s - \rho^2 - 2\sqrt{s}\rho \sinh \beta -4m_2^2)
\nonumber \\
&=&\frac{\rho}{(2\pi)^8\sqrt{s}} \cosh^2 \beta d\rho d\beta d\Omega
\delta \! \left[ \rho- \sqrt{s - 2(m_1^2 + m_2^2 )}\right]
\delta \! \left[ \sinh \beta - \frac{m_1^2 - m_2^2}{\sqrt{s}\rho}
\right]
\label{eqn:7.54}
\end{eqnarray}
Changing variables as
\begin{equation}
\rho \cosh^2 \beta d\beta = \rho \cosh \beta d (\sinh \beta)
%
\label{eqn:7.55}
\end{equation}
and recognizing $\rho\cosh\beta$ as $|\vec{p} \, '| = 2|\vec{p}_f|$,
this becomes,
\begin{equation}
dR_2 = \frac{2|\vec{p}_f|}{\sqrt{s}} d\Omega
\label{eqn:7.56}
\end{equation}
with
\begin{equation}
\rho = \sqrt{s - 2(m_1^2 + m_2^2 )}
\label{eqn:7.57}
\end{equation}
so that
\begin{equation}
|\vec{p}_f| = \frac{\rho}{2}\sqrt{1+
\left[\frac{m_1^2 -m_2^2}{\sqrt{s}\rho}\right]^2}
= \frac{1}{2\sqrt{s}} \sqrt{s^2 - 2s(m_1^2 + m_2^2) +
(m_1^2 -m_2^2)^2} =
\frac{\lambda^{\frac{1}{2}}(s,m_1^2,m_2^2)}{2\sqrt{s}}  
\label{eqn:7.58}
\end{equation}
in agreement with the usual derivation \cite{I-Z}.

Returning to the off-shell theory, we first consider scattering
of indistinguishable particles, which means that we may take
the invariant mass parameter $M$ to be the same for all events.
The invariant measure for the matter field is given by
\begin{eqnarray}
dR_2 &=& \delta^4(p_1 + p_2 - p_{\rm T} - p_{\rm B}  )
\delta \! \left( \frac{(p_1)^2}{2M} + \frac{(p_2)^2}{2M} -
\frac{(p_{\rm B})^2}{2M} - \frac{(p_{\rm T})^2}{2M} \right)
\frac{d^4p_1}{(2\pi)^4} \frac{d^4p_2}{(2\pi)^4}  
\nonumber \\
&=& \frac{1}{2(2\pi)^8} \delta^4 (P-P') d^4 P' \ \
2M \delta \! \left((p_1)^2 + (p_2)^2 - (p_{\rm B})^2 -
(p_{\rm T})^2 \right) d^4p'
\nonumber \\
&=& \frac{2M}{(2\pi)^8}
\delta \! \left( p^2 - p' \, ^2  \right) d^4p'
\label{eqn:7.59}
\end{eqnarray}
where we have used
\begin{eqnarray}
(p_1)^2 + (p_2)^2 &=& P' \, ^2 + \frac{1}{4} p' \, ^2 + P'\cdot p'
+ P'\, ^2 + \frac{1}{4} p' \, ^2 - P'\cdot p'
\nonumber \\
&=& 2P' \, ^2 + \frac{1}{2} p' \, ^2 
\label{eqn:7.60}\\
(p_{\rm B})^2 + (p_{\rm T})^2 &=& 2P \, ^2 + \frac{1}{2} p^2 
\label{eqn:7.61}
\end{eqnarray}
Now using the parameterization in (\ref{eqn:7.51}), we find
\begin{eqnarray}
dR_2 &=& \frac{2M}{(2\pi)^8}  \delta \! \left( p^2 - \rho^2
\right)
\rho^3 \cosh^2 \beta d\rho d\beta d\Omega
\nonumber \\
&=& \frac{2M}{(2\pi)^8} \frac{1}{2\rho} \delta \! \left(
\sqrt{p^2} - \rho \right)
\rho^3 \cosh^2 \beta d\rho d\beta d\Omega
\nonumber \\
&=& \frac{M}{(2\pi)^8} \delta \! \left( \sqrt{p^2} - \rho
\right)
(\rho \cosh \beta)^2 d\rho d\beta d\Omega
\nonumber \\
&=& \frac{M}{(2\pi)^8} \delta \! \left( \sqrt{p^2} - \rho
\right)
(2|\vec{p}_f|)^2 d\rho d\beta d\Omega
\label{eqn:7.62}
\end{eqnarray}
where the absolute value of
\begin{equation}
\vec{p}_f = \vec{p}_1 = - \vec{p}_2 = \frac{1}{2} \rho \cosh
\beta \hat{n}
\label{eqn:7.63}
\end{equation}
is undetermined in the off-shell theory, because the value of
$\beta$ is not fixed by conservation of mass-momentum.  But
notice that $\rho$ is determined by
\begin{eqnarray}
\rho^2 &=& p^2 = (p_{\rm B} - p_{\rm T} )^2
= (p_{\rm B})^2 + (p_{\rm T})^2 -2 p_{\rm B} \cdot p_{\rm T}
\nonumber \\
&=& -(p_{\rm B})^2 - (p_{\rm T})^2 -2 p_{\rm B} \cdot p_{\rm T}
+ 2[(p_{\rm B})^2 + (p_{\rm T})^2]
\nonumber \\
&=& -(p_{\rm B} + p_{\rm T})^2 + 2[(p_{\rm B})^2 + (p_{\rm T})^2]
\nonumber \\
&=& s- 2[(m_{\rm B})^2 + (m_{\rm T})^2]
\label{eqn:7.64}
\end{eqnarray}
which is identical to the usual on-shell value given in
(\ref{eqn:7.57}).  Combining (\ref{eqn:7.62}) with
(\ref{eqn:7.40}), we have for the scattering cross-section
(see also \cite{scattering})
\begin{equation}
d\sigma^{(3)} = \frac{(2\pi)^5 }{(2\pi)^8} 
\frac{|\langle p_1 p_2 | \T | p_{\rm T} p_{\rm B}
\rangle|^2 }{\left|\vec{u}_{\rm B} - \vec{u}_{\rm T}\right|}
\ 4M |\vec{p}_f|^2 d\beta d\Omega .
\label{eqn:7.65}
\end{equation}
Since $p = Mu$ for the asymptotic free matter field, we may write
this in the form
\begin{equation}
d\sigma^{(3)} = \frac{1}{(2\pi)^3}  M^2
\frac{\left|\vec{p}_f \right|^2}
{\left|\vec{p}_i \right|}
|\langle p_1 p_2 | \T | p_{\rm T} p_{\rm B}\rangle|^2 
d\beta d\Omega .
\label{eqn:7.66}
\end{equation}
where $\vec{p}_i = \vec{p}_{\rm T}  - \vec{p}_{\rm B} $,
and we must keep in mind that $\vec{p}_f$ is dependent on $\beta$.

We denote by $\Delta m^2 = (m_1)^2 - (m_2)^2 $, and notice
that from (\ref{eqn:7.52}),
\begin{equation}
\Delta m^2 = (p_2)^2 - (p_1)^2 = \sqrt{s} \ \rho \sinh \beta
\label{eqn:7.67}
\end{equation}
so that
\begin{equation}
d (\Delta m^2) = \sqrt{s} \ \rho \cosh \beta d\beta =
\sqrt{s} 2 |\vec{p}_f| d\beta .
\label{eqn:7.68}
\end{equation}
Therefore, the cross-section can be written as
\begin{equation}
d\sigma^{(3)} = \frac{1}{(2\pi)^3}  M^2
\frac{\left|\vec{p}_f \right|}
{2\sqrt{s} \left|\vec{p}_i \right|}
|\langle p_1 p_2 | \T | p_{\rm T} p_{\rm B}\rangle|^2 
d\Omega d(\Delta m^2) .
\label{eqn:7.69}
\end{equation}
We may compare the cross-section with the comparable expression
in the usual on-shell theory, which can be written in the center
of mass as \cite{I-Z}
\begin{equation}
\frac{d\sigma^{(2)}}{d\Omega} = \frac{1}{64\pi^2}
\frac{\left|\vec{p}_f \right|}{s\left|\vec{p}_i \right|}
|\langle p_1 p_2 | \T | p_{\rm T} p_{\rm B}\rangle|^2 .
\label{eqn:7.70}
\end{equation}
Putting (\ref{eqn:7.69}) into a form similar to (\ref{eqn:7.70}),
we have
\begin{equation}
\frac{d\sigma^{(3)}}{d\Omega}  = \frac{1}{64\pi^2}  
\frac{\left|\vec{p}_f \right|}
{s \left|\vec{p}_i \right|}
|\langle p_1 p_2 | \T | p_{\rm T} p_{\rm B}\rangle|^2 
d(\frac{4 M^2 \sqrt{s}}{\pi} \Delta m^2) .
\label{eqn:7.71}
\end{equation}
Examining the Feynman rules for the single photon interaction
in Section 6, we see that the usual vertex factor of $e$ is
replaced by $e_0 / 2M = \lambda e /2M$ and photon propagator has
an extra factor of $1/\lambda$, so that the squared transition
amplitude will have a factor of $\lambda^2 /16 M^4$ in relation to
the usual on-shell case.  Combining this factor with the extra
factor in (\ref{eqn:7.71}), we have an overall extra factor of
\begin{equation}
\frac{\lambda^2 }{16 M^4} d(\frac{4 M^2 \sqrt{s}}{\pi} \Delta m^2) =
d(\frac{\sqrt{s} \lambda^2 }{4\pi M^2}\Delta m^2) 
\label{eqn:7.72}
\end{equation}
which we see has the expected units of length.  We may understand
this factor in the following way:  During the scattering, the
event will propagate a distance $\Delta x \sim (p/M) \Delta \tau$,
so that the uncertainty relation tells us that
\begin{equation}
\Delta \tau \sim \frac{\Delta x}{(p/M)} = \frac{M}{p} \Delta x
\sim \frac{1}{2 \Delta p} \frac{M}{p}
\label{eqn:7.73}
\end{equation}
So from $(\vec{p})^2 = E^2 - m^2$, we find that
\begin{equation}
\vec{p} \cdot d\vec{p} \sim m \ dm \sim \frac{1}{2} \Delta (m^2)
\label{eqn:7.74}
\end{equation}
and combining with (\ref{eqn:7.73}),
\begin{equation}
\Delta \tau \sim \frac{M}{\Delta (m^2)} .
\label{eqn:7.75}
\end{equation}
Shnerb and Horwitz \cite{nadav} have shown that for $\Delta \tau
> \lambda$, the photon-current interaction becomes uncorrelated,
so we may take $\lambda \sim \Delta \tau$.  Thus, the factor in
(\ref{eqn:7.72}) becomes
\begin{eqnarray}
d\left(\frac{\sqrt{s} \lambda^2 }{4\pi M^2} \Delta (m^2)\right)
&\sim&
d(\frac{\sqrt{s} (\Delta \tau)^2}{4\pi M}
\frac{\Delta (m^2)}{M}) \sim
d(\frac{\sqrt{s}(\Delta \tau)^2}{4\pi M} \frac{1}{\Delta \tau}) =
d(\frac{\sqrt{s}(\Delta \tau)}{4\pi M})
\nonumber \\
&\sim&
d( \Delta \tau \frac{E}{4\pi M} ) \sim
\frac{1}{4\pi }\frac{dt}{d\tau} d\tau  \sim dt_{\rm shift}
\ ,
\label{eqn:7.76}
\end{eqnarray}
where we understand $dt_{\rm shift}$ as the change in the
relative time coordinate of the two particles during the
scattering.
Thus, the usual on-shell scattering cross-section corresponds to
a specific $dt_{\rm int}$, and we may compare the off-shell
cross-sections with the usual results through the expression.
\begin{equation}
\frac{d\sigma^{(3)}}{d\Omega \ dt_{\rm int}}  = \frac{1}{64\pi^2}  
\frac{\left|\vec{p}_f \right|}
{s \left|\vec{p}_i \right|}
|\langle p_1 p_2 | \tilde{\T} | p_{\rm T} p_{\rm
B}\rangle|^2 \ ,
\label{eqn:7.77}
\end{equation}
where $\tilde{\T} = \left(\frac{\lambda}{4M^2}\right)^2 $.

%
\section{Scattering Processes}
\setcounter{equation}{0}

In this section, we will carry out explicit calculations,
in two ways, of
the transition amplitude for specific scattering processes, and
show how the results differ from the usual treatment in on-shell
quantum field theory.  The first calculation is made
explicitly from the S-matrix expansion (\ref{eqn:6.17}), and
subsequent calculations are made on the basis of the Feynman
rules.

{\bf Compton Scattering}

We begin with the interaction Hamiltonian (\ref{eqn:6.34}) and we
take the initial and final states to each consist of one electron
and one photon.  Thus,
\begin{equation}
| {\rm in } \rangle = | p \ k,s \rangle =
b^*(p) \ a^*(k,s) |0 \rangle \qquad\qquad
|{\rm out } \rangle = |p' \ k',s' \rangle = 
b^*(p') \ a^*(k',s') |0 \rangle 
\label{eqn:8.1}
\end{equation}
To lowest order (tree level), we have
\begin{eqnarray}
\langle {\rm out} |S| {\rm in} \rangle &=&
\langle {\rm out} | {\rm T} e^{-i\int d^5 y \K_I }
| {\rm in} \rangle 
\nonumber \\
&=& \langle {\rm out} | i\int d^5 y \frac{e_0^2}{2M}
a_\mu (y) a^\mu
(y) \psi (y) \psi^* (y) | {\rm in } \rangle +
\nonumber \\ &&\mbox{\qquad}
\langle {\rm out} |\frac{1}{2!} \int d^5 y d^5 y'
\left[\frac{ie_0}{2M}\right]^2 {\rm T} a_\mu (y) j^\mu (y)
a_\nu (y') j^\nu (y') | {\rm in } \rangle
\label{eqn:8.2}
\end{eqnarray}
where $j^\mu$ is given in (\ref{eqn:6.36}) and $d^5y = d^4y d\tau$.
The $\tau$-ordered product in the second term gives
\begin{eqnarray}
\int d^5 y d^5 y' {\rm T} a_\mu (y) j^\mu (y)
a_\nu (y') j^\nu (y') &=& \int d^5 y d^5 y' \left[
\theta(\tau - \tau') a_\mu (y) a_\nu (y') j^\mu (y) j^\nu (y')
+\right.
\nonumber \\ && \left.
\theta(\tau' - \tau) a_\nu (y') a_\mu (y) j^\nu (y') j^\mu (y)
\right] 
\label{eqn:8.3}
\end{eqnarray}
and
\begin{eqnarray}
\langle {\rm out} && \mbox{\hspace{-.4 in}} | \int d^5 y d^5 y' 
\theta(\tau - \tau')
a_\mu (y) a_\nu (y')  j^\mu (y) j^\nu (y')| {\rm in} \rangle =
\mbox{\hspace*{2.6 true in}}
\nonumber \\ 
&=&\int d^5 y d^5 y' \theta(\tau - \tau')
\langle 0| a(k's') b(p') a_\mu (y) a_\nu (y') j^\mu (y) j^\nu (y')
b^*(p) a^*(k,s) |0 \rangle 
\nonumber \\
&=& \int d^5 y d^5 y' \theta(\tau - \tau')
\langle 0| a(k's') a_\mu (y) a_\nu (y') a^*(k,s) |0 \rangle
\langle 0| b(p') j^\mu (y) j^\nu (y') b^*(p) |0 \rangle \ .
\nonumber \\
\label{eqn:8.4}
\end{eqnarray}
Using (\ref{eqn:5.14}) for $a_\mu (y)$ and keeping in mind that
only terms with equal numbers of creation and annihilation
operators will contribute to vacuum expectation values, we have
\begin{eqnarray}
\langle 0| && \mbox{\hspace{-.3 in}} a(k's') a_\mu (y)  
a_\nu (y') a^*(k,s) |0 \rangle =
\mbox{\hspace*{1 in}}   
\nonumber \\ 
&&\mbox{\hspace*{-.3 in}}\sum_{r,r'} \int
\frac{d^4 q}{2\kappa_q} \frac{d^4 q'}{2\kappa_{q'}}
\varepsilon^\mu_r (q) \varepsilon^\nu_{r'}(q')
[ \langle 0| a(k's') a^*(q',r') a(q,r)
a^*(k,s) |0 \rangle e^{i(q\cdot y + \sigma \kappa_q \tau)}
e^{-i(q'\cdot y' + \sigma \kappa_{q'} \tau')} +
\nonumber \\ &&\mbox{\quad}
\langle 0| a(k's') a^*(q,r) a(q',r')
a^*(k,s) |0 \rangle e^{i(q'\cdot y' + \sigma \kappa_{q'} \tau')}
e^{-i(q\cdot y + \sigma \kappa_q \tau)} ].
\label{eqn:8.5}
\end{eqnarray}
Using the commutation relations (\ref{eqn:5.43}) and discarding
the disconnected terms (contracting $k$ with $k'$), we get
\begin{equation}
\langle 0| a(k's') a^*(q,r) a(q',r')
a^*(k,s) |0 \rangle =
\frac{2\kappa' g(s')}{(2\pi)^4\lambda}\delta_{rs'}\delta^4 (q-k')
\frac{2\kappa g(s)}{(2\pi)^4\lambda}\delta_{r's}\delta^4 (q'-k)
\label{eqn:8.6}
\end{equation}
so that (\ref{eqn:8.5}) becomes
\begin{eqnarray}
\langle 0| && \mbox{\hspace{-.4 in}} a(k's') a_\mu (y)  
a_\nu (y) a^*(k,s) |0 \rangle =
\sum_{r,r'}
\int \frac{d^4 q}{2\kappa_q} \frac{d^4 q'}{2\kappa_{q'}}
\varepsilon^\mu_r (q) \varepsilon^\nu_{r'} (q')
\nonumber \\&&\mbox{\quad}
[\frac{2\kappa' g(s')}{(2\pi)^4\lambda}\delta_{r's'}\delta^4 (q'-k')
\frac{2\kappa g(s)}{(2\pi)^4\lambda}\delta_{rs}\delta^4 (q-k)
e^{i(q\cdot y + \sigma \kappa_q \tau)}
e^{-i(q'\cdot y + \sigma \kappa_{q'} \tau)}+
\nonumber \\ &&\mbox{\qquad} 
\frac{2\kappa' g(s')}{(2\pi)^4\lambda}\delta_{rs'}\delta^4 (q-k')
\frac{2\kappa g(s)}{(2\pi)^4\lambda}\delta_{r's}\delta^4 (q'-k)
e^{i(q'\cdot y + \sigma \kappa_{q'} \tau)}
e^{-i(q\cdot y + \sigma \kappa_q \tau)} ]
\nonumber \\
&=& \frac{g(s) g(s')}{(2\pi)^8\lambda^2} [
\varepsilon^\mu_s (k) \varepsilon^\nu_{s'} (k')
e^{i(k\cdot y + \sigma \kappa \tau)}
e^{-i(k'\cdot y' + \sigma \kappa' \tau')} +
\nonumber \\ &&\mbox{\hspace{1.7 in}}
\varepsilon^\mu_{s'} (k') \varepsilon^\nu_s (k)
e^{i(k\cdot y' + \sigma \kappa \tau')}
e^{-i(k'\cdot y + \sigma \kappa' \tau)}] .
\label{eqn:8.7}
\end{eqnarray}
Using (\ref{eqn:4.6}) and (\ref{eqn:6.36}), the second vacuum
expectation value is given by
\begin{eqnarray}
\langle 0| b(p') && \mbox{\hspace{-.4 in}}
 j^\mu (y) j^\nu (y') b^*(p) |0 \rangle
= \mbox{\hspace*{1 in}}   
\nonumber \\ 
&=& \langle 0| b(p') [(\psi^*(y) \partial^\mu \psi (y) -
(\partial^\mu \psi^* (y)) \psi (y)) ]
\nonumber \\&&\mbox{\qquad}
[(\psi^*(y') \partial^\mu \psi (y') -
(\partial^\mu \psi^* (y')) \psi (y')) ]
b^*(p) |0 \rangle
\nonumber \\
&=& - \frac{1}{(2\pi)^{16}}\int d^4 q d^4 q' d^4 l d^4 l'
e^{-i(q\cdot y - \kappa_{q} \tau)} e^{i(q'\cdot y - \kappa_{q'} \tau)}
e^{-i(l\cdot y' - \kappa_{l}\tau')}e^{i(l'\cdot y'-\kappa_{q'} \tau')}
\nonumber \\ &&\mbox{\qquad}
(q+q')^\mu (l+l')^\nu
\langle 0| b(p') b^*(q) b(q') b^*(l) b(l')
b^*(p) |0 \rangle .
\label{eqn:8.8}
\end{eqnarray}
Using the commutation relations (\ref{eqn:4.7}), we find
\begin{eqnarray}
\langle 0| b(p') && \mbox{\hspace{-.4 in}}
 j^\mu (y) j^\nu (y') b^*(p) |0 \rangle
= \mbox{\hspace*{1 in}}   
\nonumber \\ 
&&\mbox{\hspace{-.5 in}}=
- \int \frac{d^4 q }{(2\pi)^{4}} (p+q)^\nu (p'+q)^\mu
e^{-i(p'\cdot y - \kappa_{p'} \tau)}
e^{i(q\cdot y - \kappa_{q} \tau)}
e^{-i(q\cdot y' - \kappa_{q}\tau')}
e^{i(p\cdot y'-\kappa_{p} \tau')}
\label{eqn:8.9}
\end{eqnarray}
and similarly
\begin{eqnarray}
\langle 0| b(p') && \mbox{\hspace{-.25 in}}
 j^\nu (y') j^\mu (y) b^*(p) |0 \rangle
= \mbox{\hspace*{1 in}}   
\nonumber \\ 
&& \mbox{\hspace{-.7 in}}=
- \int \frac{d^4 q }{(2\pi)^{4}} (p+q)^\mu (p'+q)^\nu
e^{-i(p'\cdot y' - \kappa_{p'} \tau')} e^{i(q\cdot y' - \kappa_{q} \tau')}
e^{-i(q\cdot y - \kappa_{q}\tau)}e^{i(p\cdot y-\kappa_{p} \tau)} .
\label{eqn:8.10}
\end{eqnarray}
Putting these terms together, we find
\begin{eqnarray}
\langle {\rm out}
&& \mbox{\hspace{-.4 in}}  
|\frac{1}{2!} \int d^5 y d^5 y'
\left[\frac{ie_0}{2M}\right]^2 {\rm T} a_\mu (y) j^\mu (y)
a_\nu (y') j^\nu (y') | {\rm in } \rangle
= \mbox{\hspace*{1.5 in}}
\nonumber \\ 
 \mbox{\hspace*{-.8 in}} &=&
-\frac{1}{2}\frac{g(s) g(s')}{(2\pi)^8\lambda^2}
\left[\frac{e_0}{2M}\right]^2  \int d^5 y d^5 y'
\frac{d^4 q }{(2\pi)^{4}}
\nonumber \\ &&\mbox{\hspace*{-.2 in}}
[\varepsilon^\mu_s (k) \varepsilon^\nu_{s'} (k')
e^{i(k\cdot y + \sigma \kappa \tau)}
e^{-i(k'\cdot y' + \sigma \kappa' \tau')} +
\varepsilon^\mu_{s'} (k') \varepsilon^\nu_s (k)
e^{i(k\cdot y' + \sigma \kappa \tau')}
e^{-i(k'\cdot y + \sigma \kappa' \tau)}]
\nonumber \\ &&\mbox{\hspace*{-.2 in}}
[ \theta(\tau - \tau')
(p+q)^\nu (p'+q)^\mu
e^{-i(p'\cdot y - \kappa_{p'} \tau)} e^{i(q\cdot y - \kappa_{q} \tau)}
e^{-i(q\cdot y' - \kappa_{q}\tau')}e^{i(p\cdot y'-\kappa_{p} \tau')}
\nonumber \\ && \mbox{\hspace*{-.2 in}}+
\theta(\tau' - \tau)
(p+q)^\mu (p'+q)^\nu
e^{-i(p'\cdot y' - \kappa_{p'} \tau')} e^{i(q\cdot y' - \kappa_{q} \tau')}
e^{-i(q\cdot y - \kappa_{q}\tau)}e^{i(p\cdot y-\kappa_{p} \tau)}
] .
\nonumber \\
\label{eqn:8.11}
\end{eqnarray}
Performing the integrations on $d^4y$ and $d^4y'$ leads to
$\delta$-functions of $p,p',k,k'$ and $q$, so that the integration
on $d^4 q$ may be performed to arrive at
\begin{eqnarray}
\langle {\rm out} && \mbox{\hspace{-.4 in}}
  |\frac{1}{2!} \int d^5 y d^5 y'
  \left[ \frac{ie_0}{2M} \right] ^2 {\rm T} a_\mu (y) j^\mu (y)
  a_\nu (y') j^\nu (y') | {\rm in } \rangle
  = \mbox{\hspace*{1.5 in}}
\nonumber \\ 
  \mbox{\hspace*{-.8 in}} &=&
  -\frac{1}{2}\frac{g(s) g(s')}{(2\pi)^4\lambda^2}
  \left[ \frac{e_0}{2M} \right] ^2  \delta^4 (p+k-p'-k')
  \int d\tau d\tau'
\nonumber \\ &&
  \left\{ \theta(\tau - \tau') [
  \varepsilon^\mu_s (k) (2p'-k)_\mu
  \varepsilon^\nu_{s'}(k')(2p-k')_\nu \right.
\nonumber \\&&\mbox{\qquad}
  e^{i[\sigma\kappa + \kappa_{p'} - (p-k')^2/2M]\tau
  -i[\sigma\kappa' + \kappa_{p} - (p-k')^2/2M]\tau' }
\nonumber \\ &&\mbox{\quad}
  \varepsilon^\mu_s (k) (2p+k)_\mu
  \varepsilon^\nu_{s'}(k')(2p'+k')_\nu
  e^{i[\sigma\kappa - \kappa_{p} + (p+k)^2/2M]\tau'
  -i[\sigma\kappa' - \kappa_{p'} + (p+k)^2/2M]\tau }
\nonumber \\ &&
  +\theta(\tau' - \tau) [
  \varepsilon^\mu_s (k) (2p+k)_\mu
  \varepsilon^\nu_{s'}(k')(2p'+k')_\nu
\nonumber \\&&\mbox{\qquad}
  e^{i[\sigma\kappa - \kappa_{p} + (p+k)^2/2M]\tau
  -i[\sigma\kappa' - \kappa_{p'} + (p+k)^2/2M]\tau' }
\nonumber \\ &&\mbox{\quad}
  \left. \varepsilon^\mu_s (k) (2p'-k)_\mu
  \varepsilon^\nu_{s'}(k')(2p-k')_\nu
  e^{i[\sigma\kappa + \kappa_{p'} - (p-k')^2/2M]\tau'
  -i[\sigma\kappa' + \kappa_{p} - (p-k')^2/2M]\tau } \right\}
\nonumber \\
\label{eqn:8.12}
\end{eqnarray}
Treating the integrations on $d\tau$ and $d\tau'$ as in Section
4, we find that
\begin{eqnarray}
\int d\tau d\tau' \theta (\tau -\tau') e^{iA\tau} e^{-iB\tau'}
&=& \int d\tau e^{iA\tau} \frac{i}{B} e^{-iB\tau}
\nonumber \\
&=& (2\pi) \frac{i}{B} \delta(A-B)
\label{eqn:8.13}
\end{eqnarray}
where we have taken $B\rightarrow B+i\epsilon$ for the
integration.  So, (\ref{eqn:8.12}) becomes
\begin{eqnarray}
\langle {\rm out}
&& \mbox{\hspace{-.4 in}}  
|\frac{1}{2!} \int d^5 y d^5 y'
\left[\frac{ie_0}{2M}\right]^2 {\rm T} a_\mu (y) j^\mu (y)
a_\nu (y') j^\nu (y') | {\rm in } \rangle
= \mbox{\hspace*{1.5 in}}
\nonumber \\ 
 \mbox{\hspace*{-.8 in}} &=&
i\frac{g(s) g(s')}{(2\pi)^3\lambda^2}
\left[\frac{e_0}{2M}\right]^2  \delta^4 (p+k-p'-k')
\delta(\kappa_{p} - \kappa_{p'} -\sigma \kappa + \sigma
\kappa') 
\nonumber \\ &&\left\{
\frac{\varepsilon^\mu_s (k) (2p'-k)_\mu
\varepsilon^\nu_{s'}(k')(2p-k')_\nu}
{\frac{1}{2M} (p-k')^2 -(P+\sigma \kappa')} +
\frac{\varepsilon^\mu_s (k) (2p+k)_\mu
\varepsilon^\nu_{s'}(k')(2p'+k')_\nu}
{\frac{1}{2M} (p+k)^2 -(P-\sigma \kappa')} \right\} .
\nonumber \\
\label{eqn:8.14}
\end{eqnarray}

The first term in (\ref{eqn:8.2}) is
\begin{eqnarray}
\langle {\rm out} | && \mbox{\hspace{-.4 in}}
i\int d^5 y \frac{e_0^2}{2M} a_\mu (y) a^\mu
(y) \psi (y) \psi^* (y) | {\rm in } \rangle 
= \mbox{\hspace*{1.5 in}}
\nonumber \\ 
&=& i \frac{e_0^2}{2M} \int d^5 y
\langle 0| a(k's')a_\mu (y) a^\mu (y) a^*(k,s) |0 \rangle
\langle 0| b(p') \overline{\phi}(y)\phi (y)b^*(p) |0 \rangle
\nonumber \\
&=& i \frac{e_0^2}{M} \int d^5 y
\langle 0| a(k's')a_\mu (y) |0 \rangle
\langle 0| a^\mu (y) a^*(k,s) |0 \rangle
\langle 0| b(p') \overline{\phi}(y)|0 \rangle
\langle 0| \phi (y)b^*(p) |0 \rangle
\nonumber \\
&=& i \frac{e_0^2}{M} \sum_{r,r'} \int d^5 y 
\frac{d^4 q}{2\kappa_q} \frac{d^4 q'}{2\kappa_{q'}}
\frac{d^4 l}{(2\pi)^4} \frac{d^4 l'}{(2\pi)^4}
\varepsilon^\mu_r (q) \varepsilon_{\mu r'}(q')
\nonumber \\ &&\mbox{\qquad}
\langle 0| a(k's') [a(q,r)e^{i(q\cdot y + \sigma \kappa_q \tau)}
+ a^*(q,r)e^{-i(q\cdot y + \sigma \kappa_q \tau)} ]
|0 \rangle
\nonumber \\ &&\mbox{\qquad}
\langle 0| [a(q',r')
e^{i(q'\cdot y + \sigma \kappa_{q'} \tau)} +
a^*(q',r') e^{-i(q'\cdot y + \sigma \kappa_{q'} \tau)} ]
a^*(k,s) |0 \rangle
\nonumber \\ &&\mbox{\qquad}
\langle 0| b(p') b^*(l)
e^{-i(l\cdot y - \kappa_{l} \tau)}|0 \rangle
\langle 0| b(l') e^{i(l'\cdot y-\kappa_{l'} \tau)} b^*(p)
|0 \rangle
\nonumber \\
&=& i \frac{e_0^2}{M} \sum_{r,r'} \int
\frac{d^4 q}{2\kappa_q} \frac{d^4 q'}{2\kappa_{q'}}
\frac{d^4 l}{(2\pi)^4} \frac{d^4 l'}{(2\pi)^4}
\varepsilon^\mu_r (q) \varepsilon_{\mu r'}(q')
\nonumber \\&&\mbox{\qquad}
(2\pi)^5 \delta^4 (q-q'+l-l') \delta (\sigma\kappa_{l} -
\sigma\kappa_{l'} - \kappa_{q} + \kappa_{q'})
\nonumber \\ &&\mbox{\qquad}
\frac{2\kappa g(s)}{(2\pi)^4\lambda} \delta^4 (q'-k) \delta_{sr'}
\frac{2\kappa' g(s')}{(2\pi)^4\lambda} \delta^4 (q-k') \delta_{s'r}
(2\pi)^4 \delta^4 (p'-l) (2\pi)^4 \delta^4 (p-l')
\nonumber \\
&=& i \frac{e_0^2}{M} \frac{g(s)g(s')}
{(2\pi)^3\lambda^2} \varepsilon^\mu_s (k) \varepsilon_{\mu s'}(k')
\delta^4 (k-k'+p-p') \delta (\sigma\kappa_{k} -
\sigma\kappa_{k'} - \kappa_{p}+ \kappa_{p'})
\label{eqn:8.15}
\end{eqnarray}

By following the scattering matrix Feynman rules of Section 6,
we obtain for the first two diagrams
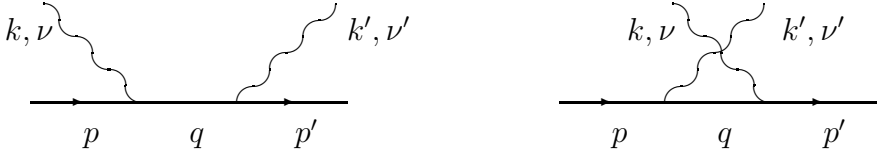
\newline
\begin{picture}(28000,13000)(0,10000)

\drawline\fermion[\W\REG](10000,16000)[6000]
  \global\advance\pmidx by -1000
  \drawarrow[\E\ATTIP](\pmidx,\pmidy)
  \global\advance\pmidy by -1500
  \put(\pmidx,\pmidy){$p$}
  \global\advance\pmidx by  2200
  \global\advance\pmidy by  1500
\drawline\photon[\NW\REG](\pmidx,\pmidy)[6]
  \global\advance\pmidy by   500
  \global\advance\pmidx by -3300
  \put(\pmidx,\pmidy){$k,\nu$}
\drawline\fermion[\E\REG](10000,16000)[6000]
  \global\advance\pmidx by  1000
  \drawarrow[\E\ATTIP](\pmidx,\pmidy)
  \global\advance\pmidy by -1500
  \put(\pmidx,\pmidy){$p'$}
  \global\advance\pmidx by -2200
  \global\advance\pmidy by  1500
\drawline\photon[\NE\REG](\pmidx,\pmidy)[6]
  \global\advance\pmidy by   500
  \global\advance\pmidx by  2300
  \put(\pmidx,\pmidy){$k',\nu'$}
\put(10000,14500){$q$}

\drawline\fermion[\W\REG](30000,16000)[6000]
  \global\advance\pmidx by -1000
  \drawarrow[\E\ATTIP](\pmidx,\pmidy)
  \global\advance\pmidy by -1500
  \put(\pmidx,\pmidy){$p$}
  \global\advance\pmidx by  2000
  \global\advance\pmidy by  1500
\drawline\photon[\NE\REG](\pmidx,\pmidy)[6]
  \global\advance\pmidy by   500
  \global\advance\pmidx by -3300
  \put(\pmidx,\pmidy){$k,\nu$}
\drawline\fermion[\E\REG](30000,16000)[6000]
  \global\advance\pmidx by  1000
  \drawarrow[\E\ATTIP](\pmidx,\pmidy)
  \global\advance\pmidy by -1500
  \put(\pmidx,\pmidy){$p'$}
  \global\advance\pmidx by -2000
  \global\advance\pmidy by  1500
\drawline\photon[\NW\REG](\pmidx,\pmidy)[6]
  \global\advance\pmidy by   500
  \global\advance\pmidx by  2300
  \put(\pmidx,\pmidy){$k',\nu'$}
\put(30000,14500){$q$}

\end{picture}
\newline
\begin{eqnarray}
  \int d^5 q && \mbox{\hspace{-.4 in}}
  (1) \times \left[ \frac{-ig(s)}{(2\pi)^4 \lambda}
  \varepsilon^\mu_s (k) \right] \times
  \left[ \frac{ie_0}{2M} (p+q)_\mu (2\pi)^5 \delta^4 (k+p-q)
  \delta (\sigma\kappa_{k} - \kappa_{p}+ \kappa_{q}) \right]
\nonumber \\ &&\mbox{\qquad}
  \times
  \left[ \frac{-i}{\frac{1}{2M}q^2 - \kappa_p -i\epsilon } \right]
\nonumber \\&&\mbox{\qquad}
  \times
  \left[\frac{ie_0}{2M} (p'+ q)_\nu (2\pi)^5 \delta^4 (k'+p'-q)
  \delta (\sigma\kappa_{k'} - \kappa_{p'}+ \kappa_{q}) \right]
\nonumber \\ &&\mbox{\qquad}
  \times \left[ \frac{ig(s')}{(2\pi)^4 \lambda}
  \varepsilon^\nu_{s'} (k') \right] \times (1) + (k,s)
  \leftrightarrow (k',s')
\nonumber \\ 
  &=& i \frac{g(s) g(s')}{(2\pi)^3\lambda^2}
  \left[\frac{e_0}{2M}\right]^2  \delta^4 (p+k-p'-k')
  \delta(\kappa_{p} - \kappa_{p'} -\sigma \kappa + \sigma \kappa') 
\nonumber \\ &&
  \left\{ \frac{\varepsilon^\mu_s (k) (2p'-k)_\mu
  \varepsilon^\nu_{s'}(k')(2p-k')_\nu}
  {\frac{1}{2M} (p-k')^2 -(P+\sigma \kappa')} +
  \frac{\varepsilon^\mu_s (k) (2p+k)_\mu
  \varepsilon^\nu_{s'}(k')(2p'+k')_\nu}
  {\frac{1}{2M} (p+k)^2 -(P-\sigma \kappa')} \right\}
\nonumber \\
\label{eqn:8.16}
\end{eqnarray}
which we see is identical to (\ref{eqn:8.14}).  Similarly, the
second diagram
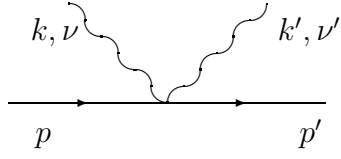
\newline
\begin{picture}(18000,13000)(0,10000)

  \drawline\photon[\NE\REG](20000,16000)[6]
  \global\advance\pmidy by   500
  \global\advance\pmidx by  2300
  \put(\pmidx,\pmidy){$k',\nu'$}
  \drawline\photon[\NW\REG](20000,16000)[6]
  \global\advance\pmidy by   500
  \global\advance\pmidx by -3300
  \put(\pmidx,\pmidy){$k,\nu$}
  \drawline\fermion[\W\REG](\photonfrontx,\photonfronty)[6000]
\drawarrow[\E\ATTIP](\pmidx,\pmidy)
  \global\advance\pmidy by -1500
  \global\advance\pmidx by -2000
  \put(\pmidx,\pmidy){$p$}
  \drawline\fermion[\E\REG](\photonfrontx,\photonfronty)[6000]
\drawarrow[\E\ATTIP](\pmidx,\pmidy)
  \global\advance\pmidy by -1500
  \global\advance\pmidx by  2000
  \put(\pmidx,\pmidy){$p'$}

\end{picture}
\newline
contributes
\begin{eqnarray}
(1)&& \mbox{\hspace{-.4 in}}
 \times \left[ \frac{-ig(s)}{(2\pi)^4 \lambda}
\varepsilon^\mu_s (k) \right] \times
\left[\frac{-ie_0^2}{M} g_{\mu\nu}(2\pi)^5 \delta^4 (k+p-p'-k')
\delta (\sigma\kappa_{k} - \kappa_{p}-\sigma\kappa_{k'} + \kappa_{p'})
\right] \times
\nonumber \\ &&\mbox{\qquad}
\times \left[ \frac{ig(s')}{(2\pi)^4 \lambda}
\varepsilon^\nu_{s'} (k') \right] \times (1) + (k,s)
\leftrightarrow (k',s')
\nonumber \\ 
&=& i \frac{e_0^2}{M} \frac{g(s)g(s')}
{(2\pi)^3\lambda^2} \varepsilon^\mu_s (k) \varepsilon_{\mu s'}(k')
\delta^4 (k-k'+p-p') \delta (\sigma\kappa_{k} -
\sigma\kappa_{k'} - \kappa_{p}+ \kappa_{p'})
\label{eqn:8.17}
\end{eqnarray}
which is identical to (\ref{eqn:8.15}).

{\bf M{\o}ller Scattering}

We now consider the scattering of two identical scalar particles,
to first order.  The diagrams which contribute are
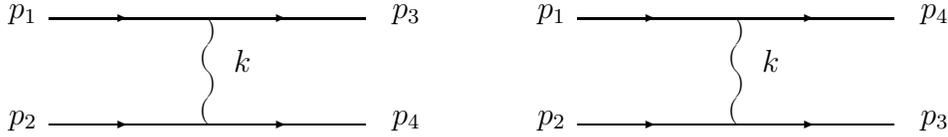
\newline
\begin{picture}(18000,13000)(0,10000)

\drawline\photon[\N\REG](10000,16000)[4]
 \global\advance\pmidx by  1000
 \put(\pmidx,\pmidy){$k$}
\drawline\fermion[\W\REG](\photonfrontx,\photonfronty)[6000]
 \drawarrow[\E\ATTIP](\pmidx,\pmidy)
 \global\advance\pmidx by -4500
 \put(\pmidx,\pmidy){$p_2$}
\drawline\fermion[\E\REG](\photonfrontx,\photonfronty)[6000]
 \drawarrow[\E\ATTIP](\pmidx,\pmidy)
 \global\advance\pmidx by  4000
 \put(\pmidx,\pmidy){$p_4$}
\drawline\fermion[\W\REG](\photonbackx,\photonbacky)[6000]
 \drawarrow[\E\ATTIP](\pmidx,\pmidy)
 \global\advance\pmidx by -4500
 \put(\pmidx,\pmidy){$p_1$}
\drawline\fermion[\E\REG](\photonbackx,\photonbacky)[6000]
 \drawarrow[\E\ATTIP](\pmidx,\pmidy)
 \global\advance\pmidx by  4000
 \put(\pmidx,\pmidy){$p_3$}

\drawline\photon[\N\REG](30000,16000)[4]
 \global\advance\pmidx by  1000
 \put(\pmidx,\pmidy){$k$}
\drawline\fermion[\W\REG](\photonfrontx,\photonfronty)[6000]
 \drawarrow[\E\ATTIP](\pmidx,\pmidy)
 \global\advance\pmidx by -4500
 \put(\pmidx,\pmidy){$p_2$}
\drawline\fermion[\E\REG](\photonfrontx,\photonfronty)[6000]
 \drawarrow[\E\ATTIP](\pmidx,\pmidy)
 \global\advance\pmidx by  4000
 \put(\pmidx,\pmidy){$p_3$}
\drawline\fermion[\W\REG](\photonbackx,\photonbacky)[6000]
 \drawarrow[\E\ATTIP](\pmidx,\pmidy)
 \global\advance\pmidx by -4500
 \put(\pmidx,\pmidy){$p_1$}
\drawline\fermion[\E\REG](\photonbackx,\photonbacky)[6000]
 \drawarrow[\E\ATTIP](\pmidx,\pmidy)
 \global\advance\pmidx by  4000
 \put(\pmidx,\pmidy){$p_4$}

\end{picture}
\newline
which contribute
\begin{eqnarray}
\langle 
&& \mbox{\hspace{-.4 in}}
3 \ 4 | iT | 1 \ 2  \rangle 
= \mbox{\hspace*{1.5 in}}
\nonumber \\ 
&=& \int d^4 q d\kappa_q  (1)^4 \left[ \frac{ie_0}{2M}
(p_1 + p_3 )^\mu (2\pi)^5 \delta^4(p_1+q-p_3) 
\delta(\kappa_{p_1} - \kappa_{p_3} -\sigma\kappa_{q}) \right]
\nonumber \\&&\mbox{\qquad}
\left[ \frac{ie_0}{2M}(p_2 + p_4 )^\nu (2\pi)^5
\delta^4(p_2-q-p_4)
\delta(\kappa_{p_2} - \kappa_{p_4} +\sigma\kappa_{q})\right]
\nonumber \\ &&\mbox{\qquad}
\frac{1}{\lambda} \frac{-i}{q^2+\sigma\kappa_{q}^2 -i\epsilon}
\P_{\mu\nu} + (3,4) \leftrightarrow (4,3)
\nonumber \\
&=& \left\{i(2\pi)^5 \delta^4(p_1+p_2-p_3-p_4)
\delta(\kappa_{p_1}+\kappa_{p_2} - \kappa_{p_3}-\kappa_{p_4} )
\right\}(2\pi)^5 \frac{e_0 e}{(2M)^2} 
\nonumber \\&&\mbox{\qquad}
\left\{ (p_1+p_3)^\mu (p_2+p_4)^\nu
\left[g_{\mu\nu} + \frac{(p_1-p_3)_\mu (p_2-p_4)_\nu}{(p_1-p_3)^2}
\right] \right.
\nonumber \\ &&\mbox{\qquad} \left.
\frac{1}{(p_1-p_3)^2+\sigma(\kappa_{p_1}-\kappa_{p_3})^2
-i\epsilon} +(3,4) \leftrightarrow (4,3) \right\} \ ,
\label{eqn:8.18}
\end{eqnarray}
where $\kappa_{p}=\frac{p^2}{2M} $.
  From (\ref{eqn:7.10}) which defines $\T$, we have
\begin{eqnarray}
  \langle 3 \ 4 | \T | 1 \ 2  \rangle 
  &=& (2\pi)^5 \frac{e_0 e}{(2M)^2}
  \left\{ \left[ (p_1+p_3)\cdot (p_2+p_4)
  + \frac{(p_1^2-p_3^2) (p_2^2-p_4^2)}{(p_1-p_3)^2}
  \right] \times \right.
\nonumber \\ && \left.
  \times \frac{1}{(p_1-p_3)^2+\sigma(\kappa_{p_1}-\kappa_{p_3})^2
  -i\epsilon} +(3,4) \leftrightarrow (4,3) \right\} .
\label{eqn:8.19}
\end{eqnarray}

At this stage it is convenient to introduce the Mandelstam
parameters
\begin{equation}
t=-(p_1 - p_3)^2 = -\frac{1}{4}(p-p')^2
\qquad\qquad u=-(p_1-p_4)^2 = -\frac{1}{4}(p+p')^2
\label{eqn:8.20}
\end{equation}
which complement the definition of $s$ in (\ref{eqn:7.47}), and
where $p,p'$ refer to the relative coordinates defined in
(\ref{eqn:7.41}) and (\ref{eqn:7.42}).  As
in the on-shell case, we have
\begin{eqnarray}
s+t+u &=& -[p_1^2 +p_2^2 +p_1^2 +p_3^2 +p_1^2 +p_4^2 +2(p_2-p_3-
p_4)\cdot p_1 ]
\nonumber \\
&=& -[p_1^2 +p_2^2 +p_3^2 +p_4^2 ]
\nonumber \\
&=& m_1^2 +m_2^2 +m_3^2 +m_4^2 .
\label{eqn:8.21}
\end{eqnarray}
Similarly, since $p_1^2 +p_2^2 = p_3^2 + p_4^2$ is guaranteed by
the `fifth' $\delta$-function, but the masses are not in general
invariant, we have
\begin{eqnarray}
(p_1+p_2)^2 &=& (p_3+p_4)^2 \Rightarrow p_1 \cdot p_2 = p_3\cdot
p_4 = -\frac{1}{2} (s+p_1^2 + p_2^2) = -\frac{1}{2} (s+p_3^2 + p_4^2)
\nonumber \\
p_1\cdot p_3 &=& \frac{1}{2} (p_1^2 +p_3^2 +t) \neq
p_2\cdot p_4 = \frac{1}{2} (p_2^2 +p_4^2 +t) 
\nonumber \\
p_1\cdot p_4 &=& \frac{1}{2} (p_1^2 +p_4^2 +u) \neq
p_2\cdot p_3 = \frac{1}{2} (p_2^2 +p_3^2 +u) 
\label{eqn:8.22}
\end{eqnarray}
Using these relations among the Mandelstam parameters, we may
re-write the terms of (\ref{eqn:8.19}) in the form,
\begin{eqnarray}
(p_1+p_3)\cdot(p_2+p_4) &=& p_1 \cdot p_2 + p_1 \cdot p_4 +
p_3 \cdot p_2 + p_3 \cdot p_4
\nonumber \\
&=& -\frac{1}{2}(s+p_1^2 + p_2^2)+\frac{1}{2} (p_1^2 +p_4^2 +u)
\nonumber \\ &&\mbox{\qquad}
+\frac{1}{2} (p_2^2 +p_3^2 +u)  -\frac{1}{2} (s+p_3^2 + p_4^2)
\nonumber \\
&=& -s + u
\label{eqn:8.23}
\end{eqnarray}
and similarly
\begin{equation}
(p_1+p_4)\cdot(p_2+p_3) = -s + t
\label{eqn:8.24}
\end{equation}
so that
\begin{equation}
\langle 3\ 4 | \T |1\ 2  \rangle 
=(2\pi)^5 \frac{e_0 e}{(2M)^2} \left\{  
\frac{ s - u -\frac{(p_1^2 - p_3^2)^2}{t} }
{t -\sigma(\kappa_{p_1}
-\kappa_{p_3})^2 } +
\frac{ s-t - \frac{(p_1^2-p_4^2)^2}{u} }
{u -\sigma(\kappa_{p_1}-\kappa_{p_4})^2 }
\right\}
\label{eqn:8.25}
\end{equation}
We notice that for the case of on-shell scattering
($m_1^2=m_3^2$), the scattering amplitude becomes
\begin{equation}
\langle 3\ 4 | \T |1\ 2  \rangle 
=(2\pi)^5 \frac{e_0 e}{(2M)^2} \left\{  
\frac{s - u }{t} + \frac{ s-t }{u} \right\}
\label{eqn:8.26}
\end{equation}

which has the form of the amplitude for the usual scattering of
identical Klein-Gordon particles \cite{I-Z}.

Consider the amplitude (\ref{eqn:8.25}) in the case of scattering
of identical particles with $m_1=m_2=m$.  In this case, in the
center of mass frame, $E(p_1)=E(p_2)$, so that
\begin{equation}
p = p_1 - p_2 = (0,  2 \vec{p}_1)
\label{eqn:8.27}
\end{equation}
and $|p| = |p'| =\rho = 2|\vec{p}_1|$.  In terms of the
parameterization (\ref{eqn:7.51}), the $\beta$ of the incoming
system is just zero.  The relative momentum of the outgoing
system is then
\begin{equation}
p' = p_3 - p_4 = 2|\vec{p}_1|(\sinh\beta,\cosh\beta \hat{n})
\label{eqn:8.28}
\end{equation}
and so the Mandelstam parameters become
\begin{eqnarray}
t &=& -\frac{1}{4} (p-p')^2
\nonumber \\
&=& -\frac{1}{4} (p^2 + p' \, ^2 -2p\cdot p')
\nonumber \\
&=& -\frac{1}{4} (2\rho^2 -2\rho^2 \cosh\beta \ \cos\theta)
\nonumber \\
&=& -2|\vec{p}_1|^2 (1-\cosh\beta \ \cos\theta) .
\label{eqn:8.29}
\end{eqnarray}
Similarly,
\begin{equation}
u= -\frac{1}{4} (p+p')^2 = -2|\vec{p}_1|^2 (1+\cosh\beta \
\cos\theta).
\label{eqn:8.30}
\end{equation}
These expressions agree with the usual on-shell expressions for
$t$ and $u$ when $\beta=0$.  Since $p_1^2 = p_2^2$, we may write
\begin{equation}
p_1^2+p_2^2 = 2p_1^2 = p_3^2 + p_4^2 
\label{eqn:8.31}
\end{equation}
so that
\begin{eqnarray}
p_1^2-p_3^2 &=& \frac{1}{2} (p_3^2 + p_4^2) -p_3^2
\nonumber \\
&=& -\frac{1}{2} (p_3^2 - p_4^2)
\nonumber \\
&=& - \frac{1}{2} (p_3-p_4)\cdot (p_3+p_4)
\nonumber \\
&=& - \frac{1}{2} p' \cdot P'
\nonumber \\
&=& \frac{1}{4} \sqrt{s} \rho \sinh \beta
\label{eqn:8.32}
\end{eqnarray}
where we have used $P'=P$, (\ref{eqn:7.46}) and (\ref{eqn:7.51}).
We see again that $\beta=0$ corresponds to on-shell scattering.
The amplitude in (\ref{eqn:8.25}) contains the term
\begin{equation}
\frac{(p_1^2-p_3^2)^2}{t} = -\frac{\frac{1}{16} s\rho^2 \sinh^2\beta}
{\frac{1}{2} \rho^2(1-\cosh\beta\cos\theta)}=
-\frac{s\sinh^2\beta}{8(1-\cosh\beta\cos\theta)} 
\label{eqn:8.33}
\end{equation}
and the denominator
\begin{equation}
t-\sigma(\kappa_{p_1} - \kappa_{p_3})^2 =
t-\frac{\sigma}{4M^2} (p_1^2 - p_3^2)^2 =
-\frac{\rho}{2} \left[ (1-\cosh\beta\cos\theta) +
\frac{\sigma s}{32M^2} \sinh^2 \beta \right]
\label{eqn:8.34}
\end{equation}
Therefore while the usual on-shell scattering amplitude has the
single forward direction pole (at $\cos\theta = 1 \Rightarrow t=0$),
the off-shell scattering amplitude has two forward direction
poles, one at
\begin{equation}
t=0 \qquad \Rightarrow \qquad \cos\theta = \frac{1}{\cosh\beta} 
\label{eqn:8.35}
\end{equation}
and the second at
\begin{equation}
t-\sigma(\kappa_{p_1} - \kappa_{p_3})^2 =0 \qquad \Rightarrow
\qquad \cos\theta = \frac{1}{\cosh\beta}
\left[ 1+\frac{\sigma s}{32M^2} \sinh^2 \beta \right].
\label{eqn:8.36}
\end{equation}
Thus, the appearance of two close, but distinct, poles in the
forward and
backward directions in M{\o}ller scattering would be a consequence
of $\Delta m^2$ not strictly vanishing and would provide a
signature for off-shell phenomena.  From the optical
theorem, we see that the finiteness of the transition amplitude
at $\theta=0$ implies finiteness of the total scattering
cross section, for a given $\beta > 0$ (the $-i\epsilon$ term in
the denominator of the transition amplitude pushes the pole
off the real axis, so that the integral over $\theta$ may be
performed).

%
\section{Renormalization}
\setcounter{equation}{0}

Unlike conventional relativistic quantum field theories, the
off-shell matter field corresponds to an underlying evolution
mechanics, and the field undergoes retarded propagation from
$\tau_1$ to $\tau_2 > \tau_1$.  Therefore, there can be no matter
field loops in off-shell QED, since a closed loop requires the
field to propagate from $\tau_1$ to $\tau_2$ and then from
$\tau_2$ to $\tau_1$.  The absence of matter field loops leads us
to expect that the charge $e_0$ will not be renormalized, and as
we demonstrate below, this follows from the absence of wave
function renormalization for the photon.

{\bf The Vector Ward Identity}  

The Ward identity in on-shell (Klein-Gordon) scalar quantum
electrodynamics expresses the symmetry associated with the
conservation of the four-current as a relationship between
the vertex function of the 3-particle interaction and the
single particle propagators.  Since the Ward identity is
preserved at all orders of perturbation theory, it
leads to the universality of charge renormalization.

In off-shell quantum electrodynamics, the conservation of
current is expressed as a vanishing {\it five-divergence}
(\ref{eqn:1.22}).  Recalling equation (\ref{eqn:6.34})
\begin{equation}
\L_{\rm int} = - \frac{ie_0}{2M} a_\mu
(\psi^* \partial^\mu \psi - (\partial^\mu \psi^*) \psi )
- \frac{e_0^2}{2M} a_\mu a^\mu |\psi|^2 \ ,
\label{eqn:L_int}
\end{equation}
we may express the interaction in terms of the current as
\begin{equation}
\L_{\rm int} = e_0 a_\mu j^\mu
- \frac{e_0^2}{2M} a_\mu a^\mu j^5
\label{eqn:L_int_j}
\end{equation}
where $(j^\mu,j^5)$ is the current for the free matter field.
Since the entire interaction Lagrangian is in the form of 
{\it photon} $\times$ {\it current}, we shall see that the
Ward identity is a relationship among the single particle
propagators and the vertex functions
of {\it both} the 3-particle interaction and the 4-particle
interaction.

We begin with the three-point Green's function
associated with the diagram
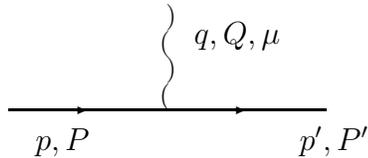
\newline
\begin{picture}(18000,13000)(0,10000)

  \drawline\photon[\N\REG](20000,16000)[4]
  \global\advance\pmidy by   500
  \global\advance\pmidx by  1000
  \put(\pmidx,\pmidy){$q,Q,\mu$}
  \drawline\fermion[\W\REG](\photonfrontx,\photonfronty)[6000]
\drawarrow[\E\ATTIP](\pmidx,\pmidy)
  \global\advance\pmidy by -1500
  \global\advance\pmidx by -2000
  \put(\pmidx,\pmidy){$p,P$}
  \drawline\fermion[\E\REG](\photonfrontx,\photonfronty)[6000]
\drawarrow[\E\ATTIP](\pmidx,\pmidy)
  \global\advance\pmidy by -1500
  \global\advance\pmidx by  2000
  \put(\pmidx,\pmidy){$p',P'$}

\end{picture}
\newline
where $p'=q+p$ and $P'=P-\sigma Q$.  To all orders in perturbation
theory, the vertex function (the amputated Green's function) is
given by $\Gamma^{(3)}_\mu (p,P;q,Q)$, where
\begin{equation}
G^{(3) \; \mu} (p,P;q,Q) = G^{(2)}(p,P) \ G^{(2)}(p',P')
\ d^{\mu\nu}(q,Q) \ \Gamma^{(3)}_\mu (p,P;q,Q) \ .
\label{eqn:def-amp}
\end{equation}
To calculate the vertex function, we may write \cite{Cheng-Li}
\begin{eqnarray}
G^{(3)}_\mu (p,P;q,Q) &=& {\cal F} \Biggl\{
\langle 0 |{\rm T} \ a_\mu(x_1,\tau_1) \ \psi^*(x_2,\tau_2)
\ \psi(x_3,\tau_3) |0 \rangle \Biggr\}
\nonumber \\
&=& {\cal F} \Biggl\{
\langle 0 |{\rm T} \ a_\mu(x_1,\tau_1) \ \psi^*(x_2,\tau_2)
\ \psi(x_3,\tau_3)
e^{ie_0 \int d^4x_4 d\tau_4 a_\nu j^\nu}|0 \rangle _{\rm free}
\Biggr\}
\nonumber \\
&=& ie_0 {\cal F} \Biggl\{
\int d^4x_4 d\tau_4
\langle 0 |{\rm T} \ a_\mu(x_1,\tau_1)
\ a_\nu (x_4,\tau_4) |0 \rangle _{\rm free}
\nonumber \\ &&\mbox{\qquad\qquad} 
\langle 0 |{\rm T} \ \psi^*(x_2,\tau_2) \ \psi(x_3,\tau_3)
\ j^\nu (x_4,\tau_4)|0 \rangle _{\rm free} \Biggr\}
\label{eqn:greens}
\end{eqnarray}
where ${\cal F}$ represents the Fourier transform.  The vertex
function becomes,
\begin{eqnarray}
\Gamma^{(3)}_\mu (p,P;q,Q) &=& \frac{1}{G^{(2)}(p) G^{(2)}(p')}
\int d^4 x d\tau d^4 x' d\tau '
e^{-i[p\cdot x' -P \tau ']} e^{-i[q\cdot x +\sigma Q \tau]}
\nonumber \\ &&\mbox{\quad}
\langle 0 |{\rm T} \ ie_0 j_\mu(x,\tau) \ \psi (x',\tau ')
 \ \psi^* (0) |0 \rangle
\label{eqn:3-pt_f}
\end{eqnarray}
where we have used translation invariance of the Green's functions
to shift one of the field points to zero.  Contracting with
$q^\mu$, we obtain
\begin{eqnarray}
q^\mu \Gamma^{(3)}_\mu (p,P;q,Q) &=&
e_0 \ \frac{1}{G^{(2)}(p) G^{(2)}(p')}
\int d^4 x d\tau d^4 x' d\tau '
\left[(-\partial^\mu_x)
\ e^{-i[p\cdot x' -P \tau ']} e^{-i[q\cdot x +\sigma Q \tau]}
\right]
\nonumber \\ &&\mbox{\qquad}
\langle 0 |{\rm T} \ j_\mu(x,\tau) \ \psi (x',\tau ')
 \ \psi^* (0) |0 \rangle 
\nonumber \\
&=&
e_0 \ \frac{1}{G^{(2)}(p) G^{(2)}(p')}
\int d^4 x d\tau d^4 x' d\tau '
e^{-i[p\cdot x' -P \tau ']} e^{-i[q\cdot x +\sigma Q \tau]}
\nonumber \\ &&\mbox{\qquad}
\langle 0 |{\rm T} \ \partial^\mu_x \ j_\mu(x,\tau) \ \psi (x',\tau ')
 \ \psi^* (0) |0 \rangle \ ,
\label{eqn:q-contract}
\end{eqnarray}
where we have performed one integration by parts.  Using
current conservation, we may make the replacement
\begin{equation}
\partial^\mu_x j_\mu(x,\tau) = -\partial_\tau j^5 (x,\tau)
\ .
\label{eqn:current}
\end{equation}
Since the products are $\tau$-ordered, we must carefully
differentiate the implied $\theta$-functions to find
\begin{eqnarray}
\partial_\tau \langle 0 |{\rm T} j^5 (x,\tau) \ \psi (x',\tau ')
\ \psi^* (0) |0 \rangle \ &=& \langle 0 |{\rm T} \partial_\tau
j^5 (x,\tau) \ \psi (x',\tau ')
\ \psi^* (0) |0 \rangle \
\nonumber \\
&&\mbox{\quad}+ \delta(\tau-\tau') \langle 0 |{\rm T}
\ [j^5 (x,\tau),\psi (x',\tau ')]
\ \psi^* (0) |0 \rangle
\nonumber \\
&&\mbox{\quad}+ \delta(\tau) \langle 0 |{\rm T}
\ [j^5 (x,\tau),\psi^* (0)]
\ \psi (x',\tau ') |0 \rangle
\nonumber \\
&=&  \langle 0 |{\rm T} \partial_\tau
j^5 (x,\tau) \ \psi (x',\tau ')
\ \psi^* (0) |0 \rangle \
\nonumber \\
&&\mbox{\quad}- \delta(\tau-\tau') \delta^4(x-x')
\langle 0 |{\rm T} \ \psi (x,\tau ) \ \psi^* (0) |0 \rangle
\nonumber \\
&&\mbox{\quad}+ \delta(\tau)\delta^4(x)
\langle 0 |{\rm T} \ \psi (x',\tau') \ \psi^* (0) |0 \rangle
\label{eqn:diff-tau}
\end{eqnarray}
where we have used the commutation relations
(\ref{eqn:4.3}).  Using (\ref{eqn:current}) and
(\ref{eqn:diff-tau}) in (\ref{eqn:q-contract}), we find
\begin{eqnarray}
q^\mu \Gamma^{(3)}_\mu (p,P;q,Q) &=&
e_0 \ \frac{1}{G^{(2)}(p) G^{(2)}(p')}
\int d^4 x d\tau d^4 x' d\tau '
\left[
\ e^{-i[p\cdot x' -P \tau ']} e^{-i[q\cdot x +\sigma Q \tau]}
\right]
\nonumber \\ &&\mbox{\qquad}
\Bigl[ -\partial_\tau
\langle 0 |{\rm T} j^5 (x,\tau) \ \psi (x',\tau ')
\ \psi^* (0) |0 \rangle
\nonumber \\ &&\mbox{\qquad}
- \delta(\tau-\tau') \delta^4(x-x')
\langle 0 |{\rm T} \ \psi (x,\tau ) \ \psi^* (0) |0 \rangle
\nonumber \\
&&\mbox{\qquad}+ \delta(\tau)\delta^4(x)
\langle 0 |{\rm T} \ \psi (x',\tau') \ \psi^* (0) |0 \rangle
\Bigr]
\nonumber \\
 &=&
e_0 \ \frac{1}{G^{(2)}(p) G^{(2)}(p')}
\int d^4 x d\tau d^4 x' d\tau '
\left[
\ e^{-i[p\cdot x' -P \tau ']} e^{-i[q\cdot x +\sigma Q \tau]}
\right]
\nonumber \\ &&\mbox{\qquad}
\Bigl[ -i\sigma Q
\ \langle 0 |{\rm T} j^5 (x,\tau) \ \psi (x',\tau ')
\ \psi^* (0) |0 \rangle
\nonumber \\ &&\mbox{\qquad}
- \delta(\tau-\tau') \delta^4(x-x') G^{(2)} (x,\tau )
+ \delta(\tau)\delta^4(x) G^{(2)} (x',\tau')
\Bigr]
\nonumber \\
\label{eqn:with_alg}
\end{eqnarray}
where we have performed one integration by parts in the last
line.  Now, carrying out the integrations, we find that
\begin{eqnarray}
\int d^4 x d\tau d^4 x' d\tau '&&\mbox{\hspace{-.4 in}}
\left[
\ e^{-i[p\cdot x' -P \tau ']} e^{-i[q\cdot x +\sigma Q \tau]}
\right]
\delta(\tau-\tau') \delta^4(x-x') G^{(2)} (x,\tau )
\nonumber \\
&=&
\int d^4 x d\tau \ e^{-i[(q+p)\cdot x -(P-\sigma Q) \tau]}
G^{(2)} (x,\tau )
\nonumber \\
&=& G^{(2)} (q+p,P-\sigma Q)
\nonumber \\
&=& G^{(2)} (p',P')
\label{eqn:first}
\end{eqnarray}
and
\begin{eqnarray}
\int d^4 x d\tau d^4 x' d\tau '
\left[
\ e^{-i[p\cdot x' -P \tau ']} e^{-i[q\cdot x +\sigma Q \tau]}
\right]
\delta(\tau) \delta^4(x) G^{(2)} (x',\tau')&=&
G^{(2)} (p,P) \ .
\label{eqn:second}
\end{eqnarray}
We notice that the remaining Green's function is proportional
to the vertex function for the 4-particle interaction,
\begin{eqnarray}
\Gamma^{(4)} (p,P;q,Q) &=&  \frac{1}{G^{(2)}(p) G^{(2)}(p')}
\int d^4 x d\tau d^4 x' d\tau '
e^{-i[p\cdot x' -P \tau ']} e^{-i[q\cdot x +\sigma Q \tau]}
\nonumber \\ &&\mbox{\quad}
\langle 0 |{\rm T} \Bigl( -i\frac{e^2_0}{2M}
j^5(x,\tau)\Bigr) \psi (x',\tau ')
\ \psi^* (0) |0 \rangle \ .
\label{eqn:4-pt}
\end{eqnarray}
Thus, we find
\begin{eqnarray}
q^\mu \Gamma^{(3)}_\mu (p,P;q,Q) &=&
e_0 \ \frac{1}{G^{(2)}(p) G^{(2)}(p')}
\int d^4 x d\tau d^4 x' d\tau '
\left[
\ e^{-i[p\cdot x' -P \tau ']} e^{-i[q\cdot x +\sigma Q \tau]}
\right]
\nonumber \\ &&\mbox{\qquad}
(-i\sigma Q) \ \langle 0 |{\rm T} j^5 (x,\tau) \ \psi (x',\tau ')
\ \psi^* (0) |0 \rangle
\nonumber \\ &&\mbox{\quad}
+ e_0 \frac{1}{G^{(2)} (p',P')} - e_0 \frac{1}{G^{(2)} (p,P)}
\nonumber \\
&=& e_0 \ (-i\sigma Q) (-\frac{2M}{ie_0^2}) \Gamma^{(4)} (p,P;q,Q,k,k')
\nonumber \\&&\mbox{\qquad}
+ e_0 \frac{1}{G^{(2)} (p',P')} - e_0 \frac{1}{G^{(2)} (p,P)}
\nonumber \\
&=& \sigma Q \frac{1}{e_0/(2M)} \Gamma^{(4)} (p,P;q,Q,k,k')
\nonumber \\&&\mbox{\qquad}
+ e_0 \frac{1}{G^{(2)} (p',P')} - e_0 \frac{1}{G^{(2)} (p,P)}
\label{eqn:near_final}
\end{eqnarray}
and the Ward identity takes the form,
\begin{equation}
e_0 \ q^\mu \Gamma^{(3)}_\mu (p,P;q,Q) -\sigma 
Q \ 2M \Gamma^{(4)} (p,P;q,Q,k,k')=
e^2_0 \left[ \frac{1}{G^{(2)} (p',P')} -
\frac{1}{G^{(2)}(p,P)}  \right] \ .
\label{eqn:final}
\end{equation}
As expected, in the case that $Q=0$,
(\ref{eqn:final}) reduces to the Ward identity for
on-shell (Klein-Gordon) scalar QED \cite{Cheng-Li}.
At the tree level, where
\begin{equation}
\Gamma^{(4)} (p,P;q,Q,k,k') = \frac{-ie_0^2}{2M} \quad
G^{(2)} (p,P) = \frac{-i}{\frac{p^2}{2M} - P} \quad
\Gamma^{(3)}_\mu (p,P;q,Q) = i\frac{e_0}{2M}(p+p')_\mu
\label{eqn:tree_vals}
\end{equation}
we verify that (\ref{eqn:final}) is satisfied:
\begin{eqnarray}
e_0(p'-p)^\mu \cdot \frac{ie_0}{2M}(p'+p)_\mu
- \sigma Q \ M \cdot \frac{-ie_0^2}{M}
- e_0^2 \left[i(\frac{p^{\prime 2}}{2M} - P') -
i(\frac{p^2}{2M} - P) \right] =
\nonumber \\
i \frac{e^2_0}{2M} \left( p^{\prime 2} - p^2 \right)
+ i e_0^2 (P-P') 
- e^2_0 \left[ i(\frac{p^{\prime 2}}{2M} - P') -
i(\frac{p^2}{2M} - P) \right] = 0
%
\label{eqn:tree}
\end{eqnarray}

Since the invariance of the Lagrangian is not changed by
multiplying each invariant term by a constant, the most general
gauge invariant form of the Lagrangian of
(\ref{eqn:2.1}), written in terms of renormalized quantities
is given by
\begin{eqnarray}
{\cal L} &=& Z_2 \; \psi^* \left(i\partial_\tau +(\frac{Z_1}{Z_2})
\; e_0 a_5 \right) \psi
\nonumber \\&&\mbox{\qquad}
- \frac{1}{2 Z_4 M} Z_2 \;
\psi^* \left(-i\partial_\mu - (\frac{Z_1}{Z_2})
\; e_0 a_\mu \right) \left(-i\partial^\mu - (\frac{Z_1}{Z_2})\;
e_0 a^\mu \right)\psi
\nonumber \\&&\mbox{\qquad}
- \frac{\lambda}{4} \left[
Z_3 \; f_{\mu\nu}f^{\mu\nu}
+2 Z_5 \; f_{5\nu}f^{5\nu} \right]
\label{eqn:fin-form}
\end{eqnarray}
where we have written the bare field operators in terms of the
renormalized field operators as
\begin{equation}
f^{\mu\nu}_{\rm B} = Z_3 f^{\mu\nu} \qquad
a^\mu_{\rm B} = Z_3^{1/2} a^\mu \qquad\qquad
f^{5\nu}_{\rm B} = Z_5 f^{5\nu} \qquad
a^5_{\rm B} = Z_5^{1/2} a^5
\label{eqn:B-def1}
\end{equation}
\begin{equation}
\psi_{\rm B} = Z_2^{1/2} \psi \ .
\label{eqn:B-def2}
\end{equation}
Any renormalization of the coupling $\lambda$ may be
absorbed into the wave function renormalizations $Z_3$ and
$Z_5$.  Consistency requires that
\begin{equation}
(\frac{Z_1}{Z_2}) \; e_0 a_5 =
(\frac{Z_1}{Z_2}) \; e_0 Z_5^{1/2} a_5 \equiv
e_0^{\rm B} a_5^{\rm B}
\quad {\rm and} \quad
(\frac{Z_1}{Z_2}) \; e_0 a_\mu =
(\frac{Z_1}{Z_2}) \; e_0 Z_3^{1/2} a_\mu \equiv 
e_0^{\rm B} a_\mu^{\rm B} 
\label{eqn:req1}
\end{equation}
so we must have
\begin{equation}
Z_5=Z_3 \qquad\qquad e_0 = \frac{Z_2 Z_3^{1/2}}{Z_1}
e_0^{\rm B} \ .
\label{eqn:req2}
\end{equation}
We may write the bare Green's functions in terms of the
renormalized Green's functions as
\begin{eqnarray}
\Gamma^{(3)}_{\mu \; {\rm B} } &=&
\frac{1}{\langle 0 |{\rm T} a_{\rm B} a_{\rm B} | 0 \rangle }
\ \frac{1}{\langle 0 |{\rm T} \psi_{\rm B} \psi^*_{\rm B}
 | 0\rangle }
\ \frac{1}{\langle 0 |{\rm T} \psi_{\rm B} \psi^*_{\rm B} | 0 \rangle }
\ \langle 0 |{\rm T} a_{\rm B} \psi_{\rm B} \psi^*_{\rm B} | 0 \rangle 
\nonumber \\
&=& \frac{1}{Z_3\langle 0 |{\rm T} \ a \ a | 0 \rangle }
\ \frac{1}{Z_2\langle 0 |{\rm T} \ \psi \ \psi^* | 0 \rangle }
\ \frac{1}{Z_2\langle 0 |{\rm T} \ \psi \ \psi^* | 0 \rangle }
\ Z^{1/2}_3 Z_2\langle 0 |{\rm T} \ a \psi \ \psi^* | 0 \rangle 
\nonumber \\
&=& \frac{1}{Z_3^{1/2} Z_2} \Gamma^{(3)}_\mu 
\label{eqn:g-3_ren}
\end{eqnarray}
and similarly
\begin{equation}
\Gamma^{(4)}_{\rm B} = \frac{1}{Z_3 Z_2} \Gamma^{(4)} \ .
\label{eqn:g-4_ren}
\end{equation}
Since the Ward identity must be valid for the renormalized
quantities as well as the unrenormalized quantities, we may
compare
%
\begin{eqnarray}
e_0 \ q^\mu \Gamma^{(3)}_\mu (p,P;q,Q) &-&\sigma 
Q \ 2M \Gamma^{(4)} (p,P;q,Q,k,k') =
\nonumber \\&&
e^2_0 \left[ \frac{1}{G^{(2)} (p',P')} -
\frac{1}{G^{(2)}(p,P)}  \right]
\nonumber \\ \nonumber \\
\frac{Z_1 Z_3^{1/2}}{Z_2} e_0^{\rm B} \ q^\mu 
Z_3^{1/2} Z_2 \Gamma^{(3)}_{\mu \; {\rm B}}  (p,P;q,Q)
&-&\sigma Q \ 2 Z_4 M_{\rm B} Z_2 Z_3
\Gamma^{(4)}_{\rm B}(p,P;q,Q,k,k') =
\nonumber \\&&
(\frac{Z_1 Z_3^{1/2}}{Z_2} e_0^{\rm B})^2 Z_2
\left[ \frac{1}{G^{(2)}_{\rm B} (p',P')} -
\frac{1}{G^{(2)}(p,P)}_{\rm B} \right]
\nonumber \\ \nonumber \\
Z_1 Z_3 e_0^{\rm B} \ q^\mu \Gamma^{(3)}_{\mu \; {\rm B}}  (p,P;q,Q)
&-& Z_2 Z_3 Z_4 \sigma Q \ 2 M_{\rm B}
\Gamma^{(4)}_{\rm B}(p,P;q,Q,k,k') =
\nonumber \\&&
\frac{Z_1^2 Z_3}{Z_2} (e_0^{\rm B})^2
\left[ \frac{1}{G^{(2)}_{\rm B} (p',P')} -
\frac{1}{G^{(2)}(p,P)}_{\rm B} \right]
\nonumber \\ \nonumber \\
e_0^{\rm B} \ q^\mu \Gamma^{(3)}_{\mu \; {\rm B}} (p,P;q,Q)
&-& \frac{Z_2 Z_4}{Z_1} \sigma Q \ 2 M_{\rm B}
\Gamma^{(4)}_{\rm B}(p,P;q,Q,k,k') =
\nonumber \\&&
\frac{Z_1}{Z_2} (e_0^{\rm B})^2
\left[ \frac{1}{G^{(2)}_{\rm B} (p',P')} -
\frac{1}{G^{(2)}(p,P)}_{\rm B} \right]
\label{eqn:w-bare}
\end{eqnarray}
and find that
\begin{equation}
Z_1 \equiv Z_2 \qquad \qquad Z_4 \equiv 1 \ .
\label{eqn:Zs}
\end{equation}
Notice that although the charge $e_0$ appears linearly and
quadratically in the Ward identity, (\ref{eqn:w-bare}) makes
no restriction on $Z_3$, which determines the charge
renormalization (by (\ref{eqn:req2}) with $Z_1=Z_2$).  Thus,
the appearance of $\Gamma^{(4)}$ in the Ward identity does
not change the universality of charge renormalization found
in on-shell QED since the conserved five-current is a consequence
of gauge invariance.  Nevertheless, since there are no
matter field loops in off-shell QED, there are no possible
contributions to photon renormalization, and we take $Z_3
\equiv 1$.  Therefore, the charge $e_0$ is not renormalized.
\newpage
{\bf Renormalizability}

The remaining renormalization factor is $Z_2$,
which derives from the renormalization of the matter field by
photon loops.  These photon loops (matter field self-energy
diagrams) will also contribute to the mass renormalization
of the matter field,
however, the mass term ($\psi^* i\partial_\tau \psi$) will
absorb these contributions.  Nevertheless, by examining the
primitive divergent self-energy diagrams of successively
higher order, it may be seen that in order to make the
theory counter-term renormalizable, a cut-off must be
applied to the integrations over mass in loop diagrams.
For example, at second order, the self-energy diagram with
two overlapping photon loops is given by
\begin{eqnarray}
G^{(2)}_{2} (p) &=& G^{(2)}_{0} (p) \ G^{(2)}_{0} (p) 
\int d^4q dQ d^4q' dQ'
\left(
(2\pi)^5 \frac{ie_0}{2M}
\right)^4
(2p-q)_\mu \ (2p-q-2q')_\nu \ d^{\mu\nu} (q)
\nonumber \\ &&\mbox{\qquad}
(2p-2q-q')_\lambda \ (2p-q')_\sigma \ d^{\lambda\sigma} (q')
\nonumber \\ &&\mbox{\qquad}
G^{(2)}_{0} (p-q) \ G^{(2)}_{0} (p-q') \ G^{(2)}_{0} (p-q-q') \ .
\label{eqn:2nd-order}
\end{eqnarray}
This expression contains the following term proportional to $p^4$
\begin{eqnarray}
&& G^{(2)}_{0} (p) \ G^{(2)}_{0} (p) 
\left(
(2\pi)^5 \frac{ie_0}{2M}
\right)^4
 16 p^4 \int d^4q dQ d^4q' dQ'
\nonumber \\&&\mbox{\qquad}
\frac{1}{\lambda^2} \frac{1}{q^2+\sigma Q^2}
\ \frac{1}{q^{\prime 2}+\sigma Q^{\prime 2}}
\ \frac{2M}{(p-q)^2 -2M(P+\sigma Q)}
\nonumber \\&&\mbox{\qquad}
\frac{2M}{(p-q')^2 -2M(P+\sigma Q')}
\ \frac{2M}{(p-q-q')^2 -2M[P+\sigma (Q+Q')]}
\label{eqn:p4-term}
\end{eqnarray}
and if the mass integrations on $dQ$ and $dQ'$ are taken to
infinity, then this term diverges.  Since the term is proportional
to $p^4$, it may not be renormalized by a counter term of a form
which appears in the original Lagrangian.  However, if the mass
integrations are cut off at a finite limit, then the integral is
seen to converge.  Moreover, the set of divergent diagrams in the
resulting theory is just the subset of the divergent diagrams in
the on-shell theory which contain no matter field loops.
Therefore, with the mass integrations made finite, the off-shell
theory is seen to be renormalizable by the same arguments given
by Rohrlich \cite{Rohrlich} for the on-shell theory.  Notice that
unlike a momentum cut-off, the mass cut-off does not affect the
invariances of the original theory.  The presense of a finite
mass cut-off has a natural interpretation in terms of
the correlation limit found by Shnerb and Horwitz \cite{nadav}.
Frastai and Horwitz \cite{jaime} showed that in the limit of
zero-mass photons, the off-shell theory contains a natural
regularization of the highest order singularities by an
effective Pauli-Villars mass integration \cite{P-V}.

Finally, the one-loop renormalization of the Green's function
by two double-photon vertices must also be mentioned.  This
diagram contributes the logarithmic divergence,
\begin{eqnarray}
G^{(4)}_2 (p_1,p_2,p_3,p_4) &=&  G^{(2)}_0 (p_1) \ G^{(2)}_0 (p_2)
\ G^{(2)}_0 (p_3) \ G^{(2)}_0 (p_4)
\nonumber \\&&\mbox{\qquad}
2 \int d^4q dQ
\left(\frac{-ie_0^2}{M} (2\pi)^5 \right)^2
d^{\mu\nu}_0 (q) \ d_{\mu\nu}^0 \ (p_2-p_1+q)
\nonumber \\
\label{eqn:g-4}
\end{eqnarray}
and as discussed by Rohrlich \cite{Rohrlich} for on-shell scalar
QED, renormalization by counter term requires the introduction of
a direct 4-point interaction term for the matter field.  However,
unlike the usual $\phi^4$ theories, this interaction does
not introduce matter field loops, because of the retarded
propagation in $\tau$.
It may be checked that the 6-point and higher
diagrams are finite.
Since only the photon loops must be considered, the
primitive divergent diagrams of the theory contribute only to the
matter field self-energy (mass and wavefunction renormalization)
and the 4-point interaction term for the matter field.

\addcontentsline{toc}{section}{References}
%
%

%

\end{document}